%% file: Thesis.tex
\documentclass[a4paper,12pt,twoside,notitlepage,openany]{book}
\usepackage[tmargin=3cm,lmargin=3cm,rmargin=2cm,
bmargin=2cm]{geometry}

\usepackage{setspace}
\usepackage{layout}
\usepackage[latin1]{inputenc}
\usepackage{amsmath}
\usepackage{amssymb}
\usepackage{float}
\usepackage{graphicx,color}
\usepackage{siunitx}
\usepackage[labelfont=bf,small]{caption}
\usepackage{booktabs}
\usepackage{subfig}
\usepackage{pdfpages}
\usepackage[sort&compress,numbers]{natbib}
\usepackage[pdfpagelabels,colorlinks=false]{hyperref}
\usepackage{fancyhdr}
\usepackage{bm}
\usepackage{indentfirst}
\usepackage{gensymb}
\usepackage{url}

\usepackage{multirow} 

\graphicspath{{./Figuras/}}

\begin{document}


\input{Titlepage_portuguese}
\includepdf[pages=1]{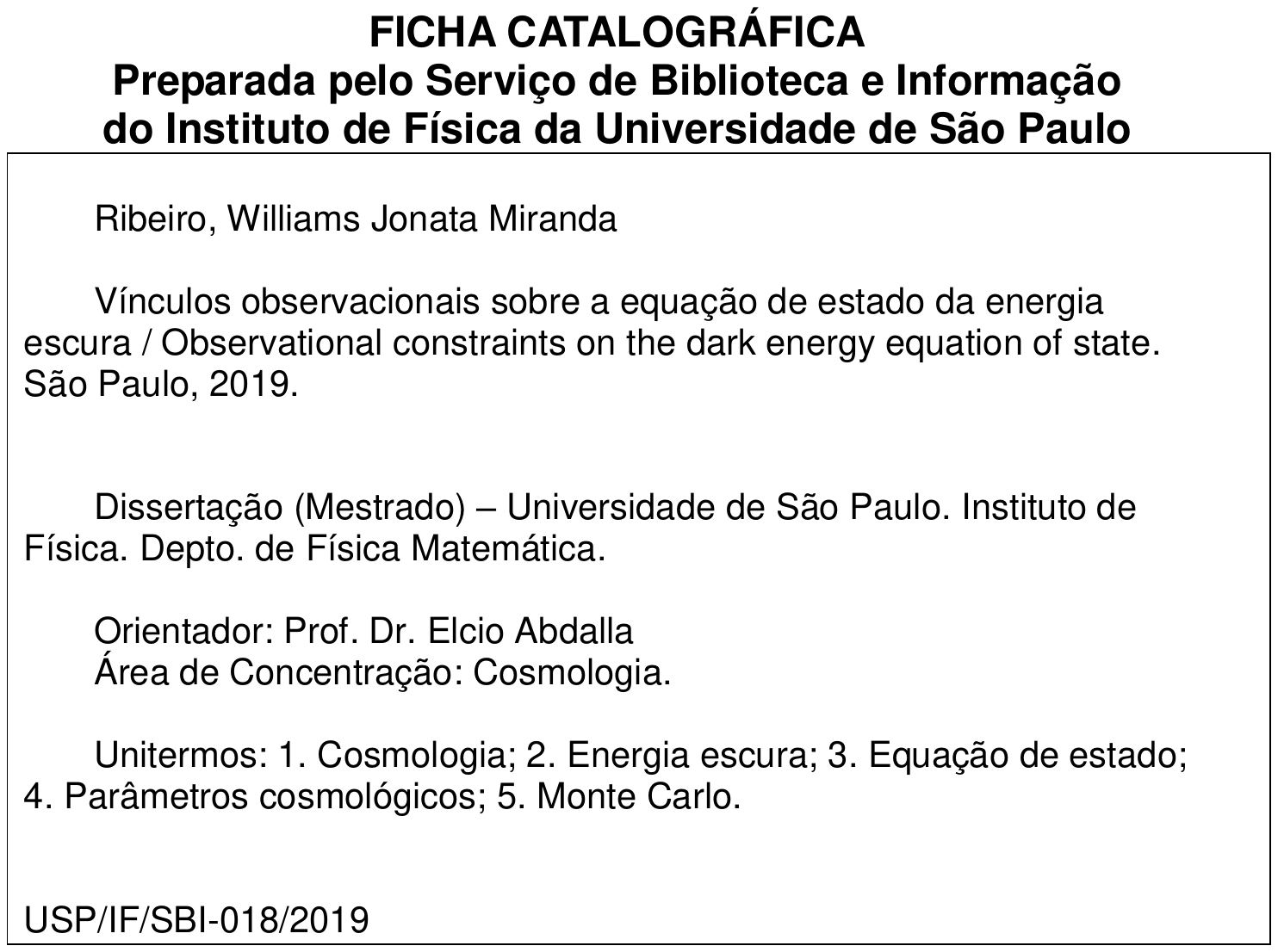}

\input{Titlepage}
\newpage
\mbox{}
\thispagestyle{empty}
\newpage
\thispagestyle{empty}

\pagenumbering{gobble}
\input{Acknowledgements}
\newpage
\mbox{}
\thispagestyle{empty}

\input{Abstract}
\newpage
\mbox{}
\thispagestyle{empty}

\input{Resumo}
\newpage
\mbox{}
\thispagestyle{empty}

\tableofcontents

\clearpage
\pagenumbering{arabic}
\input{Introduction}
\input{Chap1}

\newpage
\mbox{}
\thispagestyle{empty}
\input{Chap2}

\newpage
\mbox{}
\thispagestyle{empty}
\input{Chap3}

\input{Chap4}

\input{Conclusions}

{\setstretch{1.0}

\newpage
\renewcommand{\bibname}{References}
\bibliographystyle{apsrev4-1}
\bibliography{References}
}
\end{document}

%% file: Titlepage_portuguese.tex
\begin{titlepage}
\thispagestyle{empty}

\setlength{\voffset}{0pt}
\setlength{\hoffset}{0pt}

{\centering

{\LARGE \textsc Universidade de S\~ao Paulo}

\vspace{\stretch{0.05}}

{\LARGE \textsc Instituto de F\'isica}

\vspace{\stretch{0.8}}

{\huge \textbf{V\'inculos Observacionais sobre a Equa\c{c}\~ao de
Estado da Energia Escura}}

\vspace{\stretch{0.1}}

{\Large \textsc Williams Jonata Miranda Ribeiro}

}
\vspace{\stretch{0.5}}

\hspace{7cm}{\textsc{\bf Orientador:} Prof. Dr. Elcio Abdalla}

\vspace{\stretch{0.1}}

\hspace{7cm}{
\begin{minipage}{8cm}
{\textsc Disserta\c{c}\~ao de mestrado apresentada ao Instituto
 de F\'isica para a obten\c{c}\~ao do t\'itulo de Mestre em Ci\^encias}
\end{minipage}
}

\vspace{\stretch{0.3}}

\begin{flushleft}
{\bf Banca Examinadora:} \\

\vspace{\stretch{0.1}}

Prof. Dr. Elcio Abdalla (IF-USP) \\
Prof. Dr. Gast\~ao Cesar Bierrenbach Lima Neto (IAG-USP) \\
Prof. Dr. Carlos Alexandre Wuensche de Souza (INPE) \\
\end{flushleft}

\vspace{\stretch{0.5}}
{\centering

S\~ao Paulo \\
2018

}

\end{titlepage}

%% file: Titlepage.tex
\begin{titlepage}
\thispagestyle{empty}

\setlength{\voffset}{0pt}
\setlength{\hoffset}{0pt}

{\centering

{\LARGE \textsc University of S\~ao Paulo}

\vspace{\stretch{0.05}}

{\LARGE \textsc Institute of Physics}

\vspace{\stretch{0.8}}

{\huge \textbf{Observational Constraints on the Dark Energy Equation of State}}

\vspace{\stretch{0.1}}

{\Large \textsc Williams Jonata Miranda Ribeiro}

}
\vspace{\stretch{0.5}}

\hspace{7cm}{\textsc{\bf Advisor:} Prof. Dr. Elcio Abdalla}

\vspace{\stretch{0.1}}

\hspace{7cm}{
\begin{minipage}{8cm}
{\textsc Dissertation presented to the Institute of Physics of the University of S\~ao Paulo in partial fulfillment of the requirements for the degree of Master of Science}
\end{minipage}
}

\vspace{\stretch{0.3}}

\begin{flushleft}
{\bf Examining Committee:} \\

\vspace{\stretch{0.1}}

Prof. Dr. Elcio Abdalla (IF-USP) \\
Prof. Dr. Gast\~ao Cesar Bierrenbach Lima Neto (IAG-USP) \\
Prof. Dr. Carlos Alexandre Wuensche de Souza (INPE) \\
\end{flushleft}

\vspace{\stretch{0.5}}
{\centering

S\~ao Paulo \\
2018

}

\end{titlepage}

%% file: Acknowledgements.tex
\newpage
\thispagestyle{empty}
\chapter*{Acknowledgments}
\markboth{ACKNOWLEDGMENTS}{}
First, I would like to thank God for renewing my strength in the most difficult moments I had been through in my master's. Without His power, I would not have gone so far.

Also, I would like to thank my advisor Dr. Elcio Abdalla for accepting me to be part of his research group, even though in the beginning I had no prior knowledge about general relativity and cosmology at all.

I am very thankful to my whole family for all the support during these years of study and specially to my father Francisco Junior for his encouragement, which was essential for my personal growth.

I could not forget to mention the friends I made during these years in S\~ao Paulo (and the ones I already knew) who have assisted me in many ways in my journey, helping me either with physics or with conversations and laughs that kept me going: Bonif\'acio Lima, Christianne Moraes, Kariny Bonjorno, Naim Comar, Renan Boschetti, Riis Rhavia, Rodrigo Pinheiro and Thiago Silveira.

I would like to specially thank the postdoc Andr\'e Costa for all the assistance, tips, discussions and even time invested on my work that made me to progress so much. A gratefully thanks to my friend Caroline Guandalin for the interesting discussions about cosmology and for lending me her computer to perform the simulations for my work (thank you very much!!!). Also, thanks to the colleagues Alessandro Marins and Leonardo Werneck for the programming tips that helped me to improve my code and my computer skills.

My sincere acknowledgement to the science agency CNPq, Conselho Nacional de Desenvolvimento Cient\'ifico e Tecnol\'ogico - Brasil. This dissertation would not be possible without its financial support.

%% file: Abstract.tex
\newpage
\thispagestyle{empty}
\chapter*{Abstract}
Late-time cosmic acceleration is one of the most interesting unsolved puzzles in modern cosmology. The explanation most accepted nowadays, dark energy, raises questions about its own nature, e.g. what exactly is dark energy, and implications to the observations, e.g. how to handle fine tuning problem and coincidence problem. Hence, dark energy evolution through cosmic history, together with its equation of state, are subjects of research in many current experiments. In this dissertation, using Markov Chain Monte Carlo sampling, we try to constrain the evolution of the dark energy equation of state in a nearly model-independent approach by combining different datasets coming from observations of baryon acoustic oscillations, cosmic chronometers, cosmic microwave background anisotropies and type Ia supernovae. We found no strong evidence that could indicate deviations from $\Lambda$CDM model, which is the standard model in cosmology accepted today.

\vspace{1cm}
\noindent {\bf Keywords:} Cosmology. Dark Energy. Equation of State. Cosmological Parameters. Monte Carlo.

%% file: Resumo.tex
\newpage
\thispagestyle{empty}
\chapter*{Resumo}
A acelera\c{c}\~ao c\'osmica atual \'e um dos mais interessantes enigmas n\~ao resolvidos da cosmologia moderna. A explica\c{c}\~ao mais aceita hoje em dia, a energia escura, levanta quest\~oes acerca de sua pr\'opria natureza, como o que \'e exatamente a energia escura, e as implica\c{c}\~oes para as observa\c{c}\~oes, por exemplo como lidar com o problema do ajuste fino e o problema da coincid\^encia. Por isso, a evolu\c{c}\~ao da energia escura durante a hist\'oria c\'osmica, juntamente com sua equa\c{c}\~ao de estado, s\~ao objetos de pesquisa em in\'umeros experimentos atuais. Nesta disserta\c{c}\~ao, usando amostragem por cadeias de Markov de Monte Carlo, tentamos restringir a evolu\c{c}\~ao da equa\c{c}\~ao de estado da energia escura em uma abordagem quase independente de modelo ao combinar diferentes conjuntos de dados provenientes de observa\c{c}\~oes de oscila\c{c}\~oes ac\'usticas de b\'arions, cron\^ometros c\'osmicos, anisotropias da radia\c{c}\~ao c\'osmica de fundo e supernovas tipo Ia. N\~ao encontramos evid\^encias fortes que pudessem indicar desvios do modelo $\Lambda$CDM, o qual \'e o modelo padr\~ao aceito hoje na cosmologia.

\vspace{1cm}
\noindent {\bf Palavras-Chave:} Cosmologia. Energia Escura. Equa\c{c}\~ao de Estado. Par\^ametros Cosmol\'ogicos. Monte Carlo.

%% file: Introduction.tex
\newpage
\chapter*{Introduction}
\markboth{INTRODUCTION}{}
\addcontentsline{toc}{chapter}{Introduction}

In 1998, a remarkable discovery set up a new era in modern cosmology: through observations of type Ia supernovae, it was realized that the Universe is going through a phase of accelerated expansion \cite{riess,perlmutter}. As our common sense leads us to the idea that the expansion is decelerated due to the effects of gravitation (which was the general belief until 1997), many theories were (and are still being) developed in order to explain this acceleration. Among them, the currently accepted theory is that there exists some kind of ``exotic fluid'' with negative pressure, named \textit{dark energy}, that contributes to approximately $70\%$ of our Universe and that is the responsible for this accelerated phase. Other explanations include models of dark energy with repulsive gravity \cite{barrow}, modifications to general relativity \cite{dvali}, models of vacuum decaying \cite{LANDIM2017,alcaniz}, interacting dark matter-dark energy models \cite{abdalla,Wang2016,Costa2017,Marcondes2016}, and many others as explored in \cite{Copeland2006}. Nowadays, besides type Ia supernovae, many kinds of independent observations confirm the accelerated phase: cosmic microwave background anisotropies \cite{planck}, large-scale structure \cite{tegmark}, baryon acoustic oscillations \cite{eisenstein}, cosmic chronometers \cite{stern}, etc. Thus, in this work we are going to assume dark energy as the correct explanation for the current Universe dynamics.

One of the ways of describing dark energy is considering it as a barotropic fluid (a fluid whose density depends only on its pressure) with equation of state $w$, which is defined as the ratio between the pressure and the density of the fluid. The model that better matches the observations is named \textit{$\Lambda$CDM model}, in which $w=-1$ during all cosmic history. Despite its success in explaining many independent measurements, it is still not the final history for the fact that it does not solve problems like the \textit{cosmological constant problem} \cite{Weinberg1989} (also called fine tuning problem) and the \textit{coincidence problem} \cite{Chimento2003}. Other well known models in literature are the \textit{$w$CDM model}, which assigns a value different of $-1$ but still constant for $w$ and the \textit{CPL model} (Chevallier-Polarski-Linder), which considers $w$ as a dynamical dark energy equation of state that varies with time \cite{chevallier,linder}. There are also parametrizations that try to capture the evolution of $w$ either by performing Taylor expansions or by using something called \textit{kink parametrization}, which takes into account rapid transitions of dark energy equation of state (something not allowed by a conventional Taylor expansion) \cite{Stefano2004,Johri2004}. Several constraints in these models can be found in \cite{Tripathi2017,Chavez2016,Dinda2018,Demianski2018}.

There is no fundamental theory that can predict whether dark energy equation of state has a static value or evolves with time. In addition to the fact that this quantity cannot be measured directly, strong degeneracy with cosmological parameters makes its evolution hard to predict, specially for high redshifts \cite{ichikawa}. Also, although one can write the dark energy equation of state as a function of the background expansion (and consequently as a function of luminosity distance), accurate measurements of the latter coming from cosmic chronometers and type Ia supernovae cannot give good constrains on $w$ because in this approach there is an additional dependence on the derivative of the background expansion and because we need estimates of the present density of pressureless matter that only comes from large-scale structure methods, which actually measures the present density of clustered matter (both kinds of matter are not necessarily the same) \cite{Copeland2006,amendola}.

The importance of dark energy equation of state lies in the fact that its evolution determines the current state of the Universe, e.g. an accelerated expansion requires $w<-1/3$ \cite{amendola}, its eventual fate, e.g. a \textit{phantom dark energy} defined by $w<-1$ leads to a finite-time singularity called \textit{big-rip} singularity \cite{Caldwell2003}, and the evolution of dark energy itself via continuity equation. Nowadays, the main collaboration for studying the nature of dark energy is the Dark Energy Survey (DES) research project \cite{DES}, but there are other projects being designed for studying the dark sector such as Euclid \cite{Racca2016}, BINGO \cite{Battye2012,Dickinson2014}, J-PAS \cite{Benitez2014} and DESI \cite{Levi2013}.

Instead of testing specific models, a more robust way of extracting information from the evolution of $w$ is using different observational datasets in order to reconstruct the evolution of the dark energy equation of state in a model-independent approach (as much as possible) and check if the datasets lead to an agreement with $\Lambda$CDM model or not. To this end, we are going to use a method known as \textit{principal components analysis} (PCA) \cite{cooray,Hamilton2000,Huterer2003} along with MCMC sampling applied to several datasets in order to obtain estimates for dark energy equation of state in a few redshifts.

%% file: Chap1.tex
\newpage
\chapter{Overview}
\chaptermark{Overview}

The initial chapter defines the general terms and concepts used in modern cosmology by using the framework of the Friedmann-Lema\^itre-Robertson-Walker (FLRW) metric, which agrees with observations that display the Universe as homogeneous and isotropic at large scales. We also construct the tools in such a way that the need for dark matter and dark energy becomes clear for explaining many independent measurements.

\section{Homogeneous and isotropic Universe}
We consider the most general line element that describes a homogeneous and isotropic 4-dimensional spacetime, the FLRW metric, given by (using units of $c=1$) \cite{amendola}
\begin{equation}
\label{flrwmetric}
ds^2 = g_{\mu\nu} dx^\mu dx^\nu = -dt^2 + a^2(t) d\sigma^2,
\end{equation}
where $g_{\mu\nu}$ is the metric tensor, $a(t)$ is the scale factor which depends on the cosmic time $t$ and $d\sigma^2$ represents the spatial part of the FLRW metric with constant curvature K, given by
\begin{equation}
\label{line_element}
d\sigma^2 = \gamma_{ij} dx^i dx^j = \frac{dr^2}{1-Kr^2} + r^2(d\theta^2+\sin^2\theta d\phi^2).
\end{equation}
For curvature, $K=-1$, $K=0$ and $K=+1$ represent open, flat and closed geometries, respectively. We are also using spherical coordinates $(x^1, x^2, x^3)=(r,\theta,\phi)$ and Einstein's summation convention which states that terms with equal upper and lower indices are summed over.

Besides cosmic time $t$, we also define another useful ``time'' named \textit{conformal time} $\eta$ and defined by
\begin{equation}
\eta \equiv \int_0^t \, \frac{dt'}{a(t')},
\end{equation}
which is the amount of time a photon would take to travel from where we are to the furthest observable distance provided the universe ceased expanding. It is interesting sometimes to work with the conformal time instead of cosmic time because it may simplify the evolution equations. From now on, a prime ($'$) indicates derivative with respect to the conformal time.

The Universe evolution dynamics is described by the \textit{Einstein equations}. The first step to obtain them is to evaluate the connections from the metric tensor $g_{\mu\nu}$,
\begin{equation}
\Gamma^{\mu}_{\hphantom{\mu}\nu\lambda} = \frac{1}{2} g^{\mu\alpha} (g_{\alpha\nu,\lambda}+g_{\alpha\lambda,\nu}-g_{\nu\lambda,\alpha}).
\end{equation}

We also define the Ricci tensor and the Ricci scalar (or scalar curvature), respectively, as
\begin{equation}
\label{riccitensor}
R_{\mu\nu} = \Gamma^{\alpha}_{\hphantom{\alpha}\mu\nu,\alpha} - \Gamma^{\alpha}_{\hphantom{\alpha}\mu\alpha,\nu} + \Gamma^{\alpha}_{\hphantom{\alpha}\mu\nu}\Gamma^{\beta}_{\hphantom{\beta}\alpha\beta} - \Gamma^{\alpha}_{\hphantom{\alpha}\mu\beta}\Gamma^{\beta}_{\hphantom{\beta}\alpha\nu}
\end{equation}
and
\begin{equation}
\label{ricciscalar}
R = g^{\mu\nu} R_{\mu\nu}.
\end{equation}

From \eqref{riccitensor} and \eqref{ricciscalar} we can evaluate the Einstein tensor
\begin{equation}
\label{einsteintensor}
G_{\mu\nu} \equiv R_{\mu\nu} - \frac{1}{2}g_{\mu\nu}R,
\end{equation}
from where we obtain the cosmological dynamics by solving the Einstein equations
\begin{equation}
\label{einsteineq}
G^{\mu}_{\hphantom{\mu}\nu} = 8\pi G T^{\mu}_{\hphantom{\mu}\nu},
\end{equation}
where $T^{\mu}_{\hphantom{\mu}\nu}$ represents the energy-momentum tensor of matter components.

From the FLRW metric, the only non-vanishing connection coefficients are
\begin{align}
&\Gamma^{0}_{\hphantom{0}ij} = a^2H\gamma_{ij}, \,\,\, \Gamma^{i}_{\hphantom{i}0j} = \Gamma^{i}_{\hphantom{i}j0} = H \delta^{i}_{\hphantom{i}j}, \nonumber \\
&\Gamma^{1}_{\hphantom{1}11} = \frac{Kr}{1-Kr^2}, \,\,\, \Gamma^{1}_{\hphantom{1}22} = -r(1-Kr^2), \,\,\, \Gamma^{1}_{\hphantom{1}33} = -r(1-Kr^2)\sin^2\theta, \\
&\Gamma^{2}_{\hphantom{2}33} = -\sin\theta\cos\theta, \,\,\, \Gamma^{2}_{\hphantom{2}12} = \Gamma^{2}_{\hphantom{2}21} = \Gamma^{3}_{\hphantom{3}13} = \Gamma^{3}_{\hphantom{3}31} = \frac{1}{r}, \,\,\, \Gamma^{3}_{\hphantom{3}23} = \Gamma^{3}_{\hphantom{3}32} = \cot\theta \nonumber,
\end{align}
where $H\equiv \dot{a}/a$, a dot representing a derivative with respect to cosmic time $t$. The quantity $H$, called Hubble parameter, describes the rate of expansion of the Universe.

Using \eqref{riccitensor}, \eqref{ricciscalar} and \eqref{einsteintensor}, we can evaluate the components from the Ricci tensor, Ricci scalar and the components from the Einstein tensor, which are given by, respectively,
\begin{equation}
\begin{aligned}\normalsize
&R_{00} = -3(H^2+\dot{H}), \,\,\, R_{0i} = R_{i0} = 0, \,\,\, R_{ij} = a^2(3H^2+\dot{H}+2K/a^2)\gamma_{ij}, \\
&R = 6(2H^2+\dot{H}+K/a^2),
\end{aligned}
\end{equation}
\begin{equation}
\label{Gs}
G^{0}_{\hphantom{0}0} = -3(H^2+K/a^2), \,\,\, G^{0}_{\hphantom{0}i} = G^{i}_{\hphantom{i}0} = 0, \,\,\, G^{i}_{\hphantom{i}j} = -(3H^2+2\dot{H}+K/a^2) \delta^{i}_{\hphantom{i}j}.
\end{equation}

Considering a FLRW space-time imposes a constraint on the energy-momentum tensor, which should be represented by a perfect fluid in the form
\begin{equation}
\label{energymomentum}
T^{\mu}_{\hphantom{\mu}\nu} = (\rho+P)u^\mu u_\nu + P\delta^{\mu}_{\hphantom{\mu}\nu},
\end{equation}
where $u^\mu = (-1,0,0,0)$ is the four-velocity of the fluid in comoving coordinates, with $\rho$ and $P$ representing the energy density and pressure, respectively. The $(00)$ and $(ii)$ components of the Einstein equations \eqref{einsteineq} can be obtained using \eqref{Gs} and \eqref{energymomentum}, from which one obtains the Friedmann equations
\begin{equation}
\label{friedmann1}
H^2 = \frac{8\pi G}{3}\rho - \frac{K}{a^2}
\end{equation}
and
\begin{equation}
3H^2+2\dot{H} = -8\pi GP - \frac{K}{a^2}.
\end{equation}

The importance of these equations lies in the fact that they relate the evolution dynamics of the Universe to its own constitution and space geometry. Combining these equations by eliminating the curvature term yields
\begin{equation}
\label{twodots}
\frac{\ddot{a}}{a} = -\frac{4\pi G}{3}(\rho+3P).
\end{equation}

Multiplying \eqref{friedmann1} by $a^2$, differentiating and using \eqref{twodots}, one yields
\begin{equation}
\label{continuityeq}
\dot{\rho}+3H(\rho+P)=0,
\end{equation}
which is the \textit{continuity equation} for the cosmological components. This equation may also be derived from relations satisfied by the Einstein tensor, the Bianchi identities
\begin{equation}
\nabla_\mu G^{\mu}_{\hphantom{\mu}\nu} \equiv \frac{\partial G^{\mu}_{\hphantom{\mu}\nu}}{\partial x^\mu} + \Gamma^{\mu}_{\hphantom{\mu}\alpha\mu}G^{\alpha}_{\hphantom{\alpha}\nu} - \Gamma^{\alpha}_{\hphantom{\alpha}\nu\mu}G^{\mu}_{\hphantom{\mu}\alpha} = 0,
\end{equation}
$\nabla_\mu$ denoting the covariant derivative, which leads to the conservation of the energy-momentum tensor $\nabla_\mu T^{\mu}_{\hphantom{\mu}\nu}=0$ that implies in \eqref{continuityeq}.

We may rewrite equation \eqref{friedmann1} in a way that makes it easier to study the evolution of cosmological parameters,
\begin{equation}
\label{parameters}
\Omega_M + \Omega_K = 1,\quad \text{where}\,\,\,\Omega_M \equiv \frac{8\pi G\rho}{3H^2},\,\,\, \Omega_K \equiv -\frac{K}{(aH)^2}.
\end{equation}
Equation \eqref{parameters} defines a constraint among cosmological parameters, such that the amount of each component in the Universe must respect it at any moment of the cosmic evolution \cite{hobson}.

Referring to values today with the superscript $(0)$, let us define the density parameters $\Omega$. They are given by relativistic particles, non-relativistic matter, dark energy and curvature, respectively, as
\begin{equation}
\label{cosmo_params}
\Omega_r^{(0)} = \frac{8\pi G\rho_r^{(0)}}{3H_0^2},\,\, \Omega_m^{(0)} = \frac{8\pi G\rho_m^{(0)}}{3H_0^2},\,\, \Omega_{DE}^{(0)} = \frac{8\pi G\rho_{DE}^{(0)}}{3H_0^2},\,\, \Omega_K^{(0)} = -\frac{K}{(a_0H_0)^2},
\end{equation}
in which we can identify relativistic particles as electromagnetic radiation ($\gamma$) and neutrinos ($\nu$), non-relativistic matter as baryons ($b$) and cold dark matter ($c$), and dark energy (which we shall introduce later) as a cosmological constant ($\Lambda$) or a more general fluid ($X$). These cosmological parameters are of great interest, because if we can accurately measure or infer them from observations then it might be possible to understand the current state of the Universe, its initial moments and predict its future.

\section{Cosmological redshift, Hubble law, critical density and equation of state}
At first glance, Friedmann equations give us no information about the evolution of the scale factor in the FLRW metric: is it increasing, decreasing or is it a constant? This information comes from observations of distant galaxies that are seen to recede from us based on the displacement of their spectral lines to the red part of the spectrum \cite{hubble}. Thereby, here we relate these shifts on the spectral lines to the scale factor.

In the FLRW metric, let us define our position at the center of coordinates, without loss of generality, and consider a light ray coming toward us in the radial direction. This light ray obeys the equation $ds^2=0$, such that \eqref{flrwmetric} implies
\begin{equation}
\label{lighray}
dt=\pm a(t) \frac{dr}{\sqrt{1-Kr^2}}.
\end{equation}
Considering a light ray coming towards us from a distant source at position $r_1$ at cosmic time $t_1$, as the distance decreases the cosmic time increases and therefore we must choose the minus sign in \eqref{lighray}. Defining our position as $r=0$ at time $t_0$, one finds
\begin{equation}
\label{differential}
\int_{t_1}^{t_0} \frac{dt}{a(t)} = \int_0^{r_1} \frac{dr}{\sqrt{1-Kr^2}}.
\end{equation}

Taking \eqref{differential} in a differential form and using the fact that the radial coordinate $r_1$  is time-independent, we see that the interval $\delta t_1$ between departure of subsequent light signals is related to the interval $\delta t_0$ between arrivals of these light signals by
\begin{equation}
\frac{\delta t_1}{a(t_1)} = \frac{\delta t_0}{a(t_0)}.
\end{equation}
Considering these ``signals'' as subsequent wave fronts, the emitted frequency is $\nu_1=1/\delta t_1$ and the observed frequency is $\nu_0=1/\delta t_0$, therefore
\begin{equation}
\frac{\nu_0}{\nu_1} = \frac{a(t_1)}{a(t_0)}.
\end{equation}

If the scale factor is increasing, then $\nu_0<\nu_1 \Rightarrow \lambda_0>\lambda_1$, showing that the wavelength of photons stretches during its travel to us due to the Universe expansion. This is called \textit{redshift} and is exactly what we observe today. Conventionally, the redshift $z$ is related to the scale factor by
\begin{equation}
\label{redshift}
z \equiv \frac{\lambda_0}{\lambda_1}-1= \frac{a(t_0)}{a(t_1)}-1.
\end{equation}

In an expanding Universe, the relation between the physical distance $\bm{r}$ and the comoving distance $\bm{x}$ to an object is given by $\bm{r}=a(t)\bm{x}$. Taking the derivative of the equation $\bm{r}=a(t)\bm{x}$ with respect to cosmic time $t$, yields
\begin{equation}
\label{rdot}
\dot{\bm{r}}=H\bm{r}+a\dot{\bm{x}}.
\end{equation}
The fist term in \eqref{rdot} represents the Hubble flow, while the second term represents movements with respect to the local Hubble flow and is called \textit{peculiar velocity} $\bm{v}_p$. In the special case in which an object moves exclusively due to the Hubble flow (that is, the object moves only because of the expansion), the comoving distance does not change and the second term vanishes. In the general case, the projection of the object velocity in the radial direction is given by
\begin{equation}
\label{vprojection}
v \equiv \dot{\bm{r}} \cdot \bm{r}/r = Hr+\bm{v}_p \cdot \bm{r}/r.
\end{equation}

\begin{figure}[h!]
\centering
\includegraphics[width=0.8\textwidth]{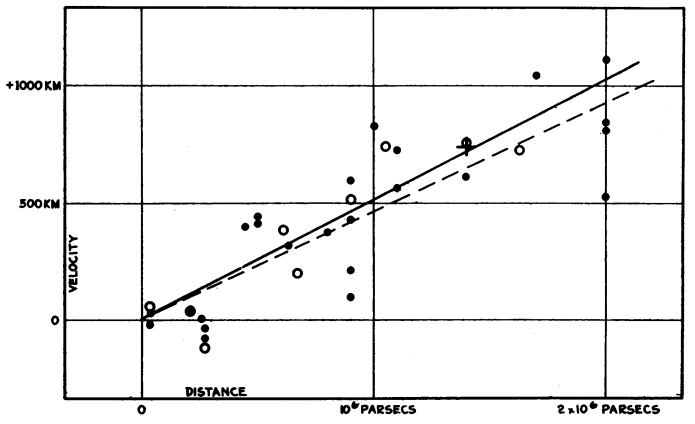}
\caption{Original Hubble results on the recessional velocity-distance relation for galaxies. Black dots and full line represent individual galaxies while circles and dashed line represent groups of galaxies. Extracted from \cite{hubble}.}
\label{hubble_law}
\end{figure}

In most cases the peculiar velocity of galaxies does not exceed $10^6 \, m/s$ \cite{amendola}. Therefore, if the second term in \eqref{vprojection} is negligible in comparison with the first one, then
\begin{equation}
\label{hubblelaw}
v \simeq H_0 r,
\end{equation}
in which the replacement of $H$ by the present value $H_0$ is justified in the limit of low redshifts ($z\ll 1$). Equation \eqref{hubblelaw} represents the \textit{Hubble law}, discovered in 1929 by Edwin Hubble when he plotted the recessional velocity $v$ of galaxies versus distance $r$ to them \cite{hubble}. Although his data were scarce and noisy, Hubble correctly concluded that the Universe was expanding. The Hubble parameter today $H_0$ is usually written in the form
\begin{equation}
H_0 = 100\, h\, km\, sec^{-1} Mpc^{-1},
\end{equation}
in which the parameter $h$ describes the uncertainty on the value of $H_0$ and we shall name it \textit{reduced Hubble parameter}. The original Hubble measurements are displayed in Fig. \ref{hubble_law}.

An important quantity of interest in cosmology is the \textit{critical density}, which is defined as the total energy density necessary for the Universe to be spatially flat \cite{hobson}. From \eqref{parameters}, one may rewrite it as
\begin{equation}
\Omega_M \equiv \Omega_m+\Omega_r+\Omega_{DE}= 1-\Omega_K,
\end{equation}
which is related to the total equivalent mass density by $\Omega_M=8\pi G \rho/(3H^2)$. Thus, for any spatially flat cosmological model, one requires $\Omega_M=1$ and is common to write the corresponding mass density as a \textit{critical density}, which is given by
\begin{equation}
\rho_c \equiv \frac{3H^2}{8\pi G}.
\end{equation}
The critical density today can be written as
\begin{equation}
\label{crit_density}
\rho_c^{(0)} \equiv \frac{3H_0^2}{8\pi G} = 1.88 \, h^2 \times 10^{-29} g \, cm^{-3},
\end{equation}
where we have used $G=6.67\times 10^{-8} cm^3g^{-1}sec^{-2}$. Therefore, if the total density of the Universe is greater than $\rho_c^{(0)}$ then the Universe is spatially closed and if the total density is smaller than $\rho_c^{(0)}$ then we live in a spatially open Universe.

Another useful definition is the \textit{equation of state} of a fluid (in cosmology, modeled as a dilute gas), which is the ratio of its pressure $P$ to its density $\rho$ and is given by (in natural units)
\begin{equation}
\label{EoS}
w \equiv P/\rho
\end{equation}
such that, as we shall see from thermodynamics, relativistic particles and non-relativistic matter have equation of state $w=1/3$ and $w=0$, respectively. This is an important quantity that governs the evolution of different species in the Universe and, when related to dark energy, it tells us if the Universe is going through an accelerated expansion or not and if it will expand forever or collapse in the near future \cite{amendola}.

Several parametrizations for dark energy equation of state have been proposed so far. The main ones are based on the expansion
\begin{equation}
    w(z)=\sum_n w_n\,x_n,
\end{equation}
where $w_n$ are parameters to be adjusted using observational datasets and $x_n$ are phenomenological dependencies on redshift (or scale factor). The cases most considered in the literature are
\begin{align}
&{\rm (i)\,constant}~w: \;\,\quad x_{0}(z)=1;~x_{n}=0\,,~~n \ge 1\,, \\
&{\rm (ii)\,redshift:} \;\;\;\quad \quad x_n(z)=z^n\,, \\
&{\rm (iii)\,scale~factor:} \quad  x_n(z)
=\left( 1-\frac{a}{a_{0}} \right)^n=
\left(\frac{z}{1+z}\right)^n \,, \\
&{\rm (iv)\,logarithmic:} \quad  x_n(z)=
\left[{\rm log}(1+z)\right]^n.
\end{align}
In case (i), if $w_0=-1$ we have a $\Lambda$CDM model and if $w_0\ne -1$ we have a $w$CDM model. In case (ii), the model in which $n\leq 1$ was studied in \cite{Turner2001,Weller2001}. Also at linear order, case (iii) is called CPL model \cite{chevallier,linder} and case (iv) was firstly introduced in \cite{Efstathiou1999}. The evolution of dark energy equation of state is the main objective of study in this work.

\section{Cosmic distances}
In order to relate observations to quantities of interest (e.g. matter density, curvature,...), it is important to define distances in cosmology. However, defining distance measures in the Universe can be confusing because light received by us from a distant galaxy was emitted when the Universe was younger and, therefore, the physical distance has increased since the time of emission because of the finite velocity $c$ of light. For that matter, defining cosmic distances in terms of the comoving distance may be a better approach.

As usual, we will work in a FLRW spacetime \eqref{flrwmetric} by setting $r=\sin\chi$ ($K=+1$), $r=\chi$ ($K=0$) and $r=\sinh\chi$ ($K=-1$) in \eqref{line_element}. Therefore, the 3-dimensional space line-element can be expressed as
\begin{equation}
d\sigma^2 = d\chi^2 + (f_K(\chi))^2 (d\theta^2 + \sin^2\theta \, d\phi^2),
\end{equation}
where
\begin{equation}
\label{fk}
 f_K(\chi) = 
  \begin{cases} 
   \sin\chi & (K=+1),\\
    \chi & (K=0), \\
    \sinh\chi & (K=-1).
  \end{cases}
\end{equation}
The function \eqref{fk} can be written in an unified way
\begin{equation}
\label{fKchi}
f_K(\chi) = \frac{1}{\sqrt{-K}} \sinh(\sqrt{-K} \chi),
\end{equation}
in which one might recover the flat case by taking the limit $K\rightarrow -0$.

Now, let us show that the comoving distance $d_c$ is a function of the Hubble parameter and, thereby, the cosmological parameters. The trajectory of a photon traveling along the $\chi$ direction follows the geodesic equation $ds^2=-c^2dt^2+a^2(t)d\chi^2=0$ (we have now restored $c$ in in the line element). Similarly to the derivation of the cosmological redshift, let us consider the case where light is emitted at time $t=t_1$ in position $\chi=\chi_1$ (redshift $z$), reaching an observer at time $t=t_0$ at position $\chi=0$ (redshift $z=0$). Integrating the geodesic equation, one recovers the comoving distance
\begin{equation}
d_c \equiv \chi_1 = \int_0^{\chi_1} d\chi = -\int_{t_0}^{t_1} \frac{c}{a(t)}dt.
\end{equation}

Combining the definition of the Hubble parameter with \eqref{redshift}, one obtains the relation $dt=-dz/[H(1+z)]$ and therefore the comoving distance is given by
\begin{equation}
\label{comovingdistance}
d_c = \frac{c}{a_0H_0} \int_0^z\frac{d\tilde{z}}{E(\tilde{z})}, \,\,\,\,\, E(z) \equiv \frac{H(z)}{H_0}.
\end{equation}

\subsection{Luminosity distance}
We define now the cosmic distance most important in the study of type Ia supernovae, which was the first observation that determined the accelerated expansion rate of the Universe. This distance indicator is called \textit{luminosity distance} ($d_L$) and is defined by
\begin{equation}
\label{d_l}
d_L^2 \equiv \frac{L_s}{4\pi \mathcal{F}},
\end{equation}
where $L_s$ is the absolute luminosity of a source (in SI units, measured in $W=Js^{-1}$) and $\mathcal{F}$ is the observed flux (in SI units, measured in $Wm^{-2}$).

As one can see, the observed luminosity $L_0$ (detected at $\chi=0$ and $z=0$) is different from the absolute luminosity $L_s$ from the source (emitted at $\chi$ with the redshift $z$). By definition, the flux is $\mathcal{F}\equiv L_0/S$, where $S=4\pi(a_0f_K(\chi))^2$ is the area of a sphere at $z=0$ considering the space curvature. Then the luminosity distance \eqref{d_l} is given by
\begin{equation}
d_L^2 = (a_0f_K(\chi))^2\frac{L_s}{L_0}.
\end{equation}

In order to rewrite the ratio $L_s/L_0$ in terms of redshift, let us define the energy of a pulse emitted at the time-interval $\Delta t_1$ as $\Delta E_1$, which implies that the absolute luminosity is $L_s=\Delta E_1/\Delta t_1$. On the same way, the observed luminosity is $L_0=\Delta E_0/\Delta t_0$, where $\Delta E_0$ is the energy of light detected at the time-interval $\Delta t_0$. As the energy of a photon is directly proportional to its frequency $\nu$ (inversely proportional to $\lambda$) we have that $\Delta E_1/\Delta E_0 = \lambda_0/\lambda_1=1+z$. Moreover, as the light signal propagates with the same velocity $c=\lambda/\Delta t$ through its path, then $\lambda_1/\Delta t_1 = \lambda_0/\Delta t_0$, where $\lambda_1$ and $\lambda_0$ represent the wavelength of light at the emission and detection points, respectively. This implies that $\Delta t_0/\Delta t_1 = \lambda_0/\lambda_1=1+z$. Hence we find that
\begin{equation}
\frac{L_s}{L_0} = \frac{\Delta E_1}{\Delta E_0} \frac{\Delta t_0}{\Delta t_1} = (1+z)^2
\end{equation}
and luminosity distance reduces to
\begin{equation}
\label{dLalmost}
d_L = a_0 f_K(\chi)(1+z).
\end{equation}

Combining \eqref{fKchi} and \eqref{comovingdistance} with \eqref{dLalmost}, then luminosity distance can finally be expressed as
\begin{equation}
\label{luminosity_dist}
d_L = \frac{c(1+z)}{H_0 \sqrt{\Omega_K^{(0)}}} \sinh \left(\sqrt{\Omega_K^{(0)}} \int_0^z\frac{d\tilde{z}}{E(\tilde{z})}\right),
\end{equation}
where $\Omega_K^{(0)}=-Kc^2/(a_0H_0)^2$, as given by \eqref{cosmo_params}. As one can see, our expression for luminosity distance is strongly dependent on the cosmology adopted because there is a dependence with the cosmological parameters ($E(z)$ depends on the $\Omega_i^{(0)}$'s, as we shall see) that affects the evolution of $d_L$.

\subsection{Angular diameter distance}
Another useful definition of cosmic distance is the \textit{angular diameter distance}, which is very popular in CMB and BAO data analysis and is defined by
\begin{equation}
\label{d_a}
d_A \equiv \frac{\Delta x}{\Delta\theta},
\end{equation}
where $\Delta\theta$ is the angle subtended by an object of actual size $\Delta x$ orthogonal to the line of sight.

In order to express \eqref{d_a} in terms of the FLRW metric, suppose we have two radial null geodesics (light paths) meeting at the observer at time $t_0$ with angular separation $\Delta\theta$, which have been emitted from a source of size $\Delta x$ at time $t_1$ at a comoving coordinate $\chi$ (we also assume that $\phi=$ constant along the photon paths). From the angular part of the FLRW metric we have
\begin{equation}
\Delta x = a(t_1)f_K(\chi)\Delta\theta.
\end{equation}

\begin{figure}[h!]
\centering
\includegraphics[width=0.8\textwidth]{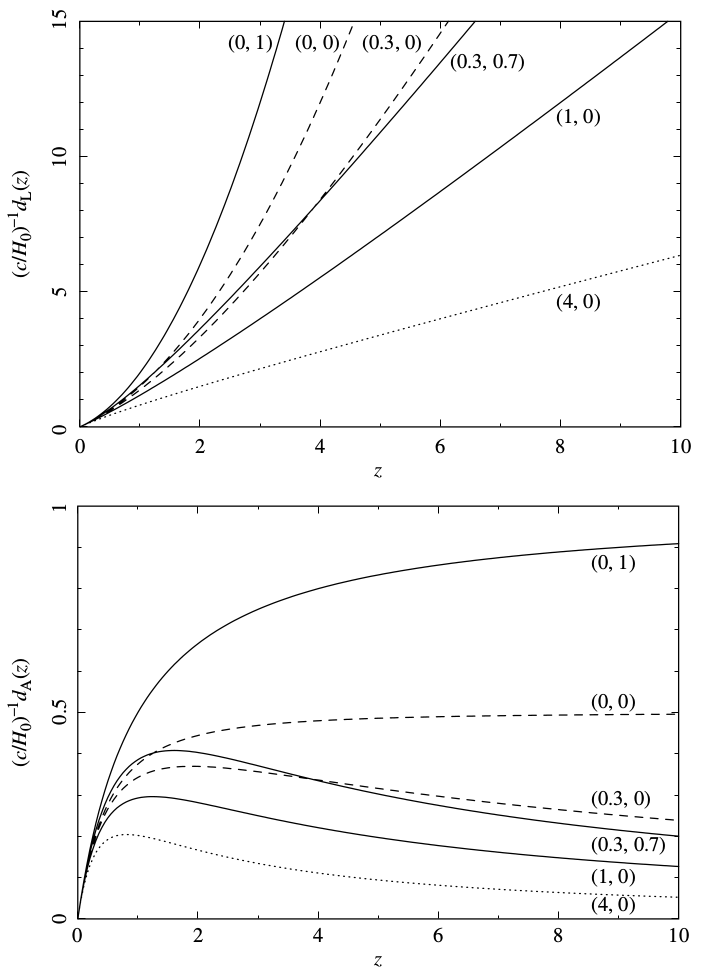}
\caption{Dimensionless luminosity distance (top panel) and angular diameter distance (bottom panel) as a function of redshift for different cosmologies. The pairs are written as $(\Omega_m^{(0)},\Omega_\Lambda^{(0)})$, with negligible radiation. Solid, dashed and dotted lines represent flat, open and closed space geometries, respectively. Extracted from \cite{hobson}.}
\label{lumin_ang_distance}
\end{figure}

Therefore, we may identify the angular diameter distance as
\begin{equation}
\label{angular_dist}
d_A = a(t_1)f_K(\chi) = \frac{a_0f_K(\chi)}{1+z} = \frac{1}{1+z} \frac{c}{H_0 \sqrt{\Omega_K^{(0)}}} \sinh \left(\sqrt{\Omega_K^{(0)}} \int_0^z\frac{d\tilde{z}}{E(\tilde{z})}\right),
\end{equation}
which also displays strong dependence on the cosmology adopted just like luminosity distance. The evolution of $d_L$ and $d_A$ for some cosmologies is shown in Fig. \ref{lumin_ang_distance}.

It is interesting to notice that combining \eqref{luminosity_dist} and \eqref{angular_dist} yields the relation
\begin{equation}
d_A = \frac{d_L}{(1+z)^2},
\end{equation}
which is called \textit{reciprocity} or \textit{duality} or \textit{Etherington relation} \cite{etherington2007}, valid for all cosmological models based on Riemannian geometry as long as flux is conserved. This relation has been tested by many authors \cite{holanda} and it seems to be in good agreement with data.

\section{Cosmic components}
In this section we shall study the species that are supposed to form the Universe today. These are generally classified into relativistic particles, non-relativistic matter and dark energy. There is an additional component, presumably a scalar field, that is supposed to have dominated the Universe during the inflationary era, but this component is not of our immediate interest here. Therefore, we will focus on the main components and analyze its thermodynamic properties of interest for the thermal history of the Universe.

During cosmic ages, many processes have happened so rapidly that the equilibrium was achieved for most of the time, with different kinds of particles sharing the same temperature. We wish to measure quantities like density and pressure in terms of this equilibrium temperature. For that matter, it is necessary to introduce the \textit{occupation number} or \textit{distribution function} or \textit{phase space occupancy} of a species, which counts the number of particles in a given region in phase space around position $\bm{x}$ and momentum $\bm{p}$ \cite{dodelson2003}. Considering a particle with momentum $\bm{p}$ and mass $m$, from special relativity the energy of this particle is $E=\sqrt{p^2+m^2}$, where $p\equiv \left|\bm{p}\right|$. Thus, the distribution function of some species in equilibrium at temperature $T$ is given by
\begin{equation}
\label{distributions}
f(\bm{x},\bm{p},t)=\frac{1}{\exp[(E-\mu)/T]\pm 1},
\end{equation}
where $\mu$ is the chemical potential for each of the species. The plus and minus signs are defined depending on which kind of particles we are dealing with: plus representing Fermi-Dirac statistics and minus the Bose-Einstein statistics.

Equation \eqref{distributions} defines the general case where the distribution may depend on the position of the species, but in an homogeneous Universe we can relax this assumption and write $f=f(|\bm{p}|)\equiv f(p)$. By Heisenberg's principle, no particle can be localized in a region of phase space smaller than $(2\pi\hbar)^3$, therefore it is a fundamental size which defines the number of phase space elements in the volume $d^3x\,d^3p$ as $d^3x\,d^3p/(2\pi\hbar)^3$. By defining $g_\ast$ as the number of internal degrees of freedom (e.g. spin states), the energy density $\rho$ and pressure $P$ are given by 
\begin{align}
\label{rho_general}
\rho &= g_\ast \int \frac{d^3p}{(2\pi\hbar)^3}E(p)f(p) = \frac{g_\ast}{2\pi^2}\int_m^\infty dE\frac{(E^2-m^2)^{1/2}}{\exp[(E-\mu)/T]\pm 1}E^2, \\
\label{P_general}
P &= g_\ast \int \frac{d^3p}{(2\pi\hbar)^3}\frac{pv}{3}f(p) = g_\ast \int\frac{d^3p}{(2\pi\hbar)^3} \frac{p^2}{3E}f(p) \nonumber \\
&= \frac{g_\ast}{6\pi^2} \int_m^\infty dE\frac{(E^2-m^2)^{3/2}}{\exp[(E-\mu)/T]\pm 1}.
\end{align}

In \eqref{rho_general} and \eqref{P_general}, the integrals are not evaluated over $d^3x$ because the energy density and pressure are defined as quantities per unit volume. In the first equality of \eqref{P_general} we have used the fact that the pressure per unit number density of particles is given by $pv/3$ ($v$ is the particle velocity), and in the second equality we used the relation $v=p/E$ (using units of $c=1$) from special relativity (which can be derived combining the equations $E=mc^2/\sqrt{1-v^2/c^2}$ and $p=mv/\sqrt{1-v^2/c^2}$). For the final expressions, we have adopted $\hbar=1$. In what follows, we particularize \eqref{rho_general} and \eqref{P_general} for different particle species.

\subsection{Relativistic species}
The limit of relativistic species is equivalent to consider $T\gg m$ in \eqref{rho_general} and \eqref{P_general}, i.e. $m\rightarrow 0$. For non-degenerate particles ($T\gg \mu$) one obtains
\begin{align}
\label{rho_boson_fermion}
 \rho &=
  \begin{cases} 
   (\pi^2/30)g_\ast T^4, & (Bosons)\\    
   (7/8)(\pi^2/30)g_\ast T^4, & (Fermions)
  \end{cases}
\\  P &= \rho/3,
\label{P_rho}
\end{align}
where we have used $\int_0^\infty dx\,x^3/(e^x-1)=\pi^4/15$ and $\int_0^\infty dx\,x^3/(e^x+1)=7\pi^4/120$. Equation \eqref{P_rho} shows that for relativistic particles without degeneracies the equation of state is $w=1/3$.

The main relativistic particle that one might want to study is the photon, which is a boson. For this kind of particle, $g_\ast=2$ to account for the two spin states. Also, we could have solved \eqref{rho_general} and \eqref{P_general} for photons simply by considering the chemical potential as zero, which is expected theoretically because in the early Universe photon number is not conserved (e.g. electrons and positrons can annihilate to produce photons) \cite{dodelson2003}. Moreover, one might safely ignore the chemical potential because precision measurements on the CMB spectrum (which shall be described in the next chapter) constrain the chemical potential to $\mu/T<9\times 10^{-5}$ \cite{fixsen}. This leads the CMB photon density to be
\begin{equation}
\label{rho_gamma}
\rho_\gamma = \frac{\pi^2}{15} T_\gamma^4.
\end{equation}

\begin{figure}[h!]
\centering
\includegraphics[width=0.7\textwidth]{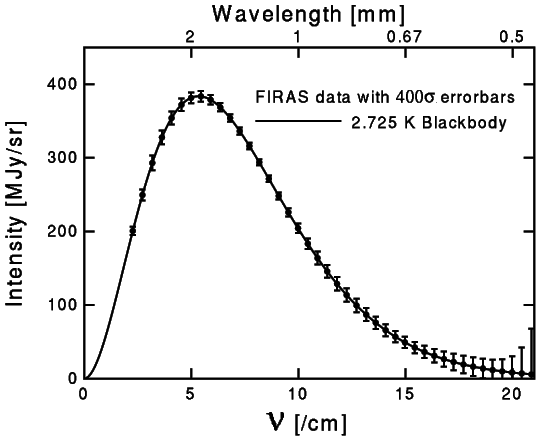}
\caption{CMB spectrum measured by COBE satellite. The agreement between measurements and Planck blackbody theory is so huge that uncertainties have been increased in size four hundred times to allow data visualization. Extracted from \cite{fixsen}.}
\label{blackbodyCMB}
\end{figure}

The COBE satellite showed that the CMB spectrum is very close to the spectrum of a black-body radiation with a temperature of $T_\gamma=2.725\pm 0.002\, K$ \cite{smoot}, as one can see in Fig. \ref{blackbodyCMB}, which has been improved with WMAP satellite to the value $T_\gamma =2.72548\pm 0.00057\, K$ \cite{fixsen2}. Sticking to the value measured by COBE satellite and using the conversion factor $1\, K^4=1.279\times 10^{-35}g\,cm^{-3}$, the energy density of CMB photons today is $\rho_\gamma^{(0)}=4.641\times 10^{-34}g\,cm^{-3}$. Therefore, the photon density parameter is
\begin{equation}
\label{omega_gamma}
\Omega_\gamma^{(0)} \equiv \frac{8\pi G\rho_\gamma^{(0)}}{3H_0^2} = \frac{\rho_\gamma^{(0)}}{\rho_c^{(0)}} = 2.469\times10^{-5} \, h^{-2},
\end{equation}
in which we have used \eqref{crit_density}. The reduced Hubble parameter has a current value close to $h=0.67$ \cite{planck2}, which implies $\Omega_\gamma^{(0)}\simeq 5.5\times 10^{-5}$. The amount of radiation in the present Universe is very low because, as we shall see later, radiation density scales as $a^{-4}$ so that it dilutes faster than other components. The scaling of radiation density leads to a scaling in the temperature of CMB photons of $T\propto a^{-1} \,(=1+z)$, which evolves this way even after the photons went out of equilibrium with matter \cite{weinberg2008}.

Another relativistic particle to be added to the cosmic inventory is the neutrino, which has a very small mass. Neutrinos are fermionic particles with zero chemical potential and there are three types of species in the standard model (electron neutrino $\nu_e$, muon neutrino $\nu_\mu$ and tau neutrino $\nu_\tau$). As each species has one spin degree of freedom and we have to account for anti-neutrinos, from \eqref{rho_boson_fermion} the neutrino energy density is given by
\begin{equation}
\label{rho_nu}
\rho_\nu = N_{eff}\frac{7\pi^2}{120}T_\nu^4,
\end{equation}
where $N_{eff}$ is the effective number of neutrino species and $T_\nu$ is the background neutrino temperature, predicted to have the value $T_\nu=(4/11)^{1/3}\,T_\gamma\simeq 1.945\, K$ (this relation comes from the conservation of entropy before and after the annihilation of electrons and positrons \cite{amendola}). Unlike cosmic microwave background, which is measured with great accuracy, cosmic neutrino background has only indirect evidences due to the fact that neutrinos interact very weakly \cite{follin}.

Although the effective number of neutrinos in the standard model is $N_{eff}=3$, the presence of relativistic degrees of freedom changes this value slightly to the value $N_{eff}=3.04$ \cite{mangano}. As the temperature of neutrinos and photons are linked via the relation $T_\nu/T_\gamma=(4/11)^{1/3}$, by using \eqref{rho_gamma} and \eqref{rho_nu} one shows that neutrino density and photon density are linked via $\rho_\nu=N_{eff}(7/8)(4/11)^{4/3}\rho_\gamma$. Hence the radiation density parameter today, which accounts for photons and relativistic neutrinos, is given by
\begin{equation}
\label{omega_radiation}
\Omega_r^{(0)} = \frac{\rho_\gamma^{(0)}+\rho_\nu^{(0)}}{\rho_c^{(0)}} = \Omega_\gamma^{(0)}(1+0.2271N_{eff}),
\end{equation}
with $\Omega_\gamma^{(0)}$ given by \eqref{omega_gamma}. Considering again $h=0.67$ and $N_{eff}=3.04$, one obtains $\Omega_r^{(0)} \simeq 9.3\times 10^{-5}$, which shows that today radiation is very diluted.

\subsection{Non-relativistic matter}
Now we consider the case of non-relativistic particles ($T\ll m$). From \eqref{rho_general} and \eqref{P_general}, one gets
\begin{align}
\rho &= g_\ast \, m \left(\frac{mT}{2\pi}\right)^{3/2}\exp[-(m-\mu)/T], \\
\label{pressure_matter}
P &= g_\ast \, T \left(\frac{mT}{2\pi}\right)^{3/2}\exp[-(m-\mu)/T] = \frac{T}{m}\rho,
\end{align}
which are valid for both bosons and fermions. Equation \eqref{pressure_matter} is in the form of an equation of state, similar to \eqref{EoS}, and thus one infers that $w\simeq 0$, as expected, because $T/m\ll 1$. The case $w=0$ is also called \textit{dust}. The above expressions show that non-relativistic matter is not simply described only in terms of temperature (as relativistic particles do), and therefore we need to measure the density of non-relativistic particles (baryons and dark matter) directly from observations \cite{amendola}.

\textit{Baryons} compose the visible matter we can observe in the Universe: mostly protons and neutrons (although we also include electrons in this classification even knowing they are leptons, because protons and neutrons are so much more massive than electrons that virtually all the mass in atoms is in the baryons). In literature there are four main techniques used to measure baryon density: observation of baryons from gas in group of galaxies \cite{navarro}; baryon counting by spectra analysis of distant quasars \cite{rauch}; sensitivity to baryon density on the amount of light elements produced during the Big Bang nucleosynthesis epoch \cite{burles}; effects of change in baryon density on the CMB anisotropy spectrum \cite{komatsu}. The tightest constraint on baryon density comes from recent measurements of Planck satellite \cite{planck2}, which constrains baryon density to the value 
\begin{equation}
\label{baryon_dens}
\Omega_b^{(0)}h^2=0.02226\pm 0.00023
\end{equation}
at $68\%$ confidence level. Using $h=0.67$, $\Omega_b^{(0)}\simeq 0.05$ for the central value in \eqref{baryon_dens}.

All of these different techniques are in quite good agreement \cite{fukugita}, ranging baryon density from $2-5\%$ with respect to the critical density. However, total matter density in the Universe is higher than this and, therefore, there must be some kind of matter which is not baryonic to account for this discrepancy.

\begin{figure}[h!]
\centering
\includegraphics[width=0.8\textwidth]{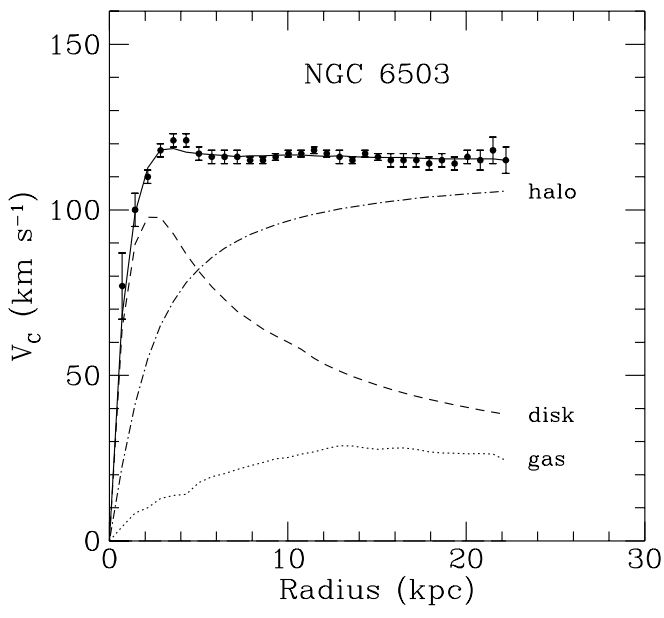}
\caption{Rotation curve for galaxy NGC6503, in which black dots represent the circular velocity measurements as a function of the distance to the galactic center. The dotted and dashed curves represent the contribution to the velocity from observed gas and disk, respectively, while dot-dashed curve represents the dark matter halo contribution that needs to be taken into account in order to match data to theory correctly. Extracted from \cite{Kamionkowski1998}.}
\label{rotation_curve}
\end{figure}

In 1933, Zwicky \cite{zwicky} studied the Coma cluster by measuring radial velocities of galaxies (which by the time were called \textit{nebulae}). He found a surprising result: galaxy velocities were highly dispersed to be explained by the ``visible'' stellar material, which indicated that the cluster density was much higher than density inferred from luminous matter only. His conclusion was that there was some kind of ``missing matter'' to account for the matter unobserved, which he dubbed \textit{dark matter}. Another evidence comes from 1970's when rotation curves of spiral galaxies showed that there was indeed some ``missing matter'' in these objects by the observation that rotation curves were flat to very large radii from the galactic center, when in fact the behavior expected was a declining one \cite{rubin}. This fact is evidenced in Fig. \ref{rotation_curve}.

These evidences and recent studies on the field (e.g. gravitational lensing \cite{tyson}, large-scale structure \cite{frenk}, Bullet Cluster \cite{clowe}) also support the existence of dark matter, suggesting it as another kind of non-relativistic matter that differs from baryons because it does not interact electromagnetically but only gravitationally. In addition, the most probable kind of dark matter is \textit{cold dark matter} (CDM), which considers dark matter as non-relativistic due to the fact that if dark matter was hot (relativistic), then structure formation would not have achieved the level of development observed today because relativistic particles hardly cluster, streaming out of overdense regions easily, and baryons would not have enough potential wells to fall into and form structures.

There are a few candidates to the origin of dark matter, which are divided into astrophysical candidates (e.g. white dwarfs, black holes and neutron stars) and particle candidates (e.g. axions and Weakly Interacting Massive Particles). We also cannot rule out new physics as non-standard gravitational effects unpredicted by general relativity, although this is unlikely due to the huge amount of independent observations that point to the existence of dark matter. The best constraint we have to date comes again from Planck satellite \cite{planck2}, which constrains dark matter density to the value 
\begin{equation}
\label{cdm_dens}
\Omega_c^{(0)}h^2=0.1186\pm 0.0020
\end{equation}
at $68\%$ confidence level. By using $h=0.67$, $\Omega_c^{(0)}\simeq 0.26$ for the central value in \eqref{cdm_dens}, which displays dark matter as a dominating component over baryonic matter.

\subsection{Dark energy}

\begin{figure}[h!]
\centering
\includegraphics[width=0.8\textwidth]{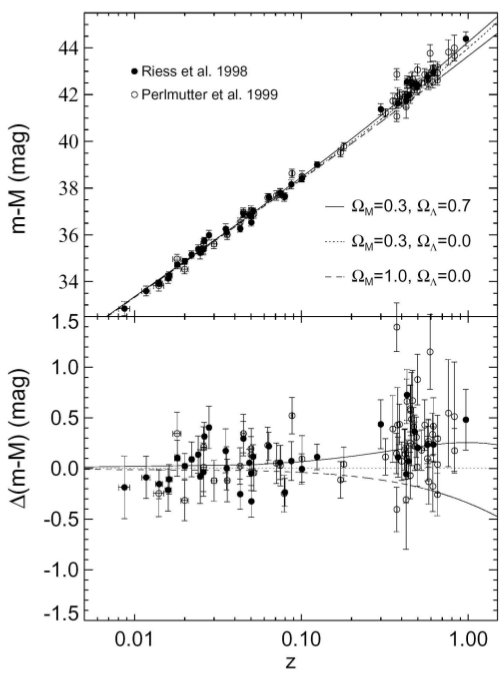}
\caption{Top panel: distance modulus evolution with redshift for Supernova Cosmology Project data (circles) and High-z Supernova Search Team data (black dots). Bottom panel: residuals from the top panel data. The curves model a set of cosmologies, indicating that data at high redshift prefer a model with dominance of dark energy. Extracted from \cite{Ryden2003}.}
\label{mod_dist}
\end{figure}

\begin{figure}[h!]
\centering
\includegraphics[width=0.8\textwidth]{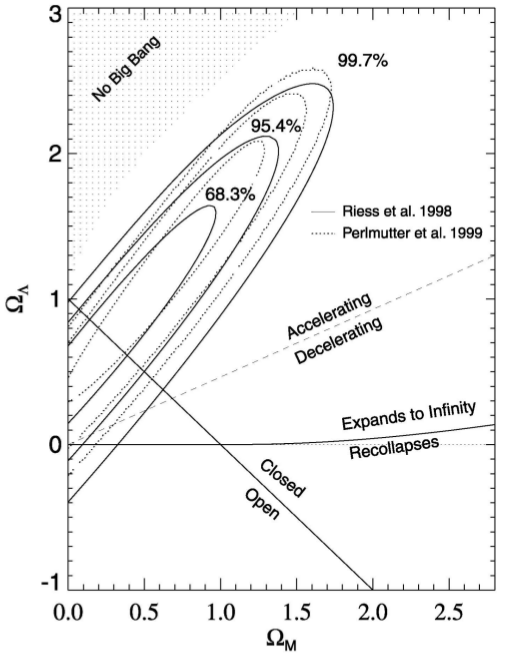}
\caption{2D confidence levels for $\Omega_m^{(0)}$ and $\Omega_\Lambda^{(0)}$ using data displayed in Fig. \ref{mod_dist}. This picture rules out at more than $3\sigma$ a flat Universe with no dark energy. Extracted from \cite{Ryden2003}.}
\label{lamb_matter_SN}
\end{figure}

Observing the previous values calculated for the density of cosmic components, it is interesting to note that baryons and dark matter account for approximately $30\%$ of the total energy budget of the Universe, while photons and neutrinos are negligible. Moreover, CMB measurements \cite{bernardis} along with inflationary scenarios \cite{mukhanov} put tight constraints on the contemporary Universe curvature, $|\Omega_K^{(0)}| \lesssim 0.01$, displaying our Universe as flat and showing that the total density is close to the critical density. As one can see, there is a lack of some cosmic component to account for the remaining $70\%$ and this is where \textit{dark energy} comes in: it gives an explanation for the lacking matter in the Universe and for the late-time cosmic acceleration measured years ago by observing type Ia supernovae \cite{riess,perlmutter}. Dark energy evidence due to type Ia supernovae can be clearly seen by analyzing Fig. \ref{mod_dist} and Fig. \ref{lamb_matter_SN}.

Planck constraint on the dark energy density, for a cosmological constant (which shall be explained latter), is given by \cite{planck2}
\begin{equation}
\label{lambda_dens}
\Omega_\Lambda^{(0)}=0.692\pm 0.012
\end{equation}
at $68\%$ confidence level, which agrees with the missing $70\%$.

\section{Radiation-matter equality, Universe dominated by dark energy and Hubble parameter evolution}
Let us consider a Universe dominated, at some point of its evolution, by a single fluid (component) with equation of state $w$. If $w$ is a constant, one can analytically solve for the evolution of the density $\rho$ and scale factor $a$ for a flat Universe. Using \eqref{friedmann1} and \eqref{continuityeq}, one obtains the solutions (with $K=0$)
\begin{equation}
\rho \propto a^{-3(1+w)},\,\,\,\,\, a\propto (t-t_i)^{2/(3(1+w))},
\end{equation}
where $t_i$ is a constant. For instance, if one considers radiation as the dominant component then the total density scales as $a^{-4}$ and the scale factor as $(t-t_i)^{1/2}$. On the other hand, in a matter-dominated era the pressure is negligible and thus the total density scales as $a^{-3}$ and the scale factor as $(t-t_i)^{2/3}$.

Consider the primitive Universe, with dominance of radiation (density $\rho_r$ and pressure $P_r=\rho_r/3$) and non-relativistic matter (density $\rho_m$ and pressure $P_m=0$). As these components scale as $\rho_r\propto a^{-4}$ and $\rho_m\propto a^{-3}$, we have
\begin{align}
\rho_r = \rho_r^{(0)} (a_0/a)^4 = \rho_r^{(0)} (1+z)^4, \\
\rho_m = \rho_m^{(0)} (a_0/a)^3 = \rho_m^{(0)} (1+z)^3.
\end{align}
The redshift of the transition from radiation-dominance to matter-dominance, which corresponds to the radiation-matter equality ($\rho_r=\rho_m$), is
\begin{equation}
1+z_{eq} = \frac{\rho_m^{(0)}}{\rho_r^{(0)}} = \frac{\Omega_m^{(0)}}{\Omega_r^{(0)}},
\end{equation}
where $\Omega_r^{(0)}$ is given by \eqref{omega_gamma} and \eqref{omega_radiation}. If one considers the effective number of neutrino species $N_{eff}=3.04$, we get
\begin{equation}
\label{rad_matter_eq}
1+z_{eq} = 2.396\times 10^4 \,\, \Omega_m^{(0)}h^2
\end{equation}
and, using Planck result $\Omega_m^{(0)}h^2=0.1415\pm 0.0019$ \cite{planck2}, one obtains $z_{eq} \simeq 3389$ for the central value, which displays the radiation-matter equality as a previous event to the decoupling of photons, as we shall see.

As mentioned in the introduction, many observations show that the Universe is going through an accelerated phase nowadays. In order to explain this framework from the point of view of a component dominance, which is supposed to drive this acceleration, we require $\ddot{a}>0$ in equation \eqref{twodots}, which leads to
\begin{equation}
P<-\rho/3 \,\,\,\, \Rightarrow \,\,\,\, w<-1/3,
\end{equation}
with $\rho$ assumed to be positive. Therefore, a component that follows this requirement and dominates the Universe today could be a simple explanation for the accelerated expansion. Such an unusual component with negative pressure is what we call \textit{dark energy}. This is our motivation here to consider dark energy as a good candidate to explain the late-time acceleration and thereby to study its equation of state evolution.

An interesting case is the so-called \textit{cosmological constant}, defined when $w=-1$, which implies from \eqref{continuityeq} that $\rho$ is a constant during all epochs. From the first Friedmann equation, one can infer that $H$ is a constant in a flat Universe and therefore the scale factor evolves exponentially as $a\propto \exp(Ht)$. Actually, this model might give an explanation for dark energy because no one knows if the acceleration will last forever or end in a near future \cite{amendola}.

From the discussion above, let us consider dark energy as a fluid with equation of state $w_{DE} = P_{DE}/\rho_{DE}$, which satisfies the equation
\begin{equation}
\dot{\rho}_{DE}+3H(\rho_{DE}+P_{DE})=0.
\end{equation}
By integrating it using the relation $dt=-dz/[H(1+z)]$, one obtains the way dark energy density evolves, given by
\begin{equation}
\rho_{DE} = \rho_{DE}^{(0)} \exp \left[\int_0^z \frac{3(1+w_{DE})}{1+\tilde{z}}d\tilde{z} \right],
\end{equation}
such that the integral form is kept because we are considering the general case in which $w$ may be time-dependent.

Expanding the first Friedmann equation in the components,
\begin{equation}
\label{fried}
H^2 = \frac{8\pi G}{3} (\rho_r+\rho_m+\rho_{DE}) - \frac{K}{a^2},
\end{equation}
as already mentioned, one can rewrite \eqref{fried} as a constraint equation for the cosmological parameters today, given by
\begin{equation}
\Omega_r^{(0)}+\Omega_m^{(0)}+\Omega_{DE}^{(0)}+\Omega_K^{(0)} = 1.
\end{equation}

If we combine \eqref{parameters} and \eqref{fried}, we get one of the main equations used in cosmology,
\begin{align}
H^2(z) = H_0^2 \biggl[ &\Omega_r^{(0)}(1+z)^4 + \Omega_m^{(0)}(1+z)^3 \nonumber \\
& + \Omega_{DE}^{(0)}\exp \left\lbrace\int_0^z \frac{3(1+w_{DE})}{1+\tilde{z}}d\tilde{z} \right\rbrace + \Omega_K^{(0)}(1+z)^2 \biggr],
\end{align}
which displays the Hubble parameter evolution as a function of redshift, the cosmological parameters today and dark energy equation of state.

%% file: Chap2.tex
\newpage
\chapter{Datasets description}
In order to constrain the evolution of the dark energy equation of state, in this section we shall describe the observational datasets we are going to use in Chapter \ref{chapt4} and the theoretical background required to model these observations. From now on, even whether not mentioned, we will stick to a flat geometry as testified by many observations \cite{planck,komatsu,Betoule2014}.

\section{Baryon acoustic oscillations}

\subsection{Overview}

The observable Universe is not composed simply by randomly-positioned galaxies all over space, but there are many structures (e.g. clusters, super-clusters, voids) arranged in a very specific way (also called \textit{cosmic web}). One of the main objectives of cosmology is to understand how these structures were developed and came to be what we observe today.

In the early Universe, temperature was so high and the Universe so dense that one can consider matter and radiation coupled as a single fluid, named \textit{photon-baryon fluid}. Such a high temperature prevented formation of atoms because electrons were constantly being ``casted out'' of hydrogen atoms, and also prevented photons of propagating freely through space because the mean free path was too low due to scattering with electrons. Moreover, the photon-baryon fluid is also a plasma.

The standard vision of gravity is that regions in space more dense than its surroundings tend to attract more matter than underdense regions, becoming even more dense, but this is not exactly what happens in the early Universe. Instead, small overdensities of baryonic matter over the whole space, coming from the inflationary era, try to grow due to gravitational instability, but they are soon counterbalanced by the pressure imbalance of the plasma caused by the photons and heat trapped in it. This pressure is so high that overdensities just oscillate on its amplitude instead of growing by gravity, with spherical sound waves being emitted from each overdensity and propagating through the Universe, similar to sound waves propagating in a fluid.

This situation goes on until approximately 380.000 years after the Big Bang, at the point the Universe has expanded enough such that matter-radiation \textit{decoupling} occurs, allowing electrons and nuclei to form neutral atoms. The plasma pressure is then released and the sound waves are frozen in, forming an overdense spherical shell of characteristic comoving radius of approximately $150$ Mpc, with center on the initial overdensity. Nevertheless, these oscillations in the primordial plasma, named \textit{baryon acoustic oscillations} (BAO), end up but leave an imprint due to the spherical shell mainly in the galaxy distribution, causing two galaxies to be more likely separated by this characteristic scale\cite{eisenstein}.

\begin{figure}[h!]
\centering
\includegraphics[width=0.9\textwidth]{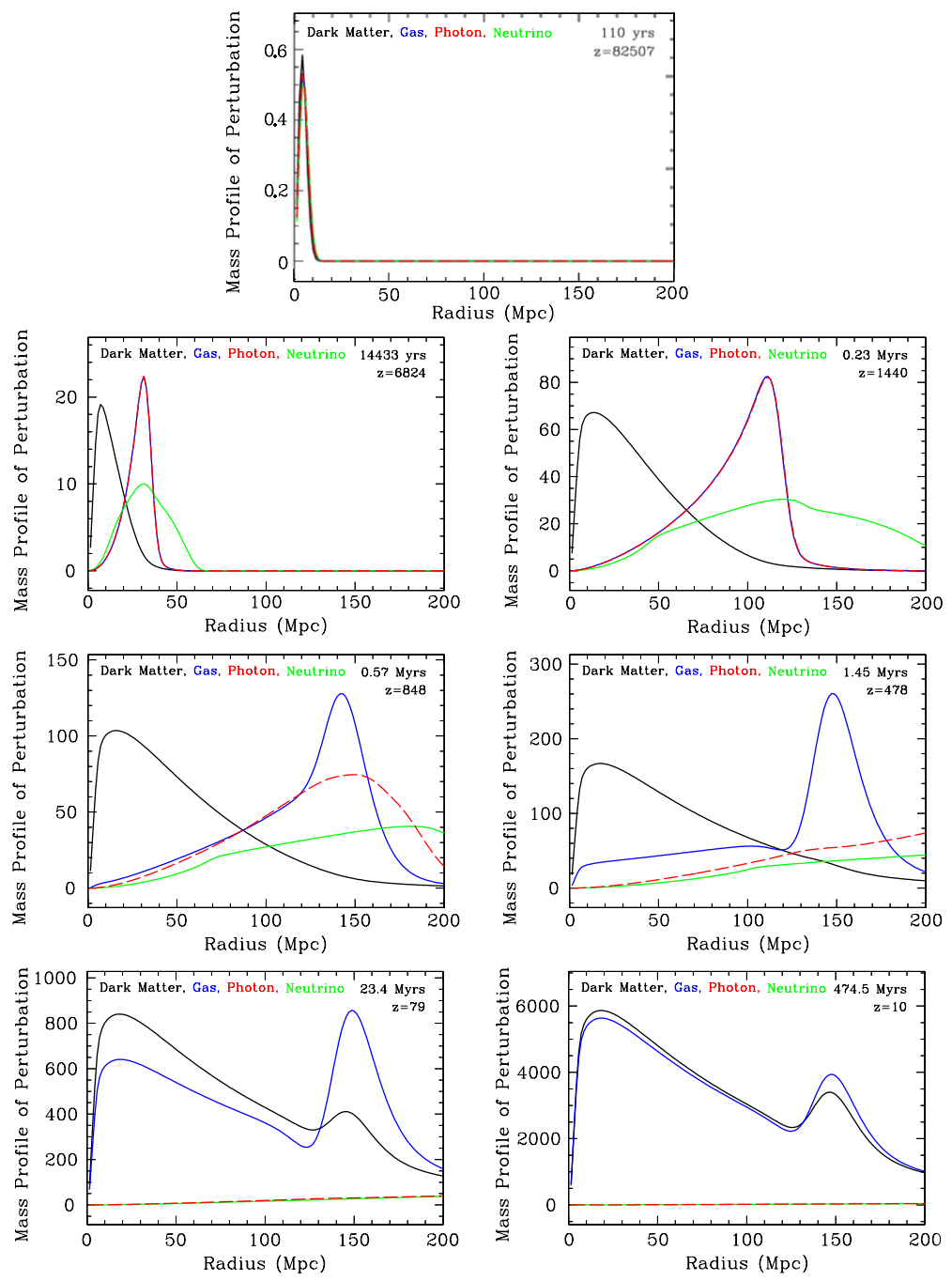}
\caption{Mass profile evolution with redshift from early to late-time Universe, showing the formation of the BAO acoustic scale. The components are dark matter (solid line in black), baryons (solid line in blue), neutrinos (solid line in green) and photons (dashed line in red). Extracted from \cite{weinberg2013}.}
\label{sound_wave_BAO}
\end{figure}

A scheme on the evolution of perturbations in the primordial era for some redshifts is displayed in Fig. \ref{sound_wave_BAO}, where we plot the density of each component times the square of the characteristic comoving radius versus the comoving radius, such that the mass in the overdensity is the area under the curve. Initial point-like overdensities for each component (dark matter, gas containing nuclei and electrons, photons and neutrinos) evolve, having in mind that these perturbations occur in many places over the Universe and their effects sum linearly because fluctuations are very small. Due to the fact that initial perturbations are adiabatic, in the beginning all species are perturbed almost by the same amount (species at $z=82507$) because the energy perturbation of neutrinos and photons is $4/3$ bigger than dark matter and gas. The first component to decouple from the others and free-stream is the neutrino because it interacts very weakly and, as a relativistic particle, it does not cluster. As dark matter interacts only gravitationally, it stands still because has no intrinsic motion for the fact that it is cold dark matter. The gas and photons are coupled to each other as already described and the spherical sound wave starts to propagate with origin in the initial overdensity (species at $z=6824$). These component perturbations grow almost nothing during radiation era: gas and photon perturbations do not grow because of the coupling and dark matter perturbations due to Meszaros effect \cite{Meszaros1974}.

As the Universe evolves, the photon-gas perturbation continues to propagate, the neutrinos have completely streamed away and dark matter perturbation remains still but gets larger due to gravitational forces that tend to attract more material (species at $z=1440$). In the meantime, the Universe is cooling to the point that photons decouple from gas and start to stream away just as neutrinos did, while baryon perturbation stops propagating and starts to grow freely (species at $z=848$). What remains is a dark matter perturbation at the original center and a baryon perturbation in a shell of radius 150 Mpc from the center (species at $z=478$), which start to attract each other and grow quickly due to combined gravitational forces (species at $z=79$). Finally, we are left with similar profiles for dark matter and gas that enhance the acoustic peak (species at $z=10$). As galaxies form in regions that are overdense, now becomes intuitive the reason why galaxies are more probable to be found separated by this characteristic scale.

\begin{figure}[h!]
\centering
\includegraphics[width=0.8\textwidth]{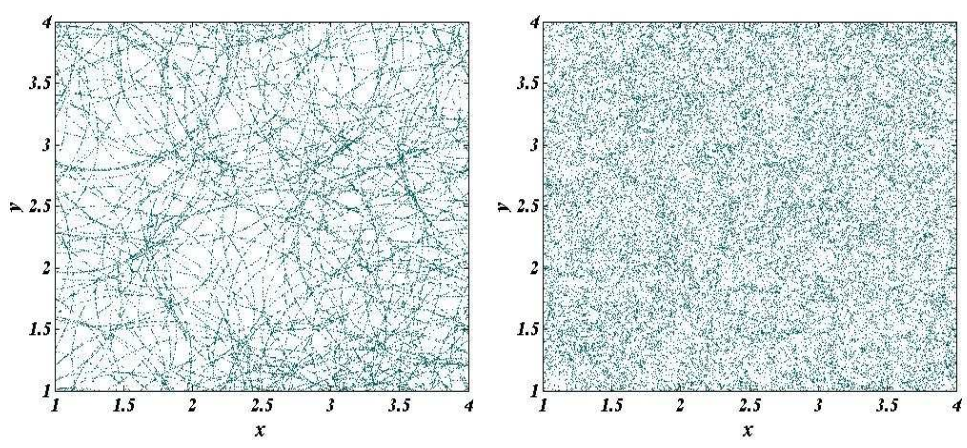}
\caption{Distribution of galaxies with a preferred scale. On the left, the scale is visible because there are many galaxies at each ring, while on the right the scale vanishes at naked eyed when there are more rings superposed, requiring statistics to recover the scale. Extracted from \cite{Bassett2009}.}
\label{rings_BAO}
\end{figure}

In order to have a good idea about the way one obtains the acoustic scale from galaxy distribution, we need to look at Fig. \ref{rings_BAO}. Representing galaxies by dots, in the left hand panel the characteristic scale is clearly visible because there are many voids and the galaxy rings are dense. On the other way, the right hand panel displays a more realistic situation where galaxy rings are less dense and the voids are filled with other galaxy rings, which visually masks the characteristic scale. Therefore, the only way of recovering the characteristic BAO radius is using statistics, which requires mapping enormous volumes of the sky to detect the BAO signal that is very weak at large scales but yet measurable from galaxy redshift surveys\cite{weinberg2013}.

\subsection{Correlation function and power spectrum}

The tool that allows one to measure the BAO peak is the \textit{correlation function}, while cosmological information is better extracted from the \textit{power spectrum}. In order to understand both, one needs to define the density contrast \cite{Peacock1999}
\begin{equation}
\delta(\bm{x}) \equiv \frac{\rho(\bm{x})-\rho_0}{\rho_0},
\end{equation}
which shows us deviations from the background density $\rho_0$. Now we define the correlation function as the spatial average
\begin{equation}
\xi(r_{ab}) \equiv \langle \delta(r_a)\delta(r_b) \rangle,
\end{equation}
made over the whole volume by taking many points ($a$ and $b$) in pairs to perform the evaluation. In this average, we are using the same separation distance $r_{ab}$ considered at various locations, which is called \textit{sample average}. The correlation function tells us if there are any correlations between overdensities separated by the distance $r_{ab}$ that could indicate some mechanism creating dependence on the distance, quantifying the excess clustering on a given scale. Also, when the correlation function $\xi(\bm{r})$ depends only on the separation $\bm{r}$ and not on the locations $\bm{r}_a$ and $\bm{r}_b$, we have \textit{statistical homogeneity} (the system has the same statistical properties everywhere). In this case, the correlation function as a sample average can be written as
\begin{equation}
\label{csi_deltas}
\xi(\bm{r}) = \frac{1}{V}\int \delta(\bm{y})\delta(\bm{y}+\bm{r}) \,dV_y,
\end{equation}
where the integration is made over all possible positions.

\begin{figure}[h!]
\centering
\includegraphics[width=0.7\textwidth]{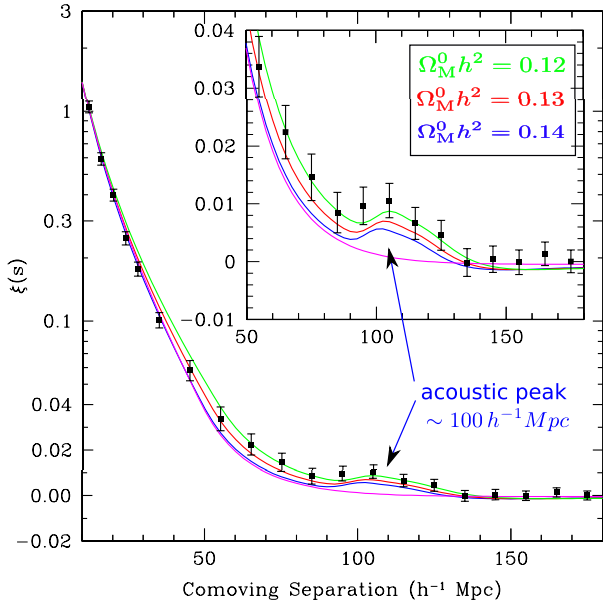}
\caption{Correlation function of SDSS luminous red galaxies sample. Models in green, red and blue are indicated by its total matter quantities with baryon corresponding to $\Omega_b^{(0)}h^2=0.024$, while model in magenta, which lacks the acoustic peak, corresponds to pure CDM (no baryons) with $\Omega_m^{(0)}h^2=0.105$. The BAO bump is clearly visible, in agreement with the predictions. Extracted from \cite{Silvestri2009}.}
\label{corr_function}
\end{figure}

In practice, for obtaining the correlation function one has to compare the real catalog of galaxies with a mock galaxy distribution generated randomly. We can therefore obtain an estimate of the correlation function as
\begin{equation}
\xi = \frac{DD}{DR}-1,
\end{equation}
where DD represents the number of galaxies at distance $r$ counted by an observer at a galaxy in the real catalog, while DR also represents the number of galaxies at distance $r$ but now counted in the random catalog. In the literature, there are other estimators for the correlation function which have been compared rigorously in \cite{Kerscher2000}. The first BAO detection was made by Eisenstein \textit{et al.} \cite{eisenstein} while analyzing a spectroscopic sample of 46748 luminous red galaxies (LRG) measured from the Sloan Digital Sky Survey (SDSS), where a peak corresponding to the BAO scale was found at $100\,h^{-1}$ Mpc, as we can see in Fig. \ref{corr_function}.

Sometimes it is useful to derive some quantities in the Fourier space. For that matter, we will apply the Fourier transform and its inverse, respectively, with the convention
\begin{align}
f_{\bm{k}} = \frac{1}{V} \int f(\bm{x}) e^{-i\bm{k}\cdot\bm{x}} d^3x, \\
f(\bm{x}) = \frac{V}{(2\pi)^3} \int f_{\bm{k}} e^{i\bm{k}\cdot\bm{x}} d^3k.
\end{align}

When studying perturbation theory in cosmology, one finds that it is interesting to work with the density contrast in Fourier space because this decouples the evolution equations, making the equations to evolve independently \cite{amendola}. Since the average of a perturbed variable is zero, the first important quantity related to perturbations is a quadratic function. Therefore, we define the power spectrum as
\begin{equation}
P(k) \equiv A \left|\delta_k\right|^2,
\end{equation}
definition that applies to any perturbed variable in Fourier space (as we will see, temperature perturbations in the CMB follow the same pattern).

The Fourier transform of the density contrast of a density field $\delta(\bm{x})$ is given by
\begin{equation}
\delta_{\bm{k}} = \frac{1}{V} \int \delta(\bm{x}) e^{-i\bm{k}\cdot\bm{x}} dV
\end{equation}
and, by using the definition of the power spectrum in the form
\begin{equation}
P(\bm{k}) = V \left|\delta_{\bm{k}}\right|^2 = V\delta_{\bm{k}}\delta_{\bm{k}}^{*},
\end{equation}
where the normalization $A$ is identified with the volume V considered, we have
\begin{equation}
P(\bm{k}) = \frac{1}{V} \int \delta(\bm{x}) \delta(\bm{y})e^{-i\bm{k}\cdot(\bm{x}-\bm{y})}\, dV_x \, dV_y.
\end{equation}

With the change $\bm{r}=\bm{x}-\bm{y}$, it follows that
\begin{equation}
\label{power_spec_eq}
P(\bm{k}) = \int \xi(\bm{r}) e^{-i\bm{k}\cdot\bm{r}}dV,
\end{equation}
with $\xi(\bm{r})$ defined by \eqref{csi_deltas}, which tells us that the power spectrum is the Fourier transform of the correlation function (Fourier pair). Therefore, it follows conversely that
\begin{equation}
\xi(\bm{r}) = (2\pi)^{-3}\int P(\bm{k}) e^{i\bm{k}\cdot\bm{r}} d^3k.
\end{equation}

If the correlation function does not depend on the direction $\bm{r}$ but only on the modulus $r=|\bm{r}|$ and, accordingly, the power spectrum depends only on $k=|\bm{k}|$, then the system has spatial isotropy and the power spectrum simplifies to
\begin{equation}
P(k)=\int \xi(r)\,r^2dr \int_0^{\pi}e^{-ikr\cos\theta}\sin\theta \, d\theta \int_0^{2\pi}d\phi = 4\pi\int\xi(r)\,\frac{\sin kr}{kr}\,r^2dr.
\end{equation}

\begin{figure}[h!]
\centering
\includegraphics[width=0.7\textwidth]{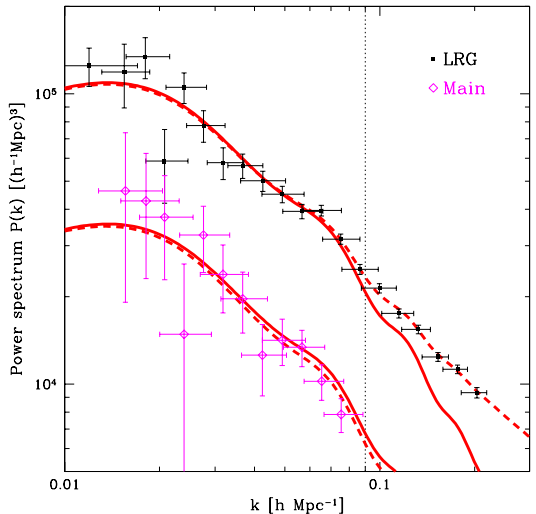}
\caption{Power spectra of two SDSS samples: black is LRG sample and magenta is the main SDSS sample. The peak in the correlation function becomes an oscillation pattern in the matter power spectrum. Solid and dashed lines represent the theoretical prediction for $\Lambda$CDM model and non-linear corrections, respectively. Extracted from \cite{Tegmark2006}.}
\label{power_spec}
\end{figure}

Because the correlation function and the power spectrum are related by a Fourier transform via \eqref{power_spec_eq}, features displayed in one have consequences on the other. As a simple example, if $\xi(\bm{r})$ has the form of a delta function centered at a scale $r_*$, the result is an oscillation pattern in the power spectrum. Although the correlation function of a real survey is not a delta but a small bump, this feature is still preserved and the acoustic peak in $\xi(\bm{r})$ induces oscillations in the power spectrum, which characterizes the baryon acoustic oscillations, as shown in Fig. \ref{power_spec}.

\subsection{Perturbation evolution in the primordial plasma}

In order to get a quantitative picture of the baryon acoustic oscillations, one needs to consider the fact that baryons are tightly coupled to photons in the primordial plasma ($\delta_b \sim \delta_\gamma$). This implies that perturbations in both components are similar until recombination and analyzing photon features is the same as analyzing baryon features. Thus, let us define the Fourier transform of $\delta T/T$, the temperature perturbation of photons, as
\begin{equation}
\Theta \equiv \Theta(k,\mu,\eta),
\end{equation}
in which $\mu \equiv (\bm{k}\cdot\bm{p})/k$ and $\bm{p}$ is the photon direction. From now on we will work in the conformal time.

Instead of working with $\Theta$ directly, one might use the $l$th \textit{multipole moment} of the temperature field, defined by
\begin{equation}
\label{multipole}
\Theta_l \equiv \frac{1}{(-i)^l} \int_{-1}^1 \frac{d\mu}{2} \mathcal{P}_l(\mu)\Theta(\mu),
\end{equation}
where $\mathcal{P}_l$ is the $l$th Legendre polynomial. The multipole moments are an average over all photon directions and the first ones have specific names: $l=0$ is the monopole, $l=1$ is the dipole, $l=2$ is the quadrupole, and so on. Besides, photon perturbations can be characterized either by $\Theta(k,\mu,\eta)$ or by all the moments $\Theta_l(k,\eta)$.

The set of equations that describe the evolution of  photon and baryon perturbations in the early Universe comes from the general Boltzmann equation
\begin{equation}
\frac{df}{dt} = C[f],
\end{equation}
which relates changes in the distribution function of species to possible collision terms that may also depend on the distribution function (e.g. Compton scattering). After long calculations and linearization of the perturbations, the final evolution equations relevant to BAO analysis are \cite{dodelson2003}
\begin{equation}
\label{theta_eq}
\Theta'+ik\mu\Theta = -\Phi' -ik\mu\Psi-\tau'[\Theta_0-\Theta+\mu v_b],
\end{equation}
\begin{equation}
\label{delta_baryon}
\delta_b'+ikv_b = -3\Phi',
\end{equation}
\begin{equation}
\label{veloc_baryon}
v'_b+\frac{a'}{a}v_b = -ik\Psi+\frac{\tau'}{R_s}[v_b+3i\Theta_1],
\end{equation}
where $v_b$ is the baryon velocity, $R_s \equiv (3\rho_b)/(4\rho_\gamma)$ is the ratio of baryon to photon density, $\tau$ is the optical depth defined by $\tau(\eta)\equiv \int_\eta^{\eta_0} d\eta' \,n_e\, \sigma_T\,a$ ($n_e$ is the electron density and $\sigma_T$ is the Thomson cross section), and $\Phi$ and $\Psi$ are perturbations about the flat FLRW metric, corresponding to the Newtonian gauge
\begin{equation}
\label{newton_gauge}
ds^2 = a^2(\eta)[-(1+2\Psi)d\eta^2 + (1+2\Phi)\delta_{ij}dx^idx^j]
\end{equation}
in which $\Psi$ is the Newtonian potential and $\Phi$ is the perturbation in the spatial curvature.

In order to derive the BAO evolution equation, it is necessary to show that multipole terms higher than the dipole are negligible in the tightly coupled regime. This limit corresponds to the scattering rate of photons being much larger than the expansion rate ($\tau >>1$). The idea now is to turn the differential equation \eqref{theta_eq} into an infinite set of coupled equations for $\Theta_l(\eta)$ with the advantage that higher moments are very small. Thus, we multiply \eqref{theta_eq} by $\mathcal{P}_l(\mu)$ and integrate over $\mu$. Combining with \eqref{multipole}, equation \eqref{theta_eq} for $l>2$ yields
\begin{equation}
\Theta'_l+\frac{k}{(-i)^{l+1}}\int_{-1}^1 \frac{d\mu}{2}\,\mu \mathcal{P}_l(\mu)\Theta(\mu) = \tau'\Theta_l.
\end{equation}

To solve the integral, we use the recurrence relation for Legendre polynomials
\begin{equation}
(l+1)\mathcal{P}_{l+1}(\mu) = (2l+1)\mu\mathcal{P}_l(\mu) -l\mathcal{P}_{l-1}(\mu)
\end{equation}
to obtain
\begin{equation}
\Theta'_l-\frac{kl}{2l+1}\Theta_{l-1}+\frac{k(l+1)}{2l+1}\Theta_{l+1} = \tau'\Theta_l.
\end{equation}

Now we analyze the order of magnitude of each term. As $\tau>>1$, the first term on the left (of order $\Theta_l/\eta$) is much smaller than the term on the right (of order $\tau\Theta_l/\eta$). If we neglect the $\Theta_{l+1}$ term for a moment, it follows that in the tightly coupled limit
\begin{equation}
\Theta_l \sim \frac{k\eta}{2\tau}\Theta_{l-1}
\end{equation}
and, for horizon size modes $k\eta \sim 1$, it implies that $\Theta_l<<\Theta_{l-1}$, which justifies neglecting the $\Theta_{l+1}$ term. The whole assumption is valid for $l>1$, such that the monopole and dipole terms dominate over all other multipoles in the primordial plasma.

The next step is to multiply \eqref{theta_eq} by $\mathcal{P}_0(\mu)$ and $\mathcal{P}_1(\mu)$ and integrate over $\mu$. One obtains
\begin{equation}
\label{monopole_eq}
\Theta'_0+k\Theta_1 = -\Phi'
\end{equation}
and
\begin{equation}
\label{dipole_eq}
\Theta'_1-\frac{k\Theta_0}{3} = \frac{k\Psi}{3}+\tau'\left[\Theta_1-\frac{iv_b}{3}\right],
\end{equation}
where we are using the fact that higher multipoles can be neglected. The above equations are supplemented by the equations governing baryon density perturbations, \eqref{delta_baryon} and \eqref{veloc_baryon}. One might rewrite the baryon velocity equation, \eqref{veloc_baryon}, as
\begin{equation}
\label{v_b}
v_b = -3i\Theta_1+\frac{R_s}{\tau'} \left[v'_b+\frac{a'}{a}v_b+ik\Psi\right].
\end{equation}
The second term in \eqref{v_b} is much smaller than the first one because it is proportional to a factor of order $\tau^{-1}$. Thus, at lowest order, $v_b = -3i\Theta_1$. One might use this information to expand everywhere in the second term using this low order expression, obtaining
\begin{equation}
v_b \simeq -3i\Theta_1+\frac{R_s}{\tau'} \left[-3i\Theta'_1-3i\frac{a'}{a}\Theta_1+ik\Psi\right].
\end{equation}

In order to get rid of $v_b$, the above expression is inserted into \eqref{dipole_eq}. By rearranging the equation, one obtains
\begin{equation}
\label{dipole_eq2}
\Theta'_1+\frac{a'}{a}\frac{R_s}{1+R_s}\Theta_1- \frac{k}{3(1+R_s)}\Theta_0 = \frac{k\Psi}{3}.
\end{equation}

With the pair \eqref{monopole_eq} and \eqref{dipole_eq2}, we have two first-order coupled equations characterizing the monopole and the dipole in the tightly-coupled limit. It is interesting to turn then into a second-order equation by differentiating \eqref{monopole_eq} and eliminating $\Theta'_1$ using \eqref{dipole_eq2}, which leads to
\begin{equation}
\Theta_0''+k\left[\frac{k\Psi}{3}-\frac{a'}{a}\frac{R_s}{1+R_s}\Theta_1+\frac{k}{3(1+R_s)}\Theta_0\right] = -\Phi''.
\end{equation}

At last, to disappear with $\Theta_1$ we use \eqref{monopole_eq} and we finally obtain
\begin{equation}
\label{bao_eq}
\Theta_0''+\frac{R_s}{1+R_s}\mathcal{H}\Theta_0' +k^2c_s^2\Theta_0 = -\frac{k^2}{3}\Psi -\frac{R_s}{1+R_s}\mathcal{H}\Phi'-\Phi'' \equiv F(k,\eta),
\end{equation}
where $\mathcal{H} \equiv a'/a$ is the Hubble parameter in the conformal time, $F(k,\eta)$ is defined as a forcing function and $c_s$ is the propagation velocity (or sound speed) of the fluid given by
\begin{equation}
\label{sound_speed}
c_s \equiv \sqrt{\frac{1}{3(1+R_s)}}.
\end{equation}

The sound speed depends on the baryon density and the photon density, which implies that in the primordial Universe, when photon density was much larger that baryon density, sound waves propagated at the relativistic velocity $c_s=c/\sqrt{3}$. As the Universe continued the expansion, baryons became more important and this velocity was reduced (baryons make the fluid heavier, lowering the sound speed). Although \eqref{bao_eq} is written for the monopole, it is also valid for baryon density perturbations because $\Theta_0 \sim \delta_b$ due to the photon-baryon coupling. Besides, equation \eqref{bao_eq} is a wave equation with friction and source terms, which is the ``heart'' of the baryon acoustic oscillations. As the terms for $\Theta_0$ and $\Phi$ have very similar forms, and noticing that $R'_s=\mathcal{H}R_s$, one may rewrite \eqref{bao_eq} as
\begin{equation}
\label{bao_eq2}
\left[\frac{d^2}{d\eta^2}+\frac{R_s'}{1+R_s} \frac{d}{d\eta}+k^2c_s^2\right](\Theta_0+\Phi) = \frac{k^2}{3} \left(\frac{1}{1+R_s}\Phi-\Psi\right).
\end{equation}

Because \eqref{bao_eq2} is an inhomogeneous second-order differential equation, it is necessary to use Green's function method to solve it. First we find the solutions to the homogeneous part and then we use these to find the particular solution. In principle, there is a damping term in \eqref{bao_eq2} and we should solve it considering this, with the right-hand side equal to zero. However, in the primordial Universe the pressure of the photon-baryon fluid induces oscillations with time scale much shorter than the expansion of the Universe (also $R_s$ is small around this epoch) \cite{dodelson2003}, and therefore it is a good first approximation to discard the damping term and consider only the oscillating solutions. Thus, the homogeneous solution is
\begin{equation}
(\Theta_0+\Phi)^{(hom)}(k,\eta)=c_1f_1(k,\eta)+ c_2f_2(k,\eta)
\end{equation}
with
\begin{equation}
f_1(k,\eta) = \sin[kr_s(\eta)]\quad \text{and} \quad f_2(k,\eta) = \cos[kr_s(\eta)],
\end{equation}
where we define the \textit{sound horizon} as
\begin{equation}
r_s(\eta) \equiv \int_0^\eta d\tilde{\eta} \,\, c_s(\tilde{\eta}),
\end{equation}
which, since $c_s$ is the sound wave propagation velocity, is the comoving distance traveled by the sound waves until the epoch $\eta$.

The general solution to a second-order equation is a linear combination of the homogeneous solutions and a particular solution. One might construct the particular solution by integrating the source term weighted by the Green's function that is a combination of the homogeneous solutions. The general result is
\begin{align}
(\Theta_0+\Phi)(k,\eta) =\, &c_1f_1(\eta)+ c_2f_2(\eta) \nonumber \\ &+\frac{k^2}{3}\int_0^\eta d\tilde{\eta}\, [\Phi(\tilde{\eta})-\Psi(\tilde{\eta})] \frac{f_1(\tilde{\eta})f_2(\eta)-f_1(\eta)f_2(\tilde{\eta})}{f_1(\tilde{\eta})f_2^{'}(\tilde{\eta})-f_1^{'}(\tilde{\eta})f_2(\tilde{\eta})}.
\end{align}

In this equation, we are considering $R_s$ very small except in the rapidly varying sines and cosines. In the $f_1$ term, for example, $kr_s$ is evaluated with $c_s$ in its complete form \eqref{sound_speed}. As the constants $c_1$ and $c_2$ are fixed by the initial conditions ($\Theta_0$ and $\Phi$ constants, $\Theta'_0=\Phi'=0$ at $\eta=0$), $c_1$ must vanish and $c_2=\Theta_0(0)+\Phi(0)$. In the integrand, the denominator is $-kc_s(\tilde{\eta})\rightarrow -k/\sqrt{3}$ in the limit we are considering and the numerator reduces to $-\sin{[k(r_s-\tilde{r}_s)]}$, leading to
\begin{align}
\label{solution_bao}
(\Theta_0+\Phi)(k,\eta) =\,&[\Theta_0(0)+\Phi(0)]\cos(kr_s) \nonumber \\
&+\frac{k}{\sqrt{3}} \int_0^\eta d\tilde{\eta}\, [\Phi(\tilde{\eta})-\Psi(\tilde{\eta})]\sin[k(r_s(\eta)-r_s(\tilde{\eta}))].
\end{align}

\begin{figure}[h!]
\centering
\includegraphics[width=0.8\textwidth]{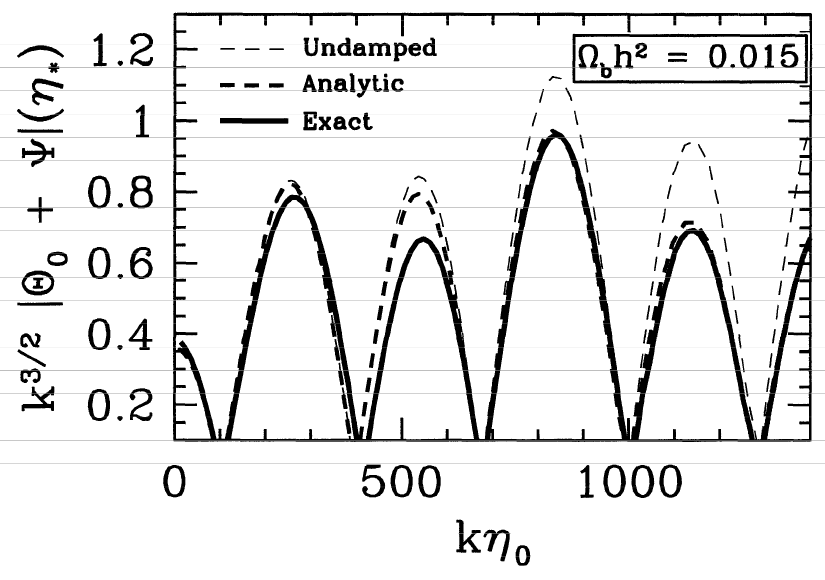}
\caption{Evolution of the monopole at recombination in CDM model. Three curves are displayed: the solid one is the exact solution obtained by solving numerically the full set of Einstein-Boltzmann equations, the light dashed line represents the solution \eqref{solution_bao}, which does not take into account diffusion damping (when the the quadrupole is non-negligible), and the heavy dashed line accounts for this damping. Extracted from \cite{dodelson2003}.}
\label{bao_oscillations}
\end{figure}

Equation \eqref{solution_bao} is an approximated solution that describes the oscillations in the baryon-photon fluid in the tightly coupled limit. This solution matches the exact solution with very good agreement, getting the peak locations correctly and the heights fairly well, as one can see in Fig. \ref{bao_oscillations}. Therefore, all we need are the external gravitational potentials from dark matter and then we can calculate the effect of these potentials on the anisotropies. Also, now we have an accurate way of finding the frequency of the oscillations and the location of the acoustic peaks. In the limit that the second term in \eqref{solution_bao} is negligible, the cosine term dominates and therefore the peaks should appear at the positions
\begin{equation}
k_p=n\pi/r_s \qquad n=1,2,3,\dots \qquad .
\end{equation}

The dipole term is also non-negligible at this point of the cosmic history, and it will be useful in the next section because it contributes to the CMB spectrum. This one is obtained by differentiating \eqref{solution_bao} and inserting it in \eqref{monopole_eq}, which leads to
\begin{align}
\Theta_1(k,\eta) =\,&\frac{1}{\sqrt{3}}[\Theta_0(0)+\Phi(0)]\sin(kr_s) \nonumber \\
&-\frac{k}{3} \int_0^\eta d\tilde{\eta}\, [\Phi(\tilde{\eta})-\Psi(\tilde{\eta})]\cos[k(r_s(\eta)-r_s(\tilde{\eta}))].
\end{align}

\subsection{BAO acoustic scale and relative BAO distance}

The BAO acoustic scale, which corresponds to the sound horizon in a specific $\eta$, can be formally defined as the comoving distance traveled by sound waves in the photon-baryon fluid until baryons are released from Compton drag of photons \cite{amendola}. This is called \textit{drag epoch} (not to be confused with recombination epoch), defined as the epoch at which baryons ``stop noticing'' photons, occurring at redshift $z_{drag}$. The sound horizon at $z=z_{drag}$ is
\begin{equation}
\label{r_s1}
r_s(z_{drag}) \equiv \int\limits_0^{\eta_{drag}} c_s(\eta)\,d\eta = \int\limits_0^{t_{drag}}\frac{c_s(t)}{a(t)}\,dt = \int\limits_{z_{drag}}^{\infty}\frac{c_s(z)}{H(z)}\,dz,
\end{equation}
where
\begin{equation}
c_s^2 \equiv \frac{\delta P_\gamma}{\delta\rho_\gamma + \delta\rho_b} = \frac{1}{3(1+R_s)}
\end{equation}
is the square of the effective sound speed of sound waves in the primordial plasma, as mentioned before. There is a fitting formula for the redshift $z_{drag}$ due to Eisenstein and Hu \cite{hu}, given by
\begin{equation}
z_{drag} = \frac{1291\omega_m^{0.251}}{1+0.659\omega_m^{0.828}} (1+b_1\omega_b^{b_2}),
\end{equation}
where
\begin{equation}
b_1 = 0.313\omega_m^{-0.419}(1+0.607\omega_m^{0.674}), \,\,\, b_2 = 0.238\omega_m^{0.223},
\end{equation}
$\omega_b\equiv \Omega_b^{(0)}h^2$ and $\omega_m\equiv \Omega_m^{(0)}h^2$.

Equation \eqref{r_s1} can be solved analytically if one takes into account the fact that dark energy is negligible for $z>z_{drag}$. Thus, one obtains
\begin{equation}
r_s(z_{drag}) = \frac{c}{\sqrt{3}a_0H_0} \int_{z_{drag}}^\infty \frac{dz}{\sqrt{1+R_s} E(z)} = \frac{c}{\sqrt{3}H_0} \frac{1}{\sqrt{\Omega_m^{(0)}}} \int_0^{a_{drag}} \frac{1}{\sqrt{1+R_s(a)}} \frac{1}{\sqrt{a+a_{eq}}}da,
\end{equation}
where $R_s(a) = (3\omega_b/4\omega_\gamma)a$ and $a_{eq}=(1+z_{eq})^{-1}$ is the scale factor at the radiation-matter equality \eqref{rad_matter_eq}. Integrating the above equation, yields
\begin{equation}
\label{sound_horizon}
r_s(z_{drag}) = \frac{4}{3} \frac{ch}{H_0} \sqrt{\frac{\omega_\gamma}{\omega_m \omega_b}} \ln \left(\frac{\sqrt{R_s^{(drag)}+R_s^{(eq)}}+\sqrt{1+R_s^{(drag)}}}{1+\sqrt{R_s^{(eq)}}}\right),
\end{equation}
where $R_s^{(drag)}\equiv R_s(a_{drag})$ and $R_s^{(eq)}\equiv R(a_{eq})$. Recent measurements from Planck Satellite \cite{planck2} constrain the BAO sound horizon value with great accuracy (approximately $1\%$) to be $r_s(z_{drag})= (147.60\pm 0.43)$ Mpc, with $z_{drag}=(1059.57\pm 0.47)$, which is the basic information for using BAO as a ``cosmological ruler'' in order to constrain cosmological parameters, especially the dark energy equation of state \cite{Blake2003}.

The power spectrum contains information about structures in the Universe and may be obtained analyzing the angular and redshift distribution of galaxies, which divides modes into perpendicular ($k_\bot$) and parallel ($k_\parallel$) to the line of sight. It can be showed \cite{amendola} that the quantities
\begin{equation}
\theta_s(z) = \frac{r_s(z_{drag})}{(1+z)d_A(z)}
\end{equation}
and
\begin{equation}
\delta z_s(z) = \frac{r_s(z_{drag}) H(z)}{c}
\end{equation}
can, in principle, be measured independently and give good estimates for the evolution of the angular diameter distance $d_A$ and the Hubble parameter $H(z)$ \cite{Shoji2009}. Just as the modes, the angle $\theta_s(z)$ corresponds to observations perpendicular to the line of sight and the redshift difference $\delta z_s(z)$ to observations in the radial direction.

The BAO data so far is not sufficient for making good measurements on $\theta_s(z)$ and $\delta z_s(z)$ independently \cite{Okumura2008}. A way around this is to define a combined distance scale ratio that takes into account the combination of two spatial dimensions perpendicular to the line of sight and one dimension along the line of sight, defined by
\begin{equation}
\label{comb_dist}
\left[\theta_s(z)^2 \delta z_s(z) \right]^{1/3} \equiv \frac{r_s(z_{drag})}{\left[(1+z)^2 d_A^2(z) c/H(z)\right]^{1/3}}.
\end{equation}

In literature, observational data is displayed in a slightly different manner from \eqref{comb_dist}, as it follows. The main information that specifies a determined galaxy survey is the \textit{relative BAO distance}, defined by
\begin{equation}
\label{r_BAO}
r_{BAO}(z) \equiv r_s(z_{drag}) / D_V(z),
\end{equation}
where
\begin{equation}
\label{D_V}
D_V(z) \equiv \left[(1+z)^2 d_A^2(z) \frac{cz}{H(z)}\right]^{1/3}
\end{equation}
is the \textit{related effective distance}. Combining \eqref{r_BAO}, \eqref{sound_horizon} and \eqref{angular_dist}, the explicit form of the relative BAO distance in a flat spatial geometry is
\begin{align}
r_{BAO}(z) = &\,\,\frac{4}{3} \sqrt{\frac{\omega_\gamma}{\Omega_m^{(0)} \omega_b}} \left[\frac{z}{E(z)}\right]^{-1/3} \left[\int_0^{z}\frac{d\tilde{z}}{E(\tilde{z})}\right]^{-2/3} \nonumber \\
&\times\ln\left(\frac{\sqrt{R_s^{(drag)}+R_s^{(eq)}}+\sqrt{1+R_s^{(drag)}}}{1+\sqrt{R_s^{(eq)}}}\right).
\end{align}

Another useful descriptor for BAO data is the acoustic parameter $A(z)$, which is independent of $h$ and is defined by \cite{eisenstein}
\begin{equation}
A(z) \equiv \frac{100\,D_V(z) \sqrt{\Omega_m h^2}}{cz} = \sqrt{\Omega_m}E(z)^{-1/3} \left[\frac{1}{z} \int_0^z \frac{dz}{E(z)} \right]^{2/3}.
\end{equation}

We shall be using BAO observational data displayed in Table \ref{tabela_bao}. These are the main BAO data found in literature, which are used extensively for testing models and they also make part of parameter estimation codes for cosmology like CosmoMC \cite{lewis} and Monte Python \cite{audren}.

\begin{table}[h]
\centering
\begin{tabular}{|c|c|c|c|}
\hline
\textbf{Survey} & \textbf{Parameter} & \textbf{Effective redshift} & \textbf{Measurement} \\ \hline
6DF  & $r_s/D_V$ & 0.106  & $0.336 \pm 0.015$    \\
SDSS-MGS & $D_V/r_s$  & 0.15 & $4.4656 \pm 0.1681$  \\
BOSS-LOWZ & $D_V/r_s$ & 0.32 & $8.250 \pm 0.170$ \\
WiggleZ  & $A$ & 0.44  & $0.474 \pm 0.034$    \\
BOSS-CMASS  & $D_V/r_s$  & 0.57  & $13.773 \pm 0.134$  \\
WiggleZ & $A$  & 0.60 & $0.442 \pm 0.020$    \\
WiggleZ & $A$ & 0.73 & $0.424 \pm 0.021$  \\ \hline
\end{tabular}
\caption{BAO data for galaxy surveys.}
\label{tabela_bao}
\end{table}

As WiggleZ data are correlated, we also show in Table \ref{covmat_bao} the inverse covariance matrix for the three WiggleZ measurements.

\begin{table}[h]
\centering
\begin{tabular}{|cccc|}
\hline
Redshift slice & $0.2<z<0.6$ & $0.4<z<0.8$ & $0.6<z<1.0$  \\ \hline
$0.2<z<0.6$  &  1040.3  &  -807.5 &  336.8   \\
$0.4<z<0.8$  &    &  3720.3 &  -1551.9   \\
$0.6<z<1.0$  &    &   &  2914.9 \\ \hline
\end{tabular}
\caption{Inverse covariance matrix for WiggleZ data. Measurements were performed in the overlapping redshifts quoted and, as the matrix is symmetric, only the upper diagonal is displayed. Extracted from \cite{Blake2011}.}
\label{covmat_bao}
\end{table}

\section{Cosmic microwave background radiation}

\subsection{Overview}

In 1963, Penzias and Wilson \cite{penzias} accidentally discovered a form of radiation that could be detected in all sky directions, which was invariant whether they changed the position of the detectors. This signal, dubbed \textit{Cosmic Microwave Background} (CMB) radiation, was first predicted by George Gamow and Robert Dicke in the 1940's, independently, and is one of the cosmic relics from which one can extract information about the primordial Universe. This discovery reinforced the Hot Big Bang model because if the Universe started in a hot, dense and opaque state, then one of the consequences would be a microwave background radiation \cite{Ryden2003}.

As previously mentioned, the basic idea about CMB radiation is that in the primordial Universe there was a strong coupling between matter and radiation until the epoch where, after the Universe had expanded enough ($z\approx 1090$), photons decoupled from baryonic matter, propagating freely ever since and having almost null interaction with matter on its path. These photons are the ones we observe today in the form of microwave radiation.

\begin{figure}[h!]
\centering
\includegraphics[width=0.5\textwidth]{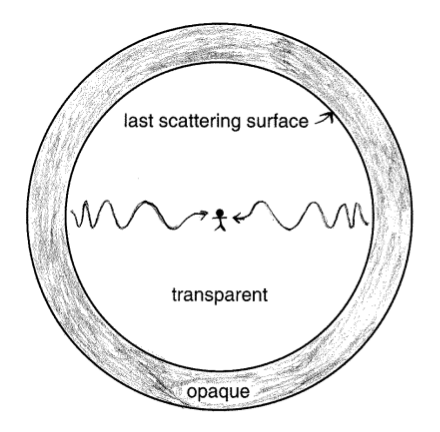}
\caption{Last scattering surface (LSS) of an observer. It must be pointed out that every observer at any position in the Universe has a spherical LSS, with photons received suffering a continuous redshift due to the expansion. Extracted from \cite{Ryden2003}.}
\label{last_scat_surf}
\end{figure}

There are three closely related epochs that need to be distinguished and that occur around this redshift: the \textit{recombination} epoch, defined as the time in which baryons stop being ionized and become neutral by combining with electrons; the \textit{photon decoupling} epoch, in which photon scattering rate with electrons becomes smaller than Hubble parameter (which is the expansion rate of the Universe), turning the Universe from opaque to transparent; and the epoch of \textit{last scattering surface} (Fig. \ref{last_scat_surf}), defined as the time in which a CMB photon last scattered off an electron. Once the expansion rate becomes larger than the scattering rate, it is very unlikely that a photon will scatter again, which makes the last scattering epoch very close to the photon decoupling epoch. That is why in the last section these moments in the history of the Universe were taken to be the same, but here they are properly defined.

\begin{figure}[h!]
\centering
\includegraphics[width=0.9\textwidth]{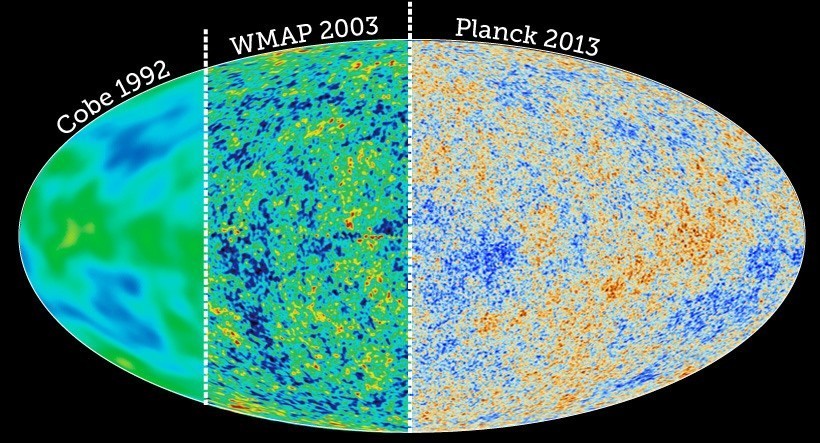}
\caption{CMB temperature anisotropies from satellites with different resolutions (COBE: $7\degree$; WMAP: $0.3\degree$; Planck: $0.07\degree$). Enhancements in the resolution lead to more accuracy on the cosmological information. Extracted from \cite{CMB_resolution}.}
\label{cmb_anisotropies}
\end{figure}

\begin{figure}[h!]
\centering
\includegraphics[width=1.0\textwidth]{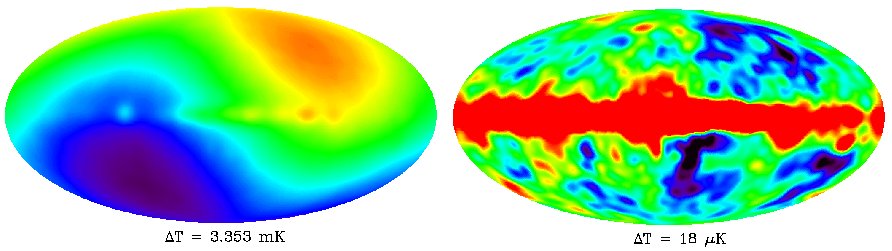}
\caption{Left: Temperature fluctuations as measured by COBE satellite, displaying a ``yin-yang'' pattern due to dipole distortion (blue region: lower temperatures; orange region: higher temperatures). Right: Temperature fluctuations after removing dipole distortion, in which the red strip in the middle represents emission from the Milky Way. In the bottom, temperature fluctuations with respect to the mean temperature are displayed. Adapted from \cite{dipole_CMB}.}
\label{dipole}
\end{figure}

The CMB measurements were performed for the first time by COBE satellite in 1992 \cite{smoot} and refined by other satellite measurements such as WMAP in 2003 \cite{spergel} and Planck in 2013 \cite{planck3}. The difference among these experiments can be seen in Fig. \ref{cmb_anisotropies}. Even though experiment improvements were huge, some of the important discoveries provided by COBE data are still valid and can be elucidated in the following terms:

\begin{enumerate}
\item To any direction in the sky (angular coordinates ($\theta,\phi$)), CMB spectrum is very close to the one of an ideal black-body (Fig. \ref{blackbodyCMB}) with mean temperature
\begin{equation}
    \langle T\rangle = \frac{1}{4\pi} \int T(\theta,\phi) \sin{\theta} d\theta d\phi \approx 2.725\,K;
\end{equation}
\item There is a \textit{dipole distortion} in the temperature maps due to Doppler effect caused by the motion of COBE satellite relative to the frame of reference where CMB is isotropic (Fig. \ref{dipole}), which means that different hemispheres from the sky are slightly blueshifted (or redshifted) to higher (or smaller) temperatures;
\item After dipole distortion removal, the remaining temperature fluctuations are very small in amplitude. This can be better understood if one defines the dimensionless temperature fluctuation across the sky as
\begin{equation}
\frac{\delta T}{T}(\theta,\phi) \equiv \frac{T(\theta,\phi)-\langle T\rangle}{\langle T\rangle}.
\end{equation}

Without dipole distortion, the root mean square temperature fluctuation (which is an average over all points in the sky except the ones in the region contaminated by our own galaxy foreground emission) for COBE data is
\begin{equation}
\left\langle \left(\frac{\delta T}{T}\right)^2 \right\rangle^{1/2} \sim 10^{-5},
\end{equation}
indicating an extraordinary closeness to isotropy.
\end{enumerate}

Due to the huge complexity of the equations involved in generating the full CMB temperature spectrum (there are nine coupled equations that account for all components and for the metric, called Einstein-Boltzmann equations), numerical simulations are necessary to solve the problem. For this purpose, open-source codes such as CAMB \cite{camb} and CLASS \cite{lesgourgues2011} were created, which are written in Fortran and C, respectively. Even with these difficulties, we can elucidate in general lines the equations that lead to CMB anisotropies and also study important parameters that describe some characteristics of the spectrum.

\subsection{Anisotropy power spectrum and Einstein-Boltzmann equations}

Let's consider temperature fluctuations $\delta T/T$ across the sky as measured from a particular experiment. Since each point in the sky has a unique temperature $T(\theta,\phi)$ defined on the surface of the celestial sphere, it is intuitive to expand the fluctuations in spherical harmonics as
\begin{equation}
\label{T_fluctuation}
\frac{\delta T}{T}(\theta,\phi) = \sum_{\ell=0}^{\infty} \sum_{m=-\ell}^\ell a_{\ell m} Y_{\ell m}(\theta, \phi),
\end{equation}
in which $Y_{\ell m}(\theta, \phi)$ represent spherical harmonic functions. Useful information can only be extracted by using statistics since we have a temperature distribution over the whole sky. Thereby, the most important statistical descriptor of $\delta T/T$ is the 2D correlation function $C(\theta)$ (and its corresponding Fourier pair, the 2D power spectrum $C_\ell$) likewise BAO information uses the 3D correlation function $\xi(r)$ (and 3D power spectrum $P(k)$).

For this purpose, consider two different directions in the sky at the last scattering surface. These points have directions $\hat{n}$ and $\hat{n}'$ separated by the angle $\theta$ such that $\cos{\theta}=\hat{n}\cdot\hat{n}'$. The correlation $C(\theta)$ between temperature fluctuations for pair of directions is defined as the product of the fluctuations averaged over all points separated by the same angle $\theta$, in the form
\begin{equation}
\label{C_correlation}
C(\theta) = \left\langle \frac{\delta T}{T}(\hat{n}) \frac{\delta T}{T}(\hat{n}') \right\rangle_{\hat{n}\cdot\hat{n}'=\cos{\theta}}
\end{equation}

If it was possible to obtain from measurements the precise form of $C(\theta)$ for angles from $\theta=0$ to $\theta=180\degree$, we would have a complete statistical description for the fluctuations over all scales. Unfortunately, there are experimental limitations that make it impossible because only a limited range of angular scales are accessible with precision \cite{Ryden2003}. Using \eqref{T_fluctuation} and \eqref{C_correlation}, the correlation function can be written as
\begin{equation}
\label{correl_CMB}
C(\theta) = \frac{1}{4\pi}\sum_{\ell=0}^\infty (2\ell+1) \, C_\ell \, P_\ell (\cos{\theta}),
\end{equation}
where $P_\ell$ are Legendre polynomials.

The important thing in the expansion \eqref{correl_CMB} are the $C_\ell$ coefficients, because now we can characterize the correlation function $C(\theta)$ by its $C_\ell$ multipole moments, which are the way CMB data are usually displayed. Generally speaking, a multipole moment is a measure of temperature fluctuations at angular scale $\theta\sim 180\degree/\ell$ \cite{dodelson2003}. Thus, it does not matter whether one prefers to use $\theta$ or $\ell$ to plot the anisotropy spectrum.

In order to obtain the $C_\ell$ coefficients, one needs to recall the definition of temperature perturbation $\Theta$ as mentioned in the last section, which comes from the equation
\begin{equation}
T(\bm{x},\hat{p},\eta) = T(\eta)\left[1+\Theta(\bm{x},\hat{p},\eta)\right].
\end{equation}
As one can see, the temperature field is defined at every point in space ($\bm{x}$) and time ($\eta$), but our observations can only be performed here ($\bm{x}_0$) and now ($\eta_0$). The information we receive is based exclusively on the direction of the incoming photons ($\hat{p}$), which means we observe temperature fluctuations in all directions but we cannot, at least at first sight, infer that the temperature in a specific direction is due to a photon that came right from the last scattering surface because a photon might have its path altered due to sources in the way (there are other effects to account for potential wells in the path between last scattering surface and us, such as Sunyaev-Zel'dovich and Sachs-Wolfe \cite{weinberg2008}). Instead of working with angular coordinates ($\theta,\phi$), we will stick to the photon direction coordinate $\hat{p}$ and rewrite the expansion \eqref{T_fluctuation} as
\begin{equation}
\label{multipoles_CMB}
\Theta(\bm{x},\hat{p},\eta) = \sum_{\ell=0}^{\infty} \sum_{m=-\ell}^\ell a_{\ell m}(\bm{x},\eta) Y_{\ell m}(\hat{p}).
\end{equation}

All the relevant information from the temperature field $T$ is encapsulated by the amplitudes $a_{\ell m}$ (which are related to $C_\ell$, as we shall see). In order to invert \eqref{multipoles_CMB} and write $a_{\ell m}$ as a function of $\Theta$, we can use the orthogonality condition for the spherical harmonics, normalized via
\begin{equation}
\label{spher_norm}
\int d\Omega \,Y_{\ell m}(\hat{p}) \,Y_{\ell' m'}^*(\hat{p}) = \delta_{\ell \ell'} \delta_{mm'},
\end{equation}
in which $\Omega$ is the solid angle covered by $\hat{p}$. Multiplying \eqref{multipoles_CMB} by $Y^*_{\ell m}(\hat{p})$, integrating it and using \eqref{spher_norm}, the result for $a_{\ell m}$ is
\begin{equation}
\label{alm_CMB}
a_{\ell m}(\bm{x},\eta) = \int \frac{d^3k}{(2\pi)^3} e^{i\bm{k}\cdot\bm{x}} \int d\Omega \,Y_{\ell m}^*(\hat{p}) \,\Theta(\bm{k},\hat{p},\eta),
\end{equation}
where we have written $\Theta(\bm{x})$ in terms of its Fourier transform $\Theta(\bm{k})$, which is the one we have equations to work with.

The $a_{\ell m}$'s cannot be exactly predicted, and as usual it is their statistical properties we are interested in order to extract information. For this purpose, the distribution from which they are drawn is the main goal of CMB measurements \cite{dodelson2003}. As the coefficients $a_{\ell m}$ are assumed to be statistically independent \cite{amendola}, we define the mean and variance of the $a_{\ell m}$'s, respectively, as 
\begin{equation}
\label{mean_var_Cl}
\langle a_{\ell m} \rangle \equiv 0 \qquad \text{and} \qquad \langle a_{\ell m}a_{\ell' m'}^* \rangle \equiv \delta_{\ell \ell'} \delta_{mm'} C_\ell,
\end{equation}
where the variance $C_\ell = \langle \left|a_{\ell m}\right|^2 \rangle$ is called \textit{CMB temperature power spectrum}.

We can now find an expression for $C_\ell$ by using \eqref{alm_CMB} and \eqref{mean_var_Cl}. For that matter, we need first to evaluate $\langle \Theta(\bm{k},\hat{p})\Theta^*(\bm{k}',\hat{p}') \rangle$, where $\eta$ dependence is implicit. To solve the problem, we need to rewrite the temperature perturbation as $\delta\times (\Theta/\delta)$, in which dark matter perturbation $\delta$ does not depend on $\hat{p}$. As the ratio $\Theta/\delta$ does not depend on the initial amplitude of a mode, it can be taken out of the average. Thus,
\begin{align}
\label{ratio_theta_delta}
\langle \Theta(\bm{k},\hat{p})\Theta^*(\bm{k}',\hat{p}') \rangle &= \langle \delta(\bm{k})\delta^*(\bm{k}') \rangle \frac{\Theta(\bm{k},\hat{p})}{\delta(\bm{k})} \frac{\Theta^*(\bm{k}',\hat{p}')}{\delta^*(\bm{k}')} \nonumber \\
&= (2\pi)^3 \delta^3(\bm{k}-\bm{k}')P(k) \frac{\Theta(k,\hat{k}\cdot\hat{p})}{\delta(k)} \frac{\Theta^*(k,\hat{k}\cdot\hat{p}')}{\delta^*(k)},
\end{align}
where in the second equality it was considered the fact that different modes are uncorrelated \cite{amendola} and that the ratio $\Theta/\delta$ depends only on the magnitude of $\bm{k}$ and the product $\hat{k}\cdot\hat{p}$ \cite{dodelson2003}. Therefore, combining \eqref{alm_CMB}, \eqref{mean_var_Cl} and \eqref{ratio_theta_delta} yields
\begin{equation}
C_\ell = \int \frac{d^3k}{(2\pi)^3} P(k) \int d\Omega \,Y_{\ell m}^*(\hat{p}) \frac{\Theta(k,\hat{k}\cdot\hat{p})}{\delta(k)} \int d\Omega' \,Y_{\ell m}(\hat{p}') \frac{\Theta^*(k,\hat{k}\cdot\hat{p}')}{\delta^*(k)}.
\end{equation}

One can expand $\Theta(k,\hat{k}\cdot\hat{p})$ and $\Theta^*(k,\hat{k}\cdot\hat{p}')$ using the inverse of \eqref{multipole}, $\Theta(k,\hat{k}\cdot\hat{p}) = \sum_\ell (-i)^\ell (2\ell+1)\,\mathcal{P}_\ell(\hat{k}\cdot\hat{p})\, \Theta_\ell(k)$, which leads to
\begin{align}
C_\ell = &\int \frac{d^3k}{(2\pi)^3} P(k) \sum_{\ell'\ell''}(-i)^{\ell'}(i)^{\ell''}(2\ell'+1)(2\ell''+1) \frac{\Theta_{\ell'}(k)\Theta_{\ell''}^*(k)}{\left|\delta(k)\right|^2} \nonumber \\
&\times \int d\Omega \, \mathcal{P}_{\ell'}(\hat{k}\cdot\hat{p}) Y_{\ell m}^*(\hat{p}) \int d\Omega' \, \mathcal{P}_{\ell''}(\hat{k}\cdot\hat{p}') Y_{\ell m}(\hat{p}').
\end{align}

The two angular integrals are equal to $4\pi Y_{\ell m}(\hat{k})/(2\ell+1)$ (or the complex conjugate) in the case $\ell'=\ell$ and $\ell''=\ell$, otherwise they are zero. The remaining $\left|Y_{\ell m}\right|^2$ combines with the angular part of the $d^3k$ integral (which is equal to one), leading to
\begin{equation}
\label{Cl_CMB}
C_\ell = \frac{2}{\pi} \int_0^\infty dk\,k^2 P(k) \left|\frac{\Theta_{\ell}(k)}{\delta(k)}\right|^2.
\end{equation}

If one uses \eqref{Cl_CMB} to evaluate CMB power spectrum, it is necessary to know the form of the matter power spectrum, dark matter overdensity and the multipole moments as a function of scale, while performing an integral over all Fourier modes. Approximate solutions for these terms can be found in \cite{dodelson2003} and \cite{weinberg2008}, but there is no general solution for the problem and, in practice, one needs a code for solving numerically the set of \textit{Einstein-Boltzmann equations}. As an example, for $\Lambda$CDM model we have nine coupled differential equations describing interactions among components and the metric of the Universe (scalar perturbations are the main source of CMB anisotropies). These equations are summarized as \cite{dodelson2003}
{\small
\begin{align}
&\Theta'+ik\mu\Theta = -\Phi' -ik\mu\Psi-\tau'[\Theta_0-\Theta+\mu v_b-\frac{1}{2}\mathcal{P}_2(\mu)\Pi] \; (\text{Boltzmann eq. for photons}),\\
&\Theta'_P + ik\mu \Theta_P = -\tau' \left[-\Theta_P + \frac{1}{2}(1-\mathcal{P}_2(\mu))\Pi \right] \quad (\text{Boltzmann eq. for polarization}),\\
&\delta'+ikv = -3\Phi' \quad (\text{Boltzmann eq. for CDM}),\\
&v'+\frac{a'}{a}v = -ik\Psi \quad (\text{Boltzmann eq. for CDM}),\\
&\delta_b'+ikv_b = -3\Phi' \quad (\text{Boltzmann eq. for baryons}),\\
&v'_b+\frac{a'}{a}v_b = -ik\Psi +\frac{\tau'}{R_s}[v_b+3i\Theta_1] \quad (\text{Boltzmann eq. for baryons}),\\
&\mathcal{N}' +ik\mu \mathcal{N} = -\Phi' -ik\mu \Psi \quad (\text{Boltzmann eq. for neutrinos}),\\
&k^2\Phi+3\frac{a'}{a}\left(\Phi'-\Psi\frac{a'}{a}\right) = 4\pi Ga^2\left[\rho\delta+\rho_b\delta_b+4\rho_\gamma\Theta_0+4\rho_\nu\mathcal{N}_0\right] \quad (\text{Einstein equation}),\\
&k^2(\Phi+\Psi) = -32\pi Ga^2\left[\rho_\gamma\Theta_2 + \rho_\nu\mathcal{N}_2\right] \quad (\text{Einstein equation}),
\end{align}
}%
and, if dark energy is not $\Lambda$CDM, a new equation must be added to account for dark energy perturbations and possible interactions with other components.

\begin{figure}[h!]
\centering
\includegraphics[width=1.0\textwidth]{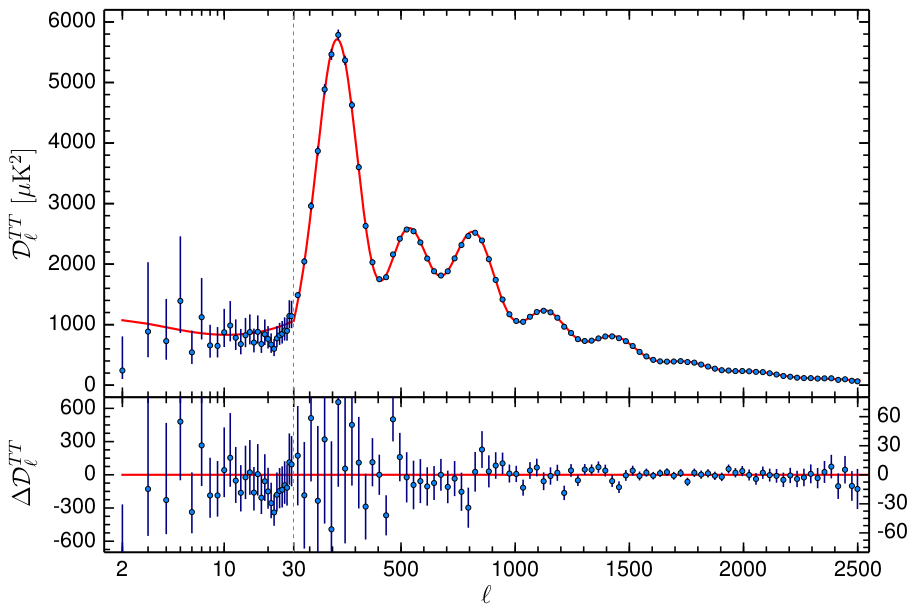}
\caption{Temperature power spectrum measurements from Planck 2015 and best-fit $\Lambda$CDM model. The plot division at $\ell=30$ represents the fact that measurements above and below this multipole are obtained using different methods (Plik cross-half-mission and Commander component-separation algorithm, respectively). In the lower panel, residuals with respect to this theoretical spectrum are displayed. Error bars show $\pm 1\sigma$ uncertainties. Extracted from \cite{planck2}.}
\label{planck_TT_spectrum}
\end{figure}

From these ones, three of them were already used in BAO section in a compact form (eqs. \eqref{theta_eq}, \eqref{delta_baryon} and \eqref{veloc_baryon}). The term $\Pi = \Theta_2 + \Theta_{P2} + \Theta_{P0}$ and $\Theta_P$ are terms due to polarization of the CMB, $\delta$ and $v$ represent CDM overdensity and velocity, respectively, and $\mathcal{N}$ is the neutrino perturbation (it is the analogue of the photon perturbation variable $\Theta$). It is now clear that either one uses approximations together with equation \eqref{Cl_CMB} or solves numerically the full set of Einstein-Boltzmann equations to find $\Theta$ (in order to get more accurate results) and combines with \eqref{alm_CMB} and \eqref{mean_var_Cl} to obtain finally the CMB power spectrum. In the literature, temperature power spectrum is usually plotted as
\begin{equation}
\mathcal{D}_\ell \equiv \frac{\ell(\ell+1)}{2\pi} C_\ell \,\,
\end{equation}
in units of $\mu K^2$, as displayed in Fig \ref{planck_TT_spectrum}.

Our approach for evaluating CMB power spectrum is numerical: we use CAMB code \cite{lewis,CAMB_website}, a code publicly available for over a decade, very well tested and improved by the community. In order to predict CMB power spectrum with high accuracy, we also take into account lensing in the code, which improves the smoothing effect on the acoustic peaks by $5\%$. This is the correct procedure as pointed out in \cite{planck2}.

\subsection{Optical depth}

An important cosmological parameter related to CMB spectrum is the \textit{Thomson scattering optical depth due to reionization} $\tau$. Before giving a definition of it, it is useful to introduce some concepts. The Universe has gone through important phases in its history, some of them having hydrogen as the main factor related to these phase transitions. We can cite, for example, recombination (as already described earlier) and \textit{reionization epoch}, the latter being the period in which the Universe goes from being neutral to becoming ionized due to the formation of the first generation of galaxies, which emitted ultraviolet radiation and started ionizing large intergalactic gas clouds \cite{Zaroubi2013}.

Even though in this epoch there was a considerable amount of helium, the significant physics can be understood simply by considering just hydrogen in its neutral form (for which we shall use the letter $H$) and its ionized form (a single proton, letter $p$). Charge neutrality requires the number density of free protons and free electrons to be the same, $n_e=n_p$. In order to quantify the ionization level of the Universe, one defines \textit{free electron fraction} (or fractional ionization) as
\begin{equation}
\label{fract_ion}
X_e \equiv \frac{n_p}{n_p+n_H} = \frac{n_p}{n_b} = \frac{n_e}{n_b},
\end{equation}
in which the subscript $b$ stands for baryons. The quantity $X_e$ ranges from $X_e=1$ when baryons are fully ionized to $X_e = 0$ when the Universe is entirely neutral.

In the reionization period, the interactions happened primarily between photons and electrons through Thomson scattering, which made these particle to exchange energy and momentum. Thus, the mean free path of a photon in this epoch (the mean distance it travels before scattering from a electron) is
\begin{equation}
\lambda = \frac{1}{n_e\sigma_T},
\end{equation}
where $\sigma_T$ is Thomson scattering cross-section ($\sigma_T = 6.65\times 10^{-29}\,m^2$). Due to the fact that photons travel at the speed of light $c$, their scattering rate is
\begin{equation}
\Gamma = \frac{c}{\lambda} = n_e\sigma_T c.
\end{equation}

Photons we observe today came from the last scattering surface and, in its path toward us, may have suffered scattering in the reionization epoch. Therefore, at the time interval $t\rightarrow t+dt$ the probability that a photon is scattered off an electron is $dP=\Gamma(t)dt$. The definition of optical depth to reionization is the integral of the scattering rate over all cosmic history, which is equivalent to the expected number of scatterings it has suffered since earlier times. Mathematically,
\begin{equation}
\tau \equiv \int_0^{t_0} \Gamma(t)\, dt = \int_0^{\eta_0} d\eta\,a \, n_e(\eta)\, \sigma_T \,c,
\end{equation}
where $n_e$ is the number density of free electrons produced by reionization. Even though the integral is over all cosmic history, we do not need to worry about the fact that there is a small residual ionization fraction coming from recombination because this one is of order $\sim 10^{-3}$, being safely neglected \cite{Lewis2008}.

Using the relation $dt = da/(aH) = -dz/[(1+z)H]$ and \eqref{fract_ion}, one can rewrite $\tau$ as
\begin{equation}
\tau = \int_0^\infty \frac{dz}{(1+z)H(z)} X_e(z)(1+z)^3 n_b^{(0)} \sigma_T \,c.
\end{equation}

Considering the fact that reionization happens at a matter-dominated epoch, $H(z)\sim (1+z)^{3/2}$ and therefore
\begin{equation}
\label{tau_CMB}
\tau \propto \int dz\, X_e(z)\sqrt{1+z} \propto \int d\left[(1+z)^{3/2}\right]X_e.
\end{equation}

Equation \eqref{tau_CMB} can be used to evaluate $\tau$ either by using some differential equation describing the evolution of $X_e(z)$ \cite{dodelson2003} or by parametrizations of $X_e$ as a function of $y\equiv (1+z)^{3/2}$, as it is standard procedure in CAMB code \cite{Lewis2008}. Recent measurements for $\tau$ come from Planck satellite, yielding the value $\tau = 0.078\pm 0.019$ \cite{planck2}.

\subsection{Primordial fluctuations: scalar spectral index and amplitude perturbation}

Inflation is the best scenario to explain why uncorrelated scales observed today have such similar temperatures, as we have seen, at the same time it solves other independent problems in cosmology and particle physics at once \cite{Guth1981}. Besides, inflation is a mechanism for generating the primordial fluctuations that led to CMB anisotropies, which is why improvements on CMB measurements could confirm inflationary theories. One of the ways of checking whether CMB points to inflation is to work with the \textit{primordial power spectrum} (scalar spectrum) that comes from the metric perturbations \eqref{newton_gauge}, which can be demonstrated to be \cite{dodelson2003} 
\begin{equation}
\label{primord_Pk}
P_\Phi(k) = \frac{8\pi}{9k^3}  \left.\frac{H^2}{\epsilon M_{Pl}^2}\right|_{aH=k} \equiv \frac{50\pi^2}{9k^3}\left(\frac{k}{H_0}\right)^{n_s-1} \delta_H^2 \left(\frac{\Omega_m^{(0)}}{D_1(a=1)}\right)^2,
\end{equation}
where $M_{Pl}$ is the Planck mass, $n_s$ is the \textit{scalar spectral index}, $\delta_H$ is the scalar amplitude at horizon crossing $aH=k$, $D_1$ is the \textit{growth function} which describes the growth of matter perturbations at late times \cite{dodelson2003}, $\Omega_m^{(0)}$ is the matter density parameter today and $\epsilon$ is the slow-roll inflation parameter defined by
\begin{equation}
\epsilon \equiv \frac{d}{dt} \left(\frac{1}{H}\right) = -\frac{H'}{aH^2},
\end{equation}
which measures changes in the Hubble rate during inflationary era that are supposed to be small due to dominance of the potential energy.

If a spectrum has the property that $k^3P_\Phi(k)$ is a constant (i.e., has no dependence on the scale), then it is called a \textit{scale-invariant} (also called scale-free or Harrison-Zel'dovich) spectrum, which is characterized by $n_s=1$. Therefore, the scalar index quantifies deviations from scale invariance and from perfect de Sitter limit \cite{Baumann2009}. For example, most models of inflation are based on the idea that a scalar field $\phi(\bm{x},t)$ (called \textit{inflaton}) runs the inflationary era, ``slowly rowling'' to the minimum of the potential energy $V(\phi)$ during the short period that inflation takes place. In these models, it can be demonstrated \cite{Baumann2009} the relation
\begin{equation}
\label{ns_slowroll}
n_s-1 = -2(2\epsilon+\delta),
\end{equation}
where $\delta$ is another slow-roll inflation parameter, given by
\begin{equation}
\delta \equiv \frac{1}{H} \frac{d^2\phi/dt^2}{d\phi/dt} = -\frac{1}{aH\phi'} \left[3aH\phi'+a^2 \frac{dV}{d\phi}\right],
\end{equation}
with $\phi$ representing now the zero-order term $\phi^{(0)}(t)$ from the space-time dependent scalar field.

From \eqref{ns_slowroll}, it becomes obvious that an inflationary scenario predicts a deviation from scale invariance, since $\epsilon$ and $\delta$ are predicted to be small but non-zero. This is exactly what is observed: Planck measurements constrain the scalar spectral index to the value $n_s=0.9655\pm 0.0062$ \cite{planck2}, giving much credit to the inflationary scenario.

It is customary to model the almost scale-invariant spectrum as a power-law close to a reference scale $k_0$ (also called \textit{pivot scale} or \textit{tilt}) as \cite{Baumann2009}
\begin{equation}
\label{spec_As}
\mathcal{P}_\mathcal{R}(k) \equiv A_s \left( \frac{k}{k_0} \right)^{n_s-1},
\end{equation}
where $A_s$ is the amplitude of the scalar spectrum measured at the pivot scale $k_0=0.05\,\text{Mpc}^{-1}$, approximately in the middle of the logarithmic range of scales probed by Planck. The choice of $k_0$ value is made to reduce degeneracy between $n_s$ and $A_s$ \cite{planck2}.

Using \eqref{primord_Pk}, it is possible to identify the spectrum amplitude as $\delta_H\sim \mathcal{P}_\mathcal{R}(k_0)^{1/2}=A_s^{1/2}$, which is constrained by Planck measurements to be $\ln{(10^{10}A_s)} = 3.089 \pm 0.036$ \cite{planck2}, where $\ln$ is the natural logarithm (the logarithm of $A_s$ is also called \textit{log power of the primordial curvature perturbations}). From \eqref{spec_As}, the scalar index can also be defined as
\begin{equation}
n_s-1 \equiv \frac{d\ln{\mathcal{P}_\mathcal{R}(k)}}{d\ln{k}}.
\end{equation}

We shall not work with tensor power spectrum that comes from tensor perturbations ($n_T=0$) neither with higher-order effects such as running of scalar ($dn_s/d\ln{k}=0$), since Planck measurements are weakly sensitive to these parameters \cite{planck2}.

While CMB theoretical power spectrum comes from CAMB code, CMB data comes from the most recent measurements of the Planck satellite. This information is encoded in the \textit{Planck likelihood} code package (also called \textit{Plik}), available at \cite{planck_lkl}, which takes as input the theoretical power spectrum and a set of ``nuisance'' parameters describing unresolved foreground and instrument calibration, and outputs the log-likelihood (which shall be described in the next chapter). In this dissertation, we use two likelihood codes: Plik TT likelihood, which uses a high-$\ell$ ($30\leq\ell\leq2508$) temperature-only dataset together with 16 ``nuisance'' parameters, and low TEB likelihood (lowP), which uses low-$\ell$ ($2\leq\ell\leq29$) temperature plus polarization datasets with a single ``nuisance'' parameter. The latter likelihood code is important to break the large degeneracy between $\tau$ and $\ln{A_s}$ \cite{Aghanim2015}, while using polarization for high-$\ell$ multipoles is not necessary because it does not effectively improve parameter constraints \cite{planck2}.

\section{Cosmic chronometers}

\subsection{Overview}

\textit{Cosmic chronometers} (or cosmic clocks) is a nearly model-independent tool developed for directly measuring the evolution of the Hubble parameter. This method is based on objects (in this case, galaxies) from which one may roughly infer the variation in the age of the Universe with the redshift \cite{jimenez2002}. Using these ``cosmological clocks'', one may infer the age difference $\Delta t$ between samples of nearby early-type galaxies (ETG) from its spectral lines and also measure the redshift difference $\Delta z$ between them. The connection between these measurements and the Hubble parameter is made by identifying the derivative $dz/dt$ with the ratio $\Delta z/\Delta t$ and by using the straightforward equation
\begin{equation}
H(z) \equiv \frac{\dot{a}}{a} = -\frac{1}{1+z} \frac{dz}{dt}
\end{equation}
to describe the Hubble parameter observationally. As this formula is very simple and has no integral nature (differently from methods based on integrated quantities, such as luminosity distance from supernovae), it has a good sensitivity for constraining a dynamical dark energy equation of state. Also, the way cosmic chronometers are applied, with a ``differential age method'', is much more reliable than working with an ``absolute age method'', which is much more susceptible to systematics and potential effects of galaxy evolution \cite{Alcaniz2001}. Therefore, the relevant physics is ranged between the redshifts where the differential is applied.

Although the idea of this method seems very simple, complications in estimating $H(z)$ are huge. For better exploration of the method, one needs to select samples of red elliptical galaxies in the form of \textit{passively evolving galaxies}, which are defined as the ones with old stellar population, low-rate stellar formation and with stellar mass $\mathcal{M}$ above $10^{11}\mathcal{M}_\odot$ \cite{moresco2012}. Besides, errors in the final estimates can be large (even more than $30\%$), requiring combination with other independent datasets to better constrain parameters.

\begin{figure}[h!]
\centering
\includegraphics[width=0.8\textwidth]{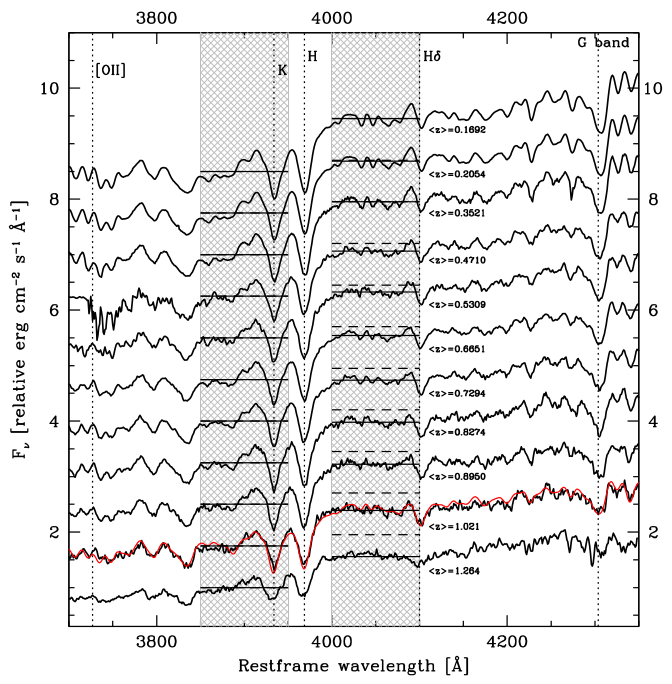}
\caption{Spectra from real and simulated ETGs. The plot shows the mean stacked spectra for various real galaxies (except for the red one, which is a simulated spectrum) taken in redshift bins, with the central redshift value of each bin displayed. The two hatched areas represent blue (3850-3950{\AA}) and red (4000-4100{\AA}) bands, while the solid and dashed segments inside both indicate the average fluxes of the stacked spectra and the average fluxes of the lowest redshift spectrum inside each bin, respectively. Extracted from \cite{moresco2012}.}
\label{clocks_spectra}
\end{figure}

One of the best ways to perform the analysis is to use the 4000 {\AA} break (which we shall name D4000) present in ETG spectra, since it is linearly related to the age of the galaxy in the case where the galaxy has an old stellar population. This break models a discontinuity in the spectrum from these galaxies close to the wavelength $\lambda_{rest}=4000${\AA}, which occurs due to metal absortion lines with amplitude that scales linearly with metallicity Z and age of the stellar population \cite{moresco2012}. Because star formation history (SFH) has a small influence only if one considers old passively evolving galaxies, the criteria adopted above must be respected. The 4000 {\AA} break feature is displayed in Fig. \ref{clocks_spectra}.

\subsection{Differential age method from galaxy surveys}

The D4000 is defined as ratio between the continuum flux densities in a red band (4050-4250{\AA}) and a blue band (3750-3950{\AA}) around the 4000{\AA} wavelength \cite{Bruzual1983}. It is written as
\begin{equation}
D4000 \equiv \frac{(\lambda_2^{blue}-\lambda_1^{blue})\int^{\lambda_2^{red}}_{\lambda_1^{red}}F_\nu\, d\lambda}{(\lambda_2^{red}-\lambda_1^{red})\int^{\lambda_2^{blue}}_{\lambda_1^{blue}}F_\nu\, d\lambda},
\end{equation}
although reference \cite{moresco2012} adopts a slightly different range for the bands in order to reduce errors due to dust reddening: 3850-3950{\AA} and 4000-4100{\AA}. From now on, this narrower band definition will be represented as $D4000_n$.

As mentioned, if the galaxy is old and passively evolving, there is a linear relation between $D4000_n$ and galaxy age (with constant metallicity) of the form
\begin{equation}
D4000_n(Z,SFH) = A(Z,SFH)\cdot age + B(Z,SFH),
\end{equation}
where $A(Z,SFH)$ is a factor to be calculated by means of stellar population models and $B(Z,SFH)$ is a ``nuisance'' parameter that must vanish by applying the differential age method as
\begin{equation}
\Delta D4000_n =  A(Z,SFH)\cdot \Delta age.
\end{equation}
Now becomes clear one of the advantages of the method: it directly traces the age evolution of ensembles of galaxies.

Thereby, Hubble evolution can be rewritten as a function of $D4000_n$ and $A(Z,SFH)$, which leads to
\begin{equation}
H(z) = -\frac{A(Z,SFH)}{1+z} \frac{dz}{dD4000_n}.
\end{equation}

\begin{figure}[h!]
\centering
\includegraphics[width=0.8\textwidth]{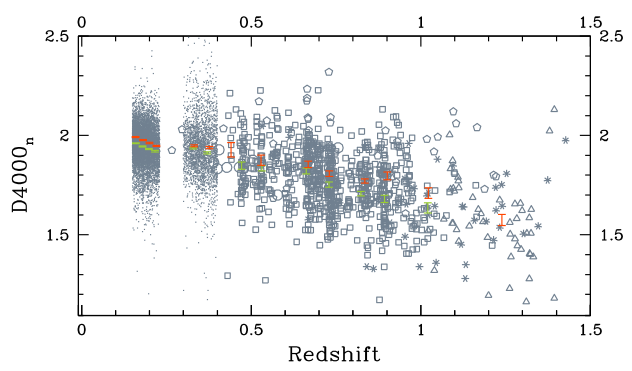}
\caption{The $D4000_n$-redshift relation. This total sample is a combination of the following subsamples (and number of galaxies): SDSS-DR6 MGS (7943), SDSS-DR7 LRGs (2459), Stern et al. sample (9), zCOSMOS 20k (746), K20 (50), GOODS-S (46), Cluster BCG (5), GDDS (16), UDS (50). Extracted from \cite{moresco2012}.}
\label{D4000n}
\end{figure}

As an example, measurements of $D4000_n$ for a sample of 11324 ETGs are displayed in Fig. \ref{D4000n} as gray dots (these data are the base for the calculation of 8 from the 31 cosmic chronometers measurements for $H(z)$ that will be presented later on). Orange and green dots represent the median value of $D4000_n$ in a determined redshift bin for different mass ranges: green for lower galaxy masses and orange for higher galaxy masses. The choice of the size of the bins depends on the amount of galaxies available to perform the statistic, and it has a crucial importance in the sampling. Also, for a redshift smaller than $0.3$, metallicity is a feature that can be estimated from the spectrum, while for a redshift larger than $0.3$ this information is no longer available and one needs to take a conservative choice for metallicity \cite{moresco2012}.

Another procedure that has to be made is to calibrate the relation between $D4000_n$ and galaxy age by using models of stellar population synthesis in order to evaluate $A(Z,SFH)$. It is important to confirm that there is no bias coming from the model, and for that matter reference \cite{moresco2012} employs two independent samples of synthetic spectra (BC03 and MaStro), finding no significant difference between them. Other details about all the procedures to obtain estimates for $H(z)$ are fully described in reference \cite{jimenez2002} and \cite{moresco2012}.

The dataset for cosmic chronometers we are going to use is a compilation of 31 measurements for $H(z)$ in the redshift range $0<z<2$, as shown in Table \ref{table_clocks}.

\begin{table}[h]
\centering
\begin{tabular}{@{}cc|cc@{}}
\toprule
$z$  &  $H(z)$  &  $z$  & $H(z)$         \\ \midrule
$0.07$                       & $69.0 \pm 19.6$                     & $0.4783$                    & $80.9 \pm 9.0$                       \\
$0.09$                       & $69.0 \pm 12.0$                     & $0.48$                      & $97.0 \pm 62.0$                      \\
$0.12$                       & $68.6 \pm 26.2$                     & $0.5929$                    & $104.0 \pm 13.0$                     \\
$0.17$                       & $83.0 \pm 8.0$                      & $0.6797$                    & $92.0 \pm 8.0$                       \\
$0.1791$                     & $75.0 \pm 4.0$                      & $0.7812$                    & $105.0 \pm 12.0$                     \\
$0.1993$                     & $75.0 \pm 5.0$                      & $0.8754$                    & $125.0 \pm 17.0$                     \\
$0.2$                        & $72.9 \pm 29.6$                     & $0.88$                      & $90.0 \pm 40.0$                      \\
$0.27$                       & $77.0 \pm 14.0$                     & $0.9$                       & $117.0 \pm 23.0$                     \\
$0.28$                       & $88.8 \pm 36.6$                     & $1.037$                     & $154.0 \pm 20.0$                     \\
$0.3519$                     & $83.0 \pm 14.0$                     & $1.3$                       & $168.0 \pm 17.0$                     \\
$0.3802$                     & $83.0 \pm 13.5$                     & $1.363$                     & $160.0 \pm 33.6$                     \\
$0.4$             & $95.0 \pm 17.0$                                & $1.43$                      & $177.0 \pm 18.0$                     \\
$0.4004$                     & $77.0 \pm 10.2$                     & $1.53$                     & $140.0 \pm 14.0$                     \\
$0.4247$                     & $87.1 \pm 11.2$                     & $1.75$                     & $202.0 \pm 40.0$                     \\
$0.4497$                     & $92.8 \pm 12.9$                     & $1.965$                     & $186.5 \pm 50.4$                     \\
$0.47$                     & $80 \pm 50$    \\

\bottomrule
\end{tabular}
\caption{Cosmic chronometers dataset. Hubble parameter is in $[km\,s^{-1}\,Mpc^{-1}]$. Extracted from \cite{adria2018}.}
\label{table_clocks}
\end{table}

\section{Supernovae}

\subsection{Overview}

\begin{figure}[h!]
\centering
\includegraphics[width=0.6\textwidth]{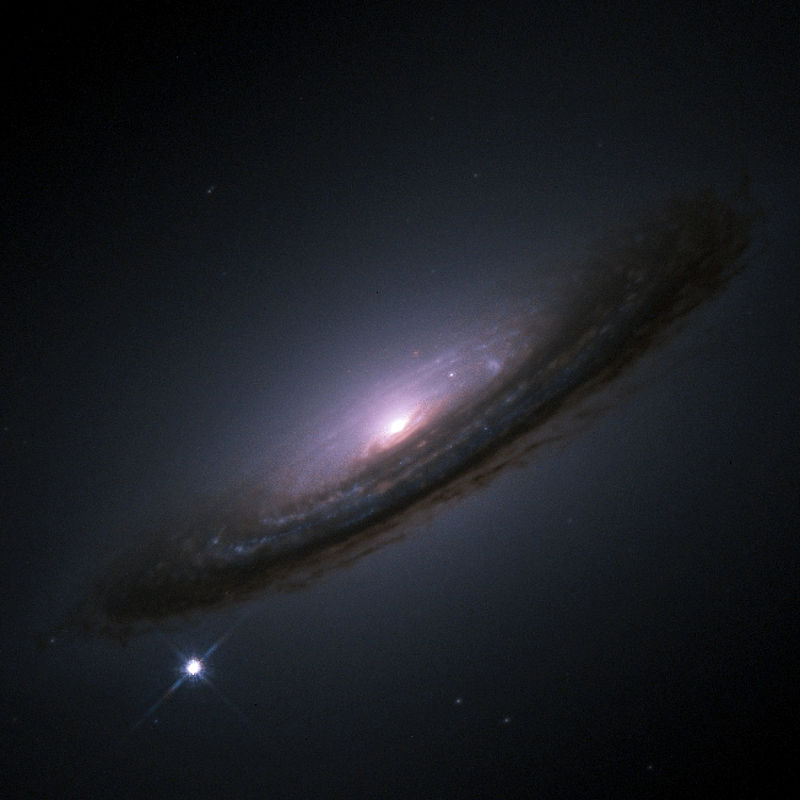}
\caption{Hubble Space Telescope image from the  Type Ia supernova 1994D (lower left) on the border of galaxy NGC 4526. Supernova explosion is an extremely luminous phenomenon that induces a burst of radiation. Extracted from \cite{SN_galaxy}.}
\label{supernova_fig}
\end{figure}

A supernova is an event that occurs due to the death of certain types of stars, such that its brightness can be so intense as the brightness of a whole galaxy, as can be seen in Fig. \ref{supernova_fig}. Historically, supernovae were divided in two classes, according to their spectra: Type I and Type II supernovae. Type I supernovae have no hydrogen absorption lines in their spectra, while Type II supernovae do \cite{Ryden2003}. For Type I supernovae there are three subdivisions: Type Ia (SN), characterized by an absorption line of singly ionized silicon; Type Ib, defined as supernovae with a line of helium in their spectra; and Type Ic, the ones that lack both lines. As Type Ib and Type Ic have similar properties to Type II supernovae (they are formed due to core collapse of the original star), we will focus on Type Ia supernovae because they have mechanisms of origin and evolution totally different of the previous ones and they are objects with ``standardizable'' properties.

\begin{figure}[h!]
\centering
\includegraphics[width=0.6\textwidth]{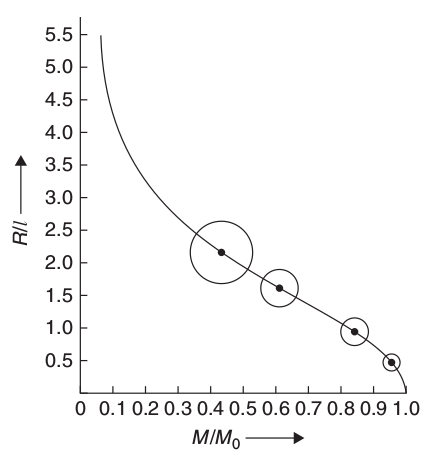}
\caption{In 1939, Chandrasekhar demonstrated that due to Pauli exclusion principle a white dwarf cannot have a mass $M$ bigger than the critical value $M_0\approx 1.44\,M_\odot$, which is interpreted as the maximum mass such that electron degeneracy pressure is able to counterbalance white dwarf self-gravity. The graphic shows the relation between white dwarf radius (in terms of the characteristic length $l\approx 3.86\times 10^8cm$) and its mass (in terms of $M_0$). Extracted from \cite{Pathria2011}.}
\label{chandrasekhar}
\end{figure}

It is believed that Type Ia supernovae occur when a white dwarf (a star whose gravity is counterbalanced with its electron degeneracy pressure) in a binary system accretes enough matter from its binary companion until it reaches Chandrasekhar limit \cite{Chandrasekhar1939}, which renders the white dwarf unstable by increasing its temperature and density, ultimately leading to a thermonuclear explosion (this can be demonstrated by applying Fermi-Dirac statistics) \cite{weinberg2008}. Because the exploding star has always a mass close to this limit (approximately $1.44\, M_\odot$, see Fig. \ref{chandrasekhar}), supernovae absolute luminosity varies for only a small amount at peak brightness, making them one of the best distance indicators we have today in cosmology (we call objects with this property \textit{standardizable candles} for fact that their luminosity can be standardized). Thus, the main information one needs to extract from a supernova is the \textit{light curve} (absolute luminosity versus time since peak brightness).

The occurrence rate of supernovae in our galaxy is one per century, approximately. Even though it is a small rate, the fact that supernovae explosion is extremely luminous makes much easier to detect them even at very large distances (up to $z\sim 1$). For some days (and sometimes months), Type Ia supernovae light can even outshine light from its host galaxy.

\begin{figure}[h!]
\centering
\includegraphics[width=1.0\textwidth]{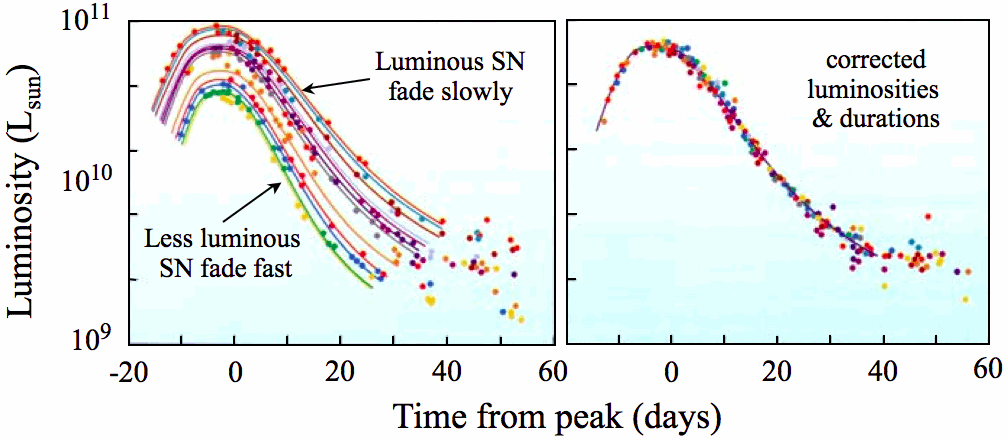}
\caption{Left: Type Ia supernovae light curves, displaying their differences at peak brightness and showing there is a relation between peak and width of each light curve. Right: Single light curve that represents the correction (calibration) for the previous ones, which allows different supernovae to be used as standard candles. Extracted from \cite{Supernova_light} and adapted from \cite{Riess1995}.}
\label{lightcurves}
\end{figure}

Although variation in the absolute magnitude of different supernovae is small, this effect needs to be taken into account and supernovae need calibration in order to use their information as a cosmological probe. For example, the average luminosity of Type Ia supernovae is, at peak brightness, $L=4\times 10^9\,L_\odot$, while Cepheid methods for determining distances show that Type Ia supernovae peak luminosity lie in range $L\approx 3\rightarrow 5\times 10^9\,L_\odot$ \cite{Ryden2003}. In order to treat this variation in magnitude, M. Phillips \cite{Phillips1993} studied a sample of nine supernovae in the B, V and I bands of the spectrum, showing there is a linear relation between peak magnitude and the amount in magnitude that the light curve decays from its peak brightness at the time interval of 15 days ($\Delta m_{15}$). For this reason, it becomes evident there is a correlation between light curve width and peak luminosity: the higher the peak brightness, the broader the light curve (Fig. \ref{lightcurves}). One of the first calibrations using this method was performed at the end of the 1990s from a local sample ($z<<1$) with the Cal\'an/Tololo survey \cite{Hamuy1996}.

\subsection{Magnitude system}

In order to relate luminosity distance \eqref{luminosity_dist} to observational data, one needs first to define the concept of \textit{magnitude} used to express luminosity and flux. The magnitude system, created by the Greek astronomer Hipparchus in the second century BC, was divided into six classes depending on the apparent brightness of visible stars: the brightest ones were of first magnitude, the faintest ones were of sixth magnitude and the intermediate ones were of second magnitude, third magnitude and so on. Some time later, it was discovered that human eye responds to a large range of light intensity, actually better described in logarithmic scale, and this suggested to define the magnitude system in terms of a logarithmic function, as follows.

We define the \textit{bolometric apparent magnitude} $m$ of a light source in terms of the source's bolometric flux $f$ as
\begin{equation}
m \equiv -2.5\log_{10}\left(\frac{f}{f_x}\right),
\end{equation}
where $f_x$ is the reference flux with value $f_x = 2.53\times 10^{-8} \,W/m^2$. The negative sign ensures that an object with a large flux has small bolometric apparent magnitude (e.g. $f_\odot=1367\,\,W/m^2 \rightarrow m_\odot=-26.8$). The reference flux was defined considering the star Vega, such that visible stars to the naked eye are typically in range $0<m<6$.

The second quantity to be defined is the \textit{bolometric absolute magnitude} $M$ from a light source in terms of its luminosity $L$, given by
\begin{equation}
M \equiv -2.5\log_{10}\left(\frac{L}{L_x}\right),
\end{equation}
where the reference luminosity is $L_x = 78.7\,L_\odot$. The quantity $M$ is defined as the bolometric apparent magnitude that a light source would have if it were at a luminosity distance of $d_L=10\,$pc from the observer (e.g. for Type Ia supernovae $M\approx -19$ at the peak brightness and, for the Sun, $M_\odot=4.74$) \cite{amendola,Ryden2003}. The reference $L_x$ is based on this definition, since it is the luminosity of a source at 10 parsecs of distance producing the reference flux $f_x$. The magnitude system seems awkward at first glance, but it is nothing more than the logarithmic measure of both flux and luminosity.

\subsection{Luminosity distance, distance modulus and B-band generalized model}

With these definitions, the relation between bolometric apparent magnitude, bolometric absolute magnitude and luminosity distance is given by, using \eqref{d_l},
\begin{equation}
M = m-5\log_{10}\left(\frac{d_L}{10\,pc}\right),
\end{equation}
which may be rewritten as
\begin{equation}
M = m-5\log_{10}\left(\frac{d_L}{1\,Mpc}\right)-25,
\end{equation}
where $d_L$ must be given in Megaparsecs.

Since measured quantities such as flux and luminosity are usually quoted in terms of magnitudes, it is convenient to write luminosity distance in terms of \textit{distance modulus} $m_{mod}$ to a light source, defined by
\begin{equation}
\label{mod_supernova}
m_{mod} \equiv m-M = 5\log_{10}\left(\frac{d_L}{1\,Mpc}\right)+25.
\end{equation}

Due to calibration of type Ia supernovae, as mentioned before, equation \eqref{mod_supernova} must be generalized to the form ($d_L$ in Megaparsecs) \cite{Betoule2014}
\begin{equation}
m_B^{mod} = 5\log_{10}d_L(z_{hel},z_{CMB})+25-\alpha X_1 +\beta\mathcal{C} + M_B,
\end{equation}
where $\alpha$, $\beta$ and $M_B$ are nuisance parameters ($M_B$ represents supernovae absolute magnitude), $m_B^{mod}$ is the apparent magnitude in the B-band (or observed peak magnitude in the rest-frame B band), $X_1$ describes the time stretching of the light curve and $\mathcal{C}$ represents the supernova color parameter at maximum brightness ($X_1$ and $\mathcal{C}$ are parameters that model the light curve). Besides, luminosity distance is defined in two different redshift measurements: $z_{hel}$ is the redshift in the Sun's reference frame and $z_{CMB}$ represents the redshift in CMB reference frame. These redshift definitions must be accounted because luminosity distance is a combination of the real comoving distance $D(z)$ (measured when peculiar velocities and gravitational redshifts are subtracted) with a factor of $(1+z)$ coming from relativistic beaming and time dilation (each one contributing with $\sqrt{1+z}$) \cite{Calcino2017}. Therefore, luminosity distance must be written as $d_L=(1+z_{hel})D(z_{CMB})$.

It was found that supernovae absolute magnitude is dependent on host galaxy properties in a way still not fully understood \cite{Sullivan2011}, but that has important implications in cosmology such that a simple model must be assumed for $M_B$ in the form
\begin{equation}
M_B=
\begin{cases}
M_B^1 \qquad \qquad \;\;\; \text{if}\: M_{stellar}<10^{10}M_\odot , \\
M_B^1 + \Delta_M \qquad \text{otherwise,}
\end{cases}
\end{equation}
where $M_{stellar}$ represents the host stellar mass.

In our analysis, supernovae samples were taken from Joint Light-curve Analysis (JLA) dataset, which contains 740 supernovae distributed in the redshift range $0<z<1.3$ and that can be obtained in \cite{JLAdata}. For JLA dataset, the covariance matrix is given by \cite{Ferreira2017}
\begin{equation}
\label{covmatSN}
\boldsymbol{C_{SN}} = \boldsymbol{D}_{stat} +\boldsymbol{C}_{stat} +\boldsymbol{C}_{sys},
\end{equation}
where
\begin{align}
\label{Dstat}
\boldsymbol{D}_{stat,ii} = &\left[\frac{5}{z_i\ln 10}\right]^2\sigma_{z,i}^2 + \sigma_{lens}^2 + \sigma_{int}^2 + \sigma_{m_B,i}^2 + \alpha^2\sigma_{X_1,i}^2 + \beta^2\sigma_{\mathcal{C},i}^2 \nonumber \\
&+ 2\alpha C_{m_BX_1,i} - 2\beta C_{m_BC,i} -2\alpha\beta C_{X_1\mathcal{C},i},
\end{align}
\begin{equation}
\label{Cstat_sys}
\boldsymbol{C}_{stat} +\boldsymbol{C}_{sys} = V_0 + \alpha^2V_a + \beta^2V_b + 2\alpha V_{0a} - 2\beta V_{0b} -2\alpha\beta V_{ab}.
\end{equation}

The first three terms in \eqref{Dstat} account for errors in redshift due to peculiar velocities, errors in the magnitude due to gravitational lensing and intrinsic errors that are not already accounted in the former terms, respectively. The fourth term accounts for errors in the observed magnitude, the fifth term takes into account errors in the stretching factor $X_1$, and the sixth accounts for colour $\mathcal{C}$ error, while the remaining terms represent their covariances. Terms in $\boldsymbol{D}_{stat}$ are all diagonal, while terms in $\boldsymbol{C}_{stat} +\boldsymbol{C}_{sys}$ have also non-diagonal elements accounting for the same thing as $\boldsymbol{D}_{stat}$ (except for the first three terms in \eqref{Dstat}) and including systematics. All the $V$ terms in \eqref{Cstat_sys} are provided in \cite{JLAdata}.

%% file: Chap3.tex
\newpage
\chapter{Statistical analysis}

In this chapter, we review some statistical tools necessary to analyze our datasets: likelihood function, Bayes' theorem, Markov chain Monte Carlo technique and principal component analysis. These tools are standard in modern cosmology, and their use is essential for a correct parameter estimation.

\section{Likelihood function and Bayes' theorem}

Suppose we know that a random variable $x$ has a probability distribution function (PDF) $f(x|\theta)$ which is a function of an unknown parameter $\theta$. The ``$|$'' distinguishes the random variable $x$ from the parameter $\theta$. This PDF is defined as a \textit{conditional probability} of having the data $x$ given the theoretical parameter $\theta$, where it is obvious that different $\theta$ will lead to different PDF's. Besides, if one samples $x$ as $f(x|\theta)$ then it is more likely that $x$ has a value that maximizes $f(x|\theta)$, while if $x$ has a particular value, then it is better to choose $\theta$ such that it maximizes the occurrence of that $x$. Therefore, the best $\theta_i$ are the parameters that maximize $f(x_1,x_2,\dots,x_n|\theta_1,\theta_2,\dots,\theta_m)$, where generally parameter space is large and one usually writes $f$ in the compact form $f(x_i,\theta_j)$. This leads us to the maximum likelihood method, which consists in finding parameters $\theta_j$ that maximize the \textit{likelihood function} $f(x_i,\theta_j)$ by solving the equations
\begin{equation}
\label{del_posterior}
\frac{\partial f(x_i|\theta_j)}{\partial \theta_j} = 0,\qquad j=1,\dots,m.
\end{equation}

In cosmology, sometimes data are pretty scarce and one has to use information from previous experiments in order to better constrain information. Besides, one is often worried about estimating the theoretical parameters $\theta_j$ from the data. The way to handle this problem is to use \textit{Bayesian statistics}: instead of working with $f(x_i|\theta_j)$, the probability of having the data given the model, it is more important to work with $L(\theta_j|x_i)$, the probability of having the model given the data. These probabilities are related through the so-called \textit{Bayes' theorem} \cite{Trotta2008}:
\begin{equation}
P(T|D) = \frac{P(D|T)P(T)}{P(D)},
\end{equation}
where data $x_i$ are denoted by $D$ and theoretical parameters $\theta_j$ are denoted by $T$. In this equation, $P(D|T)$ is the conditional probability of having the data given the theory, $P(T)$ and $P(D)$ are the probability of having the theory and the data, respectively, and $P(T|D)$ is the conditional probability of having the theory given the data. Thus,
\begin{equation}
L(\theta_j|x_i) = \frac{f(x_i|\theta_j)p(\theta_j)}{g(x_i)},
\end{equation}
in which $p(\theta_j)$ is called \textit{prior} probability for the parameters $\theta_j$, $g(x_i)$ is the PDF of the data $x_i$, $f(x_i|\theta_j)$ is the likelihood function and $L(\theta_j|x_i)$ (or simply $L(\theta_j)$) is called \textit{posterior} probability. The latter contains valuable information: it is the probability distribution for the theoretical parameters $\theta_j$ based on the observed data $x_i$ and knowledge from previous experiments.

Since $L(\theta_j|x_i)$ is a probability distribution, one can normalize it to unity as
\begin{equation}
\int L(\theta_j|x_i)\,d^n\theta_j = 1 = \frac{1}{g(x_i)}\int f(x_i|\theta_j)p(\theta_j) \,d^n\theta_j,
\end{equation}
which leads to
\begin{equation}
g(x_i) = \int f(x_i|\theta_j)p(\theta_j) \,d^n\theta_j,
\end{equation}
where $n$ is the parameter dimension.

The term $g(x_i)$ is called \textit{evidence} and is just a normalization factor, having no influence on the estimation of $\theta_j$. On the other hand, the prior $p(\theta_j)$ is an essential term and has large influence in the final estimation. For example, if a previous experiment convincingly excluded, say, $\Omega_m^{(0)}<0.1$, therefore one might safely use this information in the form $p(\Omega_m^{(0)}<0.1)=0$. As another example, if for some reason we believe that a parameter $y_{cal}$ has the value $y_{cal}=1.0000\pm 0.0025$ (this is actually the parameter describing absolute map calibration for Planck), then we could use for $p(y_{cal})$ a \textit{Gaussian} prior with mean 1 and standard deviation 0.0025. In general, one usually excludes nonphysical values, e.g. $\Omega_m^{(0)}<0$ along with using \textit{flat} priors, which is given by $p(\theta_j)=1/\Delta\theta$ in a specified range $\Delta\theta$ where $\theta_j$ is defined and $p(\theta_j)=0$ outside (such that $\int p(\theta_j)\,d\theta_j=1$). If one uses only flat priors, then
\begin{equation}
L(\theta_j|x_i) \propto \mathcal{L}(x_i|\theta_j),
\end{equation}
meaning that the posterior corresponds to the likelihood function $\mathcal{L}$ in this case.

\section{Parameter estimation}

After collecting data, it is important to translate them into a model, which is a theoretical background we assume to be true. The model has some parameters $\theta_i$ in it, which one wants to determine. Therefore, \textit{parameter estimation} is the analysis performed to obtain estimates of the parameters, errors, or ideally the complete posterior probability distribution $L(\theta_i)$, which takes into account some priors assumed. Thus, in order to find the \textit{maximum posterior estimators} $\hat{\theta}_i$, equation \eqref{del_posterior} needs to be replaced by
\begin{equation}
\frac{\partial L(\theta_i)}{\partial\theta_i} = 0,\qquad i=1,\dots,n.
\end{equation}

By discarding the evidence factor $g(x_i)$, we need to normalize the posterior with an overall normalization $N$ through the integral
\begin{equation}
N = \int L(\theta_i)\, d^n\theta_i,
\end{equation}
where this one is extended to the whole parameter range. Using the normalized posterior (i.e. $L(\theta_i)/N$, to which we shall continue to use $L(\theta_i)$), we can evaluate the \textit{regions of confidence} $R(\alpha)$ for the parameters, defined by
\begin{equation}
\int_{R(\alpha)} L(\theta_i)\, d^n\theta_i = \alpha,
\end{equation}
where $R(\alpha)$ is the region for which the integral above yields $0<\alpha<1$. One usually chooses the values $\alpha=0.683,\,0.954,\,0.997$ (denoted as 1, 2 and 3$\sigma$, respectively).

It is common to have interest in some specific subset of parameters and consider the others as ``nuisance'' which we would be happy to get rid of. For example, supernovae theory predicts for distance modulus $m_{mod}$ the equation $m_{mod} = 5\log_{10}d_L(z;\Omega_m^{(0)})+25-\alpha X_1 +\beta\mathcal{C}+M_B$, from which the main parameter of interest is $\Omega_m^{(0)}$. The general posterior depends on the full set of parameters ($\Omega_m^{(0)}$, $\alpha$, $\beta$ and $M_B$) but we can turn it into a function only of $\Omega_m^{(0)}$ by integrating out all the other parameters,
\begin{equation}
L(\Omega_m^{(0)}) \equiv \int L(\Omega_m^{(0)},\alpha, \beta,M_B)\,d\alpha\,d\beta\,dM_B,
\end{equation}
where the integration is over the domain of the ``nuisance'' parameters. This procedure is called \textit{marginalization} and allows one to translate a multidimensional posterior into a one or two-dimensional probability distribution, which makes visualization of the problem much easier. For instance, if the maximum posterior estimator of $\Omega_m^{(0)}$ is 0.3 and 
\begin{equation}
\int_R L(\Omega_m^{(0)})\,d\Omega_m^{(0)} = 0.683
\end{equation}
in the region $\Omega_m^{(0)}=[0.27, 0.33]$, the final result is quoted as $\Omega_m^{(0)}=0.30\pm 0.03$ at $68.3\%$ confidence level (or $1\sigma$).

Let us assume that the posterior $L(\bm{\theta}|\bm{x})$ ($\bm{\theta}$ and $\bm{x}$ denoting arrays of parameters $\theta_i$ and data $x_i$) is single-peaked. A common estimator for the parameters is the \textit{expected value} (or mean) $\langle \bm{\theta} \rangle$ defined by
\begin{equation}
\langle \bm{\theta} \rangle \equiv \int L(\bm{\theta}|\bm{x})\,\bm{\theta} \,d\bm{\theta} = \frac{1}{N}\sum_{t=1}^N \bm{\theta}^{(t)},
\end{equation}
and variance $\sigma^2$ defined by
\begin{equation}
\sigma^2 \equiv \int L(\bm{\theta}|\bm{x})\,(\bm{\theta}-\langle \bm{\theta} \rangle)^2 \,d\bm{\theta} = \frac{1}{N}\sum_{t=1}^N (\bm{\theta}^{(t)}-\langle \bm{\theta} \rangle)^2,
\end{equation}
where we use Markov Chain Monte Carlo method (which shall be explained in the next section) to sample the posterior $L(\bm{\theta}|\bm{x})$, in which $\bm{\theta}^{(t)}$ are taken directly from the posterior by construction \cite{Trotta2008}. Also, it is common to work with standard deviation $\sigma$ instead of variance: $\sigma\equiv\sqrt{\sigma^2}$.

It is usual to approximate the likelihood function close to the peak with a multivariate Gaussian in parameter space, in the cases where the likelihood is single-peaked and well behaved (its peak is well located, not dispersed). This can be seen whether one performs a Taylor expansion close to the peak $\bm{\theta}_0$ of the log likelihood,
\begin{equation}
\ln{\mathcal{L}}(\bm{x}|\bm{\theta}) = \ln{\mathcal{L}}(\bm{x}|\bm{\theta}_0)+\frac{1}{2}(\bm{\theta}_\alpha-\bm{\theta}_{0\alpha}) \frac{\partial^2\ln{\mathcal{L}}}{\partial\bm{\theta}_\alpha \partial\bm{\theta}_\beta} (\bm{\theta}_\beta-\bm{\theta}_{0\beta}) + \dots,
\end{equation}
which leads to
\begin{equation}
\mathcal{L}(\bm{x}|\bm{\theta}) \approx \mathcal{L}(\bm{x}|\bm{\theta}_0)\exp{\left[-\frac{1}{2}(\bm{\theta}_\alpha-\bm{\theta}_{0\alpha}) H_{\alpha\beta} (\bm{\theta}_\beta-\bm{\theta}_{0\beta})\right]},
\end{equation}
where $H_{\alpha\beta}\equiv-\frac{\partial^2\ln{\mathcal{L}}}{\partial\bm{\theta}_\alpha \partial\bm{\theta}\beta}$ is called \textit{Hessian matrix} and controls whether the estimates of $\theta_\alpha$ and $\theta_\beta$ are correlated or not \cite{Heavens2009}. Besides, in cosmology, for many problems of interest the observational data generally have a Gaussian distribution and therefore
\begin{equation}
\mathcal{L}(\bm{x}|\bm{\theta}) \propto \exp{\left[-\frac{1}{2}\sum_{ij}(x_i-x_i^{th}(\bm{\theta}))C_{ij}^{-1}(x_j-x_j^{th}(\bm{\theta}))\right]},
\end{equation}
where $x_i$ is the observed data, $x_i^{th}(\bm{\theta})$ is the theoretical prediction for $x_i$ given a set of parameters $\bm{\theta}$ and $C_{ij}^{-1}$ is the inverse covariance matrix associated with $x_i$: $C_{ij}\equiv \langle (x_i-\langle x_i\rangle)(x_j-\langle x_j\rangle)^T \rangle$. The function inside square brackets (up to the constant $-1/2$) is usually denoted as the \textit{chi-square function}
\begin{equation}
\label{chi_square_cov}
\chi^2\equiv\sum_{ij}(x_i-x_i^{th}(\bm{\theta}))C_{ij}^{-1}(x_j-x_j^{th}(\bm{\theta})),
\end{equation}
and if there is no correlation between data, then \eqref{chi_square_cov} simplifies to
\begin{equation}
\chi^2=\sum_{i}\frac{(x_i-x_i^{th}(\bm{\theta}))^2}{\sigma_i^2},
\end{equation}
where $\sigma_i^2$ stands for data variance. In this case, maximizing the likelihood function is equivalent to minimizing the $\chi^2$ function and, thus, one needs to find a set of parameters $\theta_i$ intrinsic to theory such that the latter agrees as best as it can with observations.

It is fundamental to combine different datasets for better constraining parameters because each observation has its own \textit{degenerate directions}, i.e. directions in parameter space poorly constrained by the data. Thus, pairs of datasets have much stronger power constraints than each one separately because it is only their combination that mutually breaks parameter degeneracies \cite{Trotta2008}. For any datasets, one might consider them as coming from independent experiments such that their combination happens when one performs the product of the likelihoods. Equivalently, this is the same thing as adding up the chi-square functions from all the experiments. For the datasets described in the previous chapter,
\begin{equation}
\chi^2_{total} = \chi^2_{BAO} + \chi^2_{CMB} + \chi^2_{Clocks} + \chi^2_{SN}.
\end{equation}

\section{Markov Chain Monte Carlo method: emcee}

Generally, problems in cosmology have large dimensionality in parameter space (up to 20, 30 parameters). This fact makes it unfeasible to evaluate the full likelihood function (and consequently the posterior) in a finite grid mainly because each point in the grid is in general expensive computationally, which would require a large amount of time to solve the problem completely. Besides, a considerable hypervolume inside the grid has a very small likelihood and it is thus of no interest, which turns this approach very inefficient.

In order to concentrate in the region of interest where the likelihood is high, we will work with \textit{Markov Chain Monte Carlo} (MCMC) method. This one aims to generate a set of points that sample the target density, in this case the likelihood (which leads to the posterior) in an efficient manner. MCMC draws random points in parameter space using Markov process: the next point depends on the previous one but not in the entire sampling history. By sampling the posterior with MCMC, one might estimate the quantities of interest for the parameters such as mean, variance, etc.

The main idea behind MCMC method is to sample a new point $\bm{\theta}^*$ from the present point $\bm{\theta}$, accepting the new point in the chain with probability $p(\text{acceptance})$ that depends on the ratio between the new and old target density values ($p(\bm{\theta}^*)$ and $p(\bm{\theta})$, respectively). The distribution which generates a new point based on the present one is called \textit{proposal distribution} $q(\bm{\theta}^*|\bm{\theta})$, generally taken to be a Gaussian that matches the target distribution. On the same manner, one can also define a \textit{reversible proposal distribution} $q(\bm{\theta}|\bm{\theta}^*)$, which gives a distribution to reverse the chain (reversible distributions play an important role in MCMC algorithms, such as homogeneity of the chain \cite{Sanjib2017}). A basic MCMC algorithm is the so-called \textit{Metropolis-Hastings} (M-H) algorithm, where the probability of acceptance of a new point is
\begin{equation}\label{M_H_algorithm}
p(\text{acceptance}) = min \left[1,\frac{p(\bm{\theta}^*)q(\bm{\theta}^*|\bm{\theta})}{p(\bm{\theta})q(\bm{\theta}|\bm{\theta}^*)} \right],
\end{equation}
and if the chain is reversible (symmetric proposal distribution) M-H simplifies to \textit{Metropolis} algorithm
\begin{equation}\label{M_algorithm}
p(\text{acceptance}) = min \left[1,\frac{p(\bm{\theta}^*)}{p(\bm{\theta})} \right].
\end{equation}
The basic idea of Metropolis algorithm (which can be generalized for M-H) is that if the new point $\bm{\theta}^*$ has a target density $p(\bm{\theta}^*)$ higher than the target density $p(\bm{\theta})$ of the present point $\bm{\theta}$, the new point is always accepted. However, if $\bm{\theta}^*$ has a target density smaller than the target density of $\bm{\theta}$, then the new point is accepted with probability given by \eqref{M_algorithm}, which means that in the latter case the point might be accepted or not and this makes MCMC exploration in parameter space much more efficient than grid approach. Detailed description of MCMC methods can be found in \cite{Robert2015,Verde2007,Sanjib2017}.

In cosmology, Monte Carlo methods are generally applied to data using known softwares such as CosmoMC \cite{lewis} and Monte Python \cite{audren}, which are based on Metropolis-Hastings. In this work, as an alternative approach, we will focus on a MCMC method that is mostly used in astrophysics and which has been extensively improved by the community: \textit{emcee}, a pure-Python implementation which is an \textit{affine invariant MCMC ensemble sampler} \cite{Foreman2013}. This code is designed for better exploration of parameter space, with improvements in efficiency and speed when compared to other standard MCMC methods. A faster sampler is preferable because reduces the number of likelihood evaluations, which generally have high computational cost, and thus accuracy is achieved in a minor number of steps.

Affine invariance means that by applying an affine transformation in the proposal distribution one is able to sample the target distribution efficiently even if this one is highly anisotropic, which would render sampling much more expensive computationally. For example, if one considers the skewed probability distribution in $\mathbb{R}^2$
\begin{equation}
\pi(x_1,x_2)\propto \exp{\left[-\frac{(x_1-x_2)^2}{2\epsilon}-\frac{(x_1+x_2)^2}{2} \right]}
\end{equation}
for small $\epsilon$, it follows that traditional MCMC methods have difficulty to sample this distribution. However, if one performs the affine transformation
\begin{equation}
y_1 \equiv \frac{x_1-x_2}{\sqrt{\epsilon}}\qquad \text{and} \qquad y_2\equiv x_1+x_2,
\end{equation}
the initial challenging problem turns into the easier problem of sampling the distribution
\begin{equation}
\pi_A(y_1,y_2)\propto \exp{\left[-\frac{(y_1^2+y_2^2)}{2} \right]},
\end{equation}
which is an isotropic density that even a simple Metropolis-Hastings would not have problems to solve. This example motivates an affine invariant sampler: an algorithm which samples target distributions equally well under all linear transformations, being insensitive to covariances among parameters \cite{Foreman2013}. Although target distributions in cosmology are not highly anisotropic, some parameters might be correlated and \textit{emcee} is able to take this into account.

The base algorithm for \textit{emcee} is called \textit{stretch move} and converges faster than M-H methods, therefore with superior performance \cite{Foreman2013}. It is based on evolving an ensemble of $K$ walkers $S=\{X_k\}$ where the proposal distribution for one walker $k$ depends on the position of the $K-1$ walkers in the complementary ensemble $S_{[k]}=\{X_j,\forall j\neq k\}$, in which position means the N-dimensional vector in parameter space characterizing each walker. Thus, in order to draw a new position $Y$ from the current position $X_k$, the proposal distribution is given by an affine transformation as
\begin{equation}
\label{affine_transf}
X_k(t)\rightarrow Y=X_j+ Z\left[X_k(t)-X_j \right],
\end{equation}
where $X_j$ is a walker chosen randomly from $S_{[k]}$ and $Z$ is a random variable sampled from a distribution $g(Z=z)$. This distribution is important because if it satisfies the relation
\begin{equation}
g(z^{-1})=zg(z),
\end{equation}
then the proposal \eqref{affine_transf} is symmetric, that is, the chain looks the same when run forward or backward in time \cite{Sanjib2017}, which can be stated as
\begin{equation}
p(X_k(t)\rightarrow Y) = p(Y\rightarrow X_k(t)).
\end{equation}

The appropriate acceptance probability for stretch move is
\begin{equation}
q=\text{min}\left[1,\,Z^{N-1}\frac{p(Y)}{p(X_k(t))} \right],
\end{equation}
in which $N$ is the parameter space dimension. Also, the choice made for $g(z)$ is \cite{Foreman2013}
\begin{equation}
g(z)\propto
\begin{cases}
\frac{1}{\sqrt{Z}}, \;\; \text{if}\: z \in \left[\frac{1}{a},a \right], \\
0, \quad \;\; \text{otherwise},
\end{cases}
\end{equation}
where $a$ is a scale parameter adjusted for better performance, set to 2. It is interesting to notice that the idea of using the position of other walkers to guide the next position of a specific walker is based on the fact that the complementary ensemble carries valuable information about the target density, which gives a way of adapting the trial move always aiming the target distribution (this is encoded in the affine invariant proposal), rendering the method very powerful. A more detailed description of affine invariant ensemble samplers can be found in \cite{Goodman2010}.

\section{Burn-in and convergence}

It is expected that chains generated from MCMC methods fairly sample the target distribution once convergence to a stationary distribution has been achieved, that is, no matter where the parameters were initially sampled: the ensemble of walkers are led to the peak of the target distribution and their values oscillate close to the mean value for each parameter. Thus, the first steps in the chain should not be accounted for sampling in order to remove the dependence on the starting point \cite{Heavens2009}. This period in which walkers still have a ``memory'' from where they started is called \textit{burn-in} period.

How long should we run different MCMC chains in order to get good estimates of the parameters? To answer this question, we will adopt a well-known convergence test called \textit{Gelman-Rubin convergence criterion} \cite{Gelman1992}, which basically compares variance between different chains in order to assess convergence. Suppose we are working with $m$ chains in our MCMC and that $n$ iterations were performed for each chain. Suppose also that burn-in period has already finished. Let $\theta_{i,j}$ represent a point in parameter space in position $i$ and chain $j$. We define the mean of each chain and the mean of all the chains, respectively, as
\begin{equation}
\bar{\theta}_{j}\equiv\frac{1}{n}\sum_{i=1}^{n}\theta_{i,j}
\textrm{\ \ \ \  and\ \ \ \ }
\bar{\theta}\equiv\frac{1}{m}\sum_{j=1}^{m}\bar{\theta_j},
\end{equation}
where index $i$ represents points in a chain and $j$ runs over the chains. The chain-to-chain variance $B$ and mean within-chain variance $W$ are defined, respectively, as
\begin{equation}
B=\frac{1}{m-1}\sum_{j=1}^m(\bar{\theta}_j-\bar{\theta})^2
\textrm{\ \ \ \  and\ \ \ \ }
W=\frac{1}{m}\sum_{j=1}^m \frac{1}{n-1}\sum_{i=1}^{n}(\theta_{i,j}-\bar{\theta_j})^2.
\end{equation}

The total or true variance $\hat{\sigma}^2$ for the estimator $\bar{\theta}$ is defined as the weighted average $\hat{\sigma}^2\equiv W(n-1)/n+B$. One can also define the pooled variance
\begin{equation}
\label{V_gelmann}
V\equiv\hat{\sigma}^2+\frac{B}{m}=\frac{n-1}{n}W+\frac{m+1}{m}B,
\end{equation}
which takes into account sampling variability of the estimator $\bar{\theta}$ \cite{Sanjib2017}. In the initial sampling, when the distribution is over-dispersed, $B$ is large because chains are also dispersed and variance between them is high. At the same time, $V$ overestimates $W$ because of the $B$ term in \eqref{V_gelmann}. Also, $W$ is smaller than $\hat{\sigma}^2$ because exploration of parameter space is still insufficient (that is, $B$ is still high). However, as $n\rightarrow \infty$, chain-to-chain variance $B$ goes to zero and $W$ approaches true variance $\hat{\sigma}^2$, which renders $V/W$ close to 1. Thus, the ratio $\hat{R}\equiv\sqrt{V/W}$, named \textit{potential scale reduction factor}, is the quantity of interest chosen to monitor convergence, in which a value close to one means that convergence has been achieved. The value of $\hat{R}$ used to stop sampling is chosen according to the level of precision desired, the number of parameters in the theory and the computational cost for every MCMC point, where usual choices are 1.1, 1.03 and 1.01 \cite{CosmoMC}.

\section{Principal components analysis}

Before describing the next tool, it is useful to motivate its application. In order to constrain the dark energy equation of state, we are going to use an interpolation that makes $w(z)$ to be a smooth and continuous function because $w(z)$ is integrated over redshifts whenever $H(z)$ shows up. To this purpose, dark energy equation of state is sampled in six redshift bins, $z_i\in \{0.0, 0.25, 0.50, 0.85, 1.25, 2.0\}$, with corresponding values for the equation of state $w_i\,(i=1,2,\dots,6)$. Including more than six bins does not effectively improve the constraints \cite{Said2013}. The last redshift ($z=2$) is chosen according to the fact that this is the epoch in which dark energy started to dominate the expansion of the Universe \cite{jimenez2002}. We construct $w(z)$ as
\begin{equation}
 w(z) = 
  \begin{cases} 
   w_i, & \text{if }z=z_i,\\
   \text{spline}, & \text{if }z \in \left(z_i,z_{i+1}\right), \\
    -1, & \text{if }z \ge z_6,
  \end{cases}
\end{equation}
where ``spline'' means we are using cubic spline interpolation between $z_i$'s. The parameters $w_i$ are MCMC parameters and must be constrained by data on the same way that other cosmological parameters do (e.g. $\Omega_bh^2$, $\Omega_ch^2$, ...).

Once chains containing $w_i$'s are obtained, after marginalizing over all the remaining parameters, we must face the fact that estimates for $w_i$'s are correlated because their covariance matrix is not diagonal. Therefore, one needs to perform a rotation in parameter space, which rewrites $w_i$'s in a different basis, in order to obtain new estimates that have a diagonal covariance matrix. The procedure adopted to this purpose is called \textit{principal components analysis} (PCA) \cite{cooray,Hamilton2000,Huterer2003} and it is described as follows.

Let us define $\mathcal{F}$ as the Fisher information matrix \cite{amendola,Heavens2009} of a set of estimators $\hat{\theta}$ ($w_i$'s in our problem) of parameters $\theta$ measured from observations, assuming that $\mathcal{F}$ can be approximated by the inverse covariance matrix \cite{Hamilton2000} of the estimators $\hat{\theta}$ as
\begin{equation}
\label{inv_fisher}
\mathcal{F}^{-1}=\mathcal{C} = \langle \Delta\hat{\theta}\,\Delta\hat{\theta}^\text{T} \rangle = \langle \hat{\theta}\,\hat{\theta}^\text{T}\rangle - \langle \hat{\theta}\rangle\langle \hat{\theta}^\text{T}\rangle.
\end{equation}

The \textit{decorrelation matrix} $\mathcal{W}$ is the matrix that diagonalizes the Fisher matrix $\mathcal{F}$ (or $\mathcal{C}^{-1}$) and thus satisfies the relation
\begin{equation}
\label{fisher_diag}
\mathcal{F} = \mathcal{W}^\text{T}\Lambda\mathcal{W},
\end{equation}
in which $\Lambda$ is the diagonal matrix containing the eigenvalues of $\mathcal{F}$ and the rows of $\mathcal{W}$ are the eigenvectors of $\mathcal{F}$.

The quatity of interest is $\hat{\theta}'=\mathcal{W}\hat{\theta}$ because this new set of $w_i$'s is uncorrelated due to the fact that its covariance matrix is, using \eqref{inv_fisher} and \eqref{fisher_diag},
\begin{equation}
\langle \Delta(\mathcal{W}\hat{\theta})\,\Delta(\mathcal{W}\hat{\theta})^\text{T} \rangle = \mathcal{W} \langle \Delta\hat{\theta}\,\Delta\hat{\theta}^\text{T} \rangle \mathcal{W}^{\text{T}} = \mathcal{W}\,\mathcal{F}^{-1}\,\mathcal{W}^{\text{T}} = \Lambda^{-1},
\end{equation}
which is diagonal and represents the variances of the $w_i$'s (elements of $\Lambda^{-1}$ are necessarily positive). The set $\hat{\theta}'=\mathcal{W}\hat{\theta}$ is called \textit{principal components} and the rows of $\mathcal{W}$ are weights (or window functions) that relate the principal components to the initial set $\hat{\theta}$. It is interesting to notice that one might scale the decorrelated quantities $\mathcal{W}\hat{\theta}$ to unity variance by multiplying them by the square root of the diagonal elements of $\Lambda$. Thus, relation \eqref{fisher_diag} yields
\begin{equation}
\label{fisher_unity}
\mathcal{F} = \mathcal{W}^\text{T}\,\mathcal{W}.
\end{equation}

As pointed out in \cite{Hamilton2000}, there are infinite choices of decorrelation matrices satisfying \eqref{fisher_unity}. For example, choosing an orthogonal matrix $\mathcal{O}$ and performing a rotation in $\mathcal{W}$ leads to another decorrelation matrix: $\mathcal{V}=\mathcal{O}\mathcal{W} \Rightarrow \mathcal{V}^{\text{T}}\mathcal{V}=\mathcal{W}^\text{T}\mathcal{O}^\text{T}\mathcal{O}\mathcal{W} = \mathcal{W}^\text{T}\mathcal{W} = \mathcal{F}$. There is an interesting choice for $\mathcal{W}$ which renders its weights (rows) positive almost everywhere, with very small negative contributions \cite{Hamilton2000}. This choice is the square root of the Fisher matrix $\widetilde{\mathcal{W}}\equiv\mathcal{F}^{1/2} \equiv \mathcal{C}^{-1/2}$, defined as
\begin{equation}
\widetilde{\mathcal{W}}\equiv \mathcal{W}^\text{T}\Lambda^{1/2}\mathcal{W},
\end{equation}
where we normalize $\widetilde{\mathcal{W}}$ such that its rows sum to unity. With this choice, the covariance matrix of the new parameters $\widetilde{\mathcal{W}}\hat{\theta}$ is diagonal \cite{cooray}. We adopt this choice in our work.

%% file: Chap4.tex
\newpage
\chapter{Results} \label{chapt4}

This chapter displays analyses performed via MCMC sampling using \textit{emcee} for different combinations of datasets. The first analysis accounts for BAO, cosmic clocks (CC) and supernovae (SN) datasets combined because these ones have lower constraining power than CMB data. The second analysis accounts only CMB data for constraining parameters and the third analysis considers all datasets combined in order to improve constraints. In the last section, a comparison with previous results is performed .

We developed a code in Python for sampling six cosmological parameters, six dark energy equation of state parameters, fifteen ``nuisance'' parameters from Planck 2015 data and four ``nuisance'' parameters from JLA dataset. In addition, as we are working with a time-dependent dark energy equation of state $w(z)$, when using CMB data we employ a modified version of CAMB which uses a \textit{Parameterized Post-Friedmann} (PPF) \cite{Sawicki2007,Hu2008,Fang2008} prescription for dark energy in order to cross the phantom divide \cite{amendola}, i.e. $w=-1$, multiple times. Parameter details are described in Table \ref{table_param}.

In results presented in the next sections, we have already removed burn-in and the Gelman-Rubin criterion adopted is $\hat{R}<1.01$ for the set BAO+CC+SN, $\hat{R}<1.6$ for CMB data and $\hat{R}<1.3$ for the set BAO+CC+CMB+SN. These values were chosen according to the size of parameter space and computational cost of each set (which is extremely large for the last two sets). The number of points generated in each MCMC chain is of order $\mathcal{O}(10^6)$ for BAO+CC+SN and BAO+CC+CMB+SN sets, while for CMB set is $\mathcal{O}(3\times10^5)$. Also, we generate estimates of the posterior mean of each parameter of interest together with confidence intervals ($68\%$ and $95\%$) and best fit values using \textit{GetDist} Python package \cite{GetDist}.

\begin{table}[h!]
\centering \footnotesize
\setlength\tabcolsep{1pt}
\begin{tabular}{SSSSSSS} \toprule
    {Parameter} & {Prior range} & {Central} & {Width} & {Definition} \\ \midrule
    $\omega_{b}\equiv \Omega_{b}h^{2}$ & {[0.005, 0.1]} & {0.0221} & {0.0001} & {Baryon density today} \\
    $\omega_{c}\equiv\Omega_{c}h^{2}$  & {[0.001, 0.99]} & {0.120} & {0.001} & {Cold dark matter density today} \\
    $n_{s}$  & {[0.8, 1.2]} & {0.965} & {0.004} & {Scalar spectrum index} \\
    ln${(10^{10}A_{s})}$ & {[2, 4]} & {3.100} & {0.001} & {Log power of scalar amplitude} \\
    $H_{0}$  & {[20, 100]} & {67.3} & {1.0} & {Current expansion rate in $\text{km}\, \text{s}^{-1}\text{Mpc}^{-1}$} \\
    $\tau$  & {[0.01, 0.8]} & {0.08} & {0.01} & {Thomson optical depth due to reionization} \\
    {$w_{1}\,(z=0.00)$}  & {[-10, 8]} & {-1.0} & {1.0} & {1st dark energy equation of state parameter} \\
    {$w_{2}\,(z=0.25)$}  & {[-10, 8]} & {-1.0} & {1.0} & {2nd dark energy equation of state parameter} \\
    {$w_{3}\,(z=0.50)$}  & {[-10, 8]} & {-1.0} & {1.0} & {3rd dark energy equation of state parameter} \\
    {$w_{4}\,(z=0.85)$}  & {[-10, 8]} & {-1.0} & {1.0} & {4th dark energy equation of state parameter} \\
    {$w_{5}\,(z=1.25)$}  & {[-10, 8]} & {-1.0} & {1.0} & {5th dark energy equation of state parameter} \\
    {$w_{6}\,(z=2.00)$}  & {[-10, 8]} & {-1.0} & {1.0} & {6th dark energy equation of state parameter} \\ \midrule
    $A^{\mathrm{CIB}}_{217}$  & {[0, 200]} & {65} & {10} & {CIB contamination at $\ell=3000$ (217-GHz)} \\
    $\xi^{\mathrm{tSZ}\times\mathrm{CIB}}$ & {[0, 1]} & {0} & {0.1} & {SZ$\times$CIB cross-correlation} \\
    $A^{\mathrm{tSZ}}_{143}$  & {[0, 10]} & {5} & {2} & {tSZ contamination at 143 GHz} \\
    $A^{\mathrm{PS}}_{100}$  & {[0, 400]} & {255} & {24} & {Point source contribution in 100$\times$100} \\
    $A^{\mathrm{PS}}_{143}$  & {[0, 400]} & {40} & {10} & {Point source contribution in 143$\times$143} \\
    $A^{\mathrm{PS}}_{143\times 217}$ & {[0, 400]} & {40} & {12} & {Point source contribution in 143$\times$217} \\
    $A^{\mathrm{PS}}_{217}$  & {[0, 400]} & {100} & {13} & {Point source contribution in 217$\times$217} \\
    $A^{\mathrm{kSZ}}$  & {[0, 10]} & {0} & {3} & {kSZ contamination} \\
    $A^{\mathrm{dustTT}}_{100}$  & {[0, 50]} & {7} & {2} & {Dust contamination at $\ell=200$ in 100$\times$100} \\
      & {$(7\pm2)$} &  &  &  &  \\
    $A^{\mathrm{dustTT}}_{143}$  & {[0, 50]} & {9} & {2} & {Dust contamination at $\ell=200$ in 143$\times$143} \\
      & {$(9\pm2)$} &  &  &  &  \\
    $A^{\mathrm{dustTT}}_{143\times 217}$  & {[0, 100]} & {17} & {4} & {Dust contamination at $\ell=200$ in 143$\times$217} \\
      & {$(21.0\pm8.5)$} &  &  &  &  \\
    $A^{\mathrm{dustTT}}_{217}$  & {[0, 400]} & {80} & {15} & {Dust contamination at $\ell=200$ in 217$\times$217} \\
      & {$(80\pm20)$} &  &  &  &  \\
    $c_{100}$  & {[0, 3]} & {0.999} & {0.001} & {Power spectrum calibration for the 100 GHz} \\
      & {$(0.999\pm0.001)$} &  &  &  &  \\
    $c_{217}$  & {[0, 3]} & {0.995} & {0.002} & {Power spectrum calibration for the 217 GHz} \\
      & {$(0.995\pm0.002)$} &  &  &  &  \\
    $y_{\rm cal}$  & {[0.9, 1.1]} & {1.000} & {0.002} & {Absolute map calibration for Planck} \\
      & {$(1.0000\pm0.0025)$} &  &  &  &  \\
    $\alpha$  & {[0.0, 0.3]} & {0.135} & {0.020} & {Time stretching coefficient for supernovae} \\
    $\beta$  & {[1.5, 4.0]} & {3.0} & {0.3} & {Supernovae color coefficient} \\
    $M_{B}^{1}$  & {[-25, -15]} & {-19.05} & {0.15} & {Supernovae absolute magnitude} \\
    $\Delta_{M}$  & {[-0.3, 0.3]} & {-0.05} & {0.10} & {Correction for supernovae absolute magnitude} \\ \midrule
    $\Omega_{m}$  & {-} & {-} & {-} & {Matter density today} \\
    $\Omega_{\Lambda}$  & {-} & {-} & {-} & {Dark energy density today} \\
    $\bar{\chi}^{2}$  & {-} & {-} & {-} & {Mean chi square}  \\ \bottomrule
\end{tabular}
\caption{List of parameters and definitions used in this work. First block represents cosmological parameters, second block is a set of ``nuisance'' parameters and in the third block some derived parameters are displayed. In prior range column, square brackets mean flat priors while parentheses represent Gaussian priors. There is an additional Gaussian prior which is a linear combination of two parameters: $A^{\mathrm{kSZ}}+1.6\times A^{\mathrm{tSZ}}_{143} = 9.5\pm 3.0$. Central and width columns represent an estimated value (taken from CosmoMC) for each parameter close to the peak of the posterior and a range (relative to the central value) where initial positions for walkers are randomly sampled, respectively. A detailed description of ``nuisance'' parameters from Planck 2015 is given in reference \cite{Aghanim2015}.}
\label{table_param}
\end{table}

\section{BAO+Clocks+Supernovae}

\begin{figure}[h!]
\centering
\includegraphics[width=0.8\textwidth]{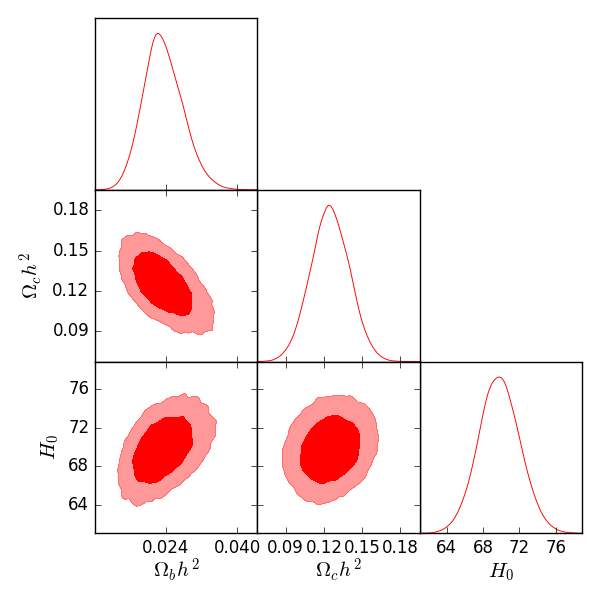}
\caption{1D and 2D cosmological parameter constraints from BAO, cosmic clocks and supernovae datasets combination. $68\%$ and $95\%$ confidence levels are displayed.}
\label{BAO_CC_SN_cosmo}
\end{figure}

\begin{figure}[h!]
\centering
\includegraphics[width=1.0\textwidth]{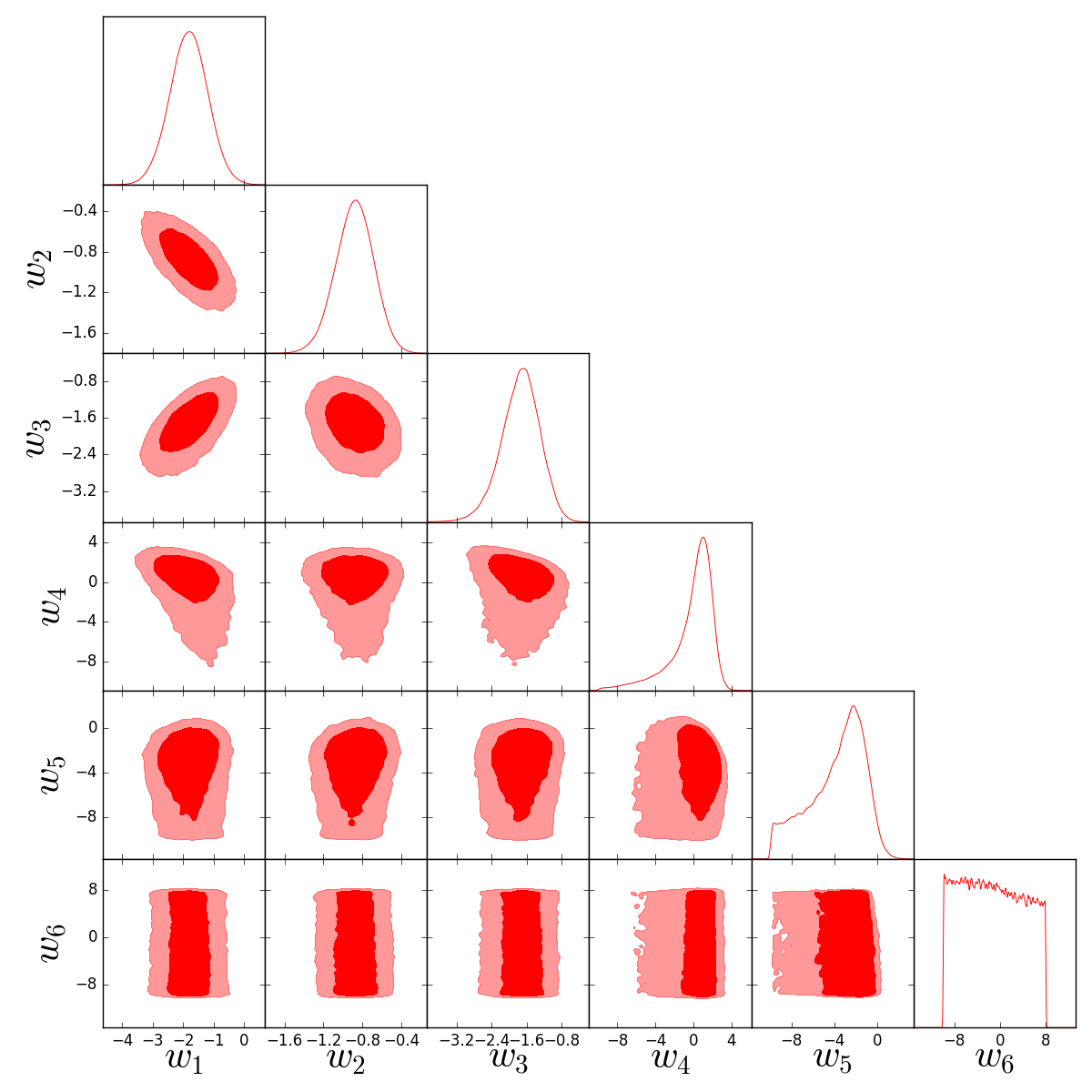}
\caption{1D and 2D dark energy equation of state parameter constraints from BAO, cosmic clocks and supernovae datasets combination. $68\%$ and $95\%$ confidence levels are displayed.}
\label{BAO_CC_SN_wis}
\end{figure}

The first investigation refers to BAO, cosmic clocks and supernovae datasets combined. The sampling results are displayed in Fig. \ref{BAO_CC_SN_cosmo}, Fig \ref{BAO_CC_SN_wis}. and Table \ref{table_BAO_CC_SN}, which show that even by combining these datasets they do not have enough power to constrain parameters completely. This is due to the fact that datasets used in this sampling are very scarce for high redshifts (only cosmic clocks has information close to $z=2.0$, while the highest redshit for BAO is $z=0.73$ and for supernovae is $z=1.3$). In particular, we tested the last three $w_i$'s even by adopting a larger prior than [-10,8] (we even used [-50,48]), but they are also not fully constrained in these conditions, which is why we have not applied PCA to dark energy equation of state using this set. On the other hand, convergence has been achieved in a high level and our cosmological parameter estimates for $\Omega_bh^2$, $\Omega_ch^2$ and $H_0$ are in good agreement (at $1\sigma$) with latest observations from Planck \cite{planck2}.

\begin{table}[h!]
\centering
\begin{tabular}{SSSSSSS} \toprule
    {Parameter} & {Best fit} & {$68\%$ limits} & {$95\%$ limits} & {Planck (TT+lowP)} \\ \midrule
    $\Omega_{b}h^{2}$ & {0.0257} & {$0.0234^{+0.0040}_{-0.0048}$} & {$0.0234^{+0.0089}_{-0.0085}$} & {$0.02222\pm 0.00023$}  \\
    $\Omega_{c}h^{2}$  & {0.106} & {$0.125\pm 0.015$} & {$0.125^{+0.030}_{-0.030}$} & {$0.1197\pm 0.0022$} \\
    $H_{0}$  & {68.2} & {$69.8\pm 2.3$} & {$69.8^{+4.6}_{-4.5}$} & {$67.31\pm 0.96$} \\
    {$w_{1}$}  & {-1.80} & {$-1.81\pm 0.63$} & {$-1.8^{+1.2}_{-1.2}$} & {-} \\
    {$w_{2}$}  & {-0.80} & {$-0.88\pm 0.20$} & {$-0.88^{+0.37}_{-0.40}$} & {-} \\
    {$w_{3}$}  & {-1.63} & {$-1.75^{+0.47}_{-0.39}$} & {$-1.75^{+0.85}_{-0.89}$} & {-} \\
    {$w_{4}$}  & {1.11} & {$<5$} & {-} & {-} \\
    {$w_{5}$}  & {-1.6} & {$<5$} & {-} & {-} \\
    {$w_{6}$}  & {2.8} & {-} & {-} & {-} \\ \midrule
    $\Omega_{m}$  & {0.282} & {$0.305\pm 0.027$} & {$0.305^{+0.055}_{-0.051}$} & {$0.315\pm 0.013$} \\
    $\Omega_{\Lambda}$  & {0.718} & {$0.695^{+0.028}_{-0.026}$} & {$0.695^{+0.051}_{-0.055}$} & {$0.685\pm 0.013$} \\
    $\bar{\chi}^{2}$  & {486.2} & {$499.2^{+4.0}_{-6.2}$} & {$499^{+11}_{-9.6}$} & {-} \\ \bottomrule
\end{tabular}
\caption{Marginalized estimates from BAO+CC+SN datasets. Best fit, $68\%$ and $95\%$ confidence levels are displayed. In these estimates, $w_i$'s are correlated. For comparison, Planck constraints for $\Lambda$CDM model at $1\sigma$ \cite{planck2} are also displayed.}
\label{table_BAO_CC_SN}
\end{table}

\section{CMB}

\begin{figure}[h!]
\centering
\includegraphics[width=1.0\textwidth]{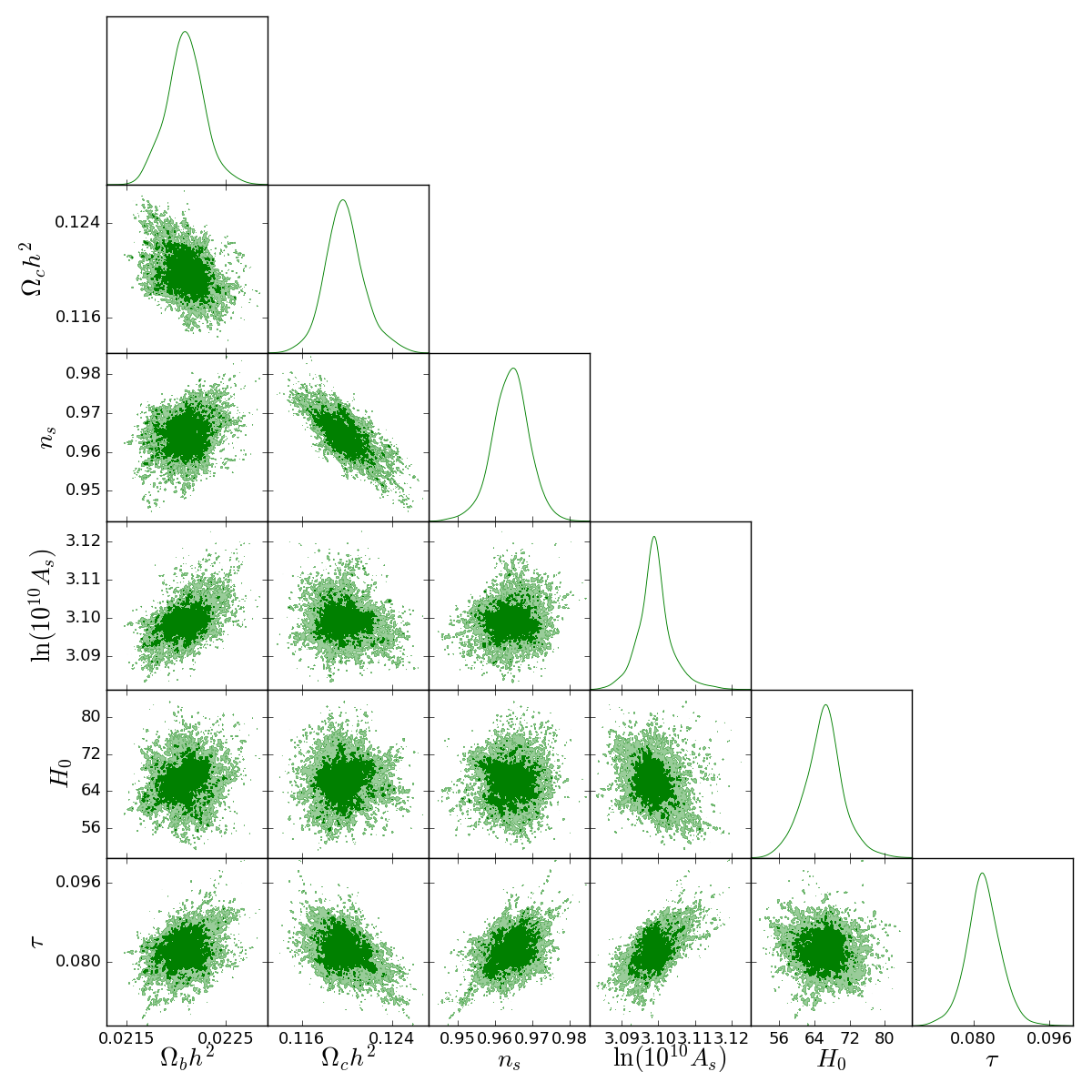}
\caption{1D and 2D cosmological parameter constraints from CMB data alone. $68\%$ and $95\%$ confidence levels are displayed.}
\label{CMB_cosmo}
\end{figure}

\begin{figure}[h!]
\centering
\includegraphics[width=1.0\textwidth]{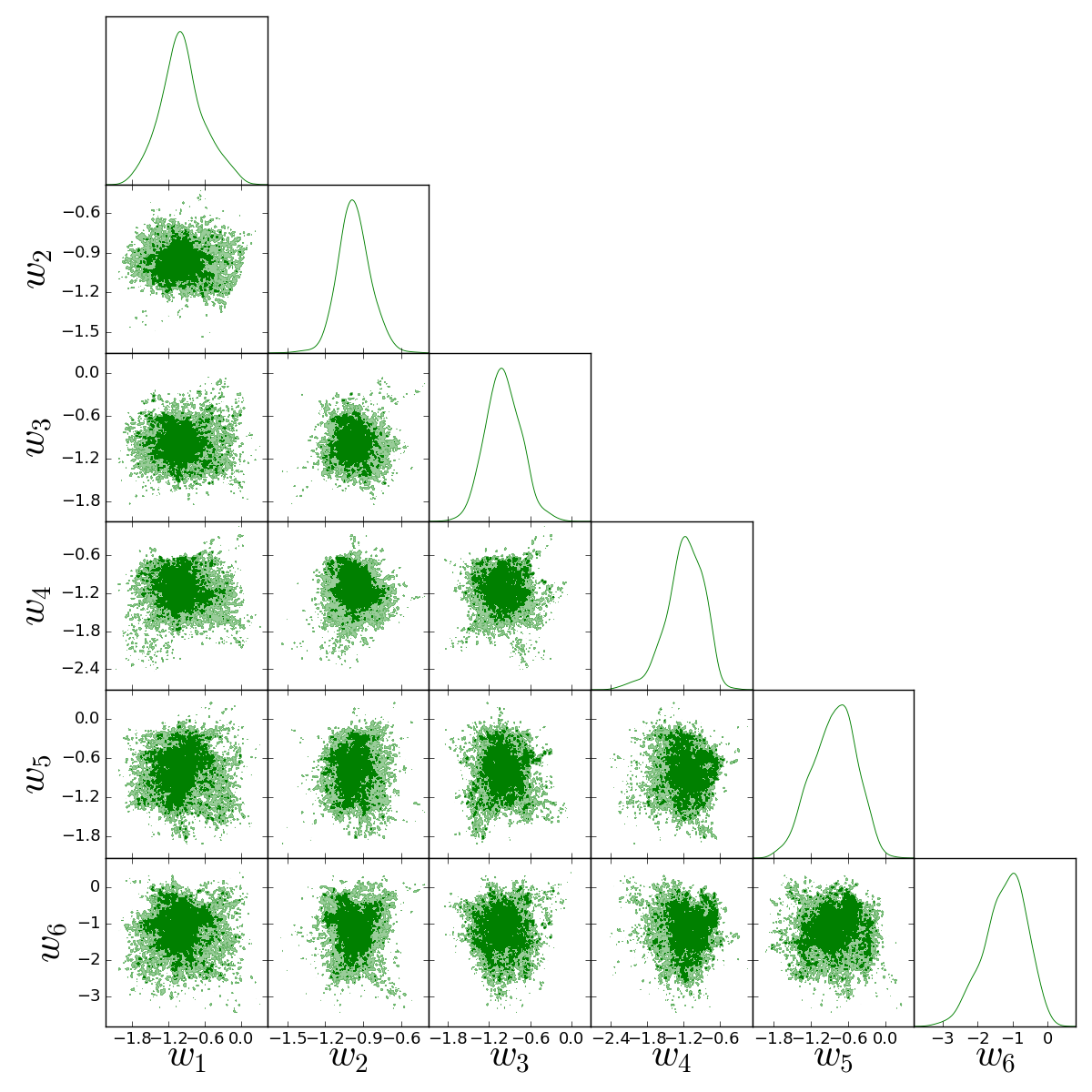}
\caption{1D and 2D dark energy equation of state parameter constraints from CMB data alone. $68\%$ and $95\%$ confidence levels are displayed.}
\label{CMB_wis}
\end{figure}

\begin{figure}[h!]
\centering
\includegraphics[width=0.8\textwidth]{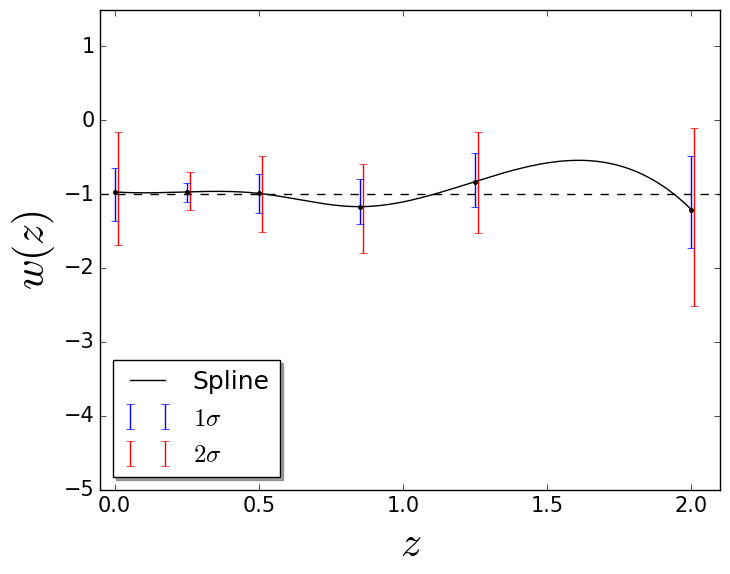}
\caption{Uncorrelated constraints on the dark energy equation of state parameters derived from CMB data alone. $68\%$ and $95\%$ confidence levels are displayed. Reconstructed $w(z)$ is represented by the dark line using cubic spline interpolation. For better visualization, $95\%$ confidence levels are slightly shifted to the right.}
\label{w_plot_planck}
\end{figure}

We shall now analyze the impact of CMB data in the parameters. Results are showed in Fig. \ref{CMB_cosmo}, Fig. \ref{CMB_wis}, Fig. \ref{w_plot_planck} and Table \ref{table_CMB}. Although this sampling has not fully converged, our cosmological parameter estimates are consistent with Planck data \cite{planck2} at $1\sigma$ (Fig. \ref{CMB_cosmo}), displaying no deviations when compared to a $\Lambda$CDM model (although CMB measurements are in strong tension with Hubble Space Telescope local measurement for the Hubble rate today, $H_0=73.48\pm 1.66 \,\text{km}\, \text{s}^{-1}\text{Mpc}^{-1}$ \cite{Riess2018}, a still unsolved problem in cosmology). When we look at the $w_i$'s (Fig. \ref{CMB_wis}), we see that Planck data alone seems to be able to constrain all dark energy equation of state parameters, differently from the previous dataset analyzed (BAO+CC+SN). Also, when we look at dark energy equation of state evolution (Fig. \ref{w_plot_planck}), we see that Planck data alone leads to an agreement with $\Lambda$CDM model at $1\sigma$. It is interesting to notice that, as opposed to the previous section where the first $w_i$'s are well constrained, in this case $w_1$ parameter has an error bar larger than other $w_i$'s (except for $w_6$), which is expected since most of CMB information comes from a high redshift ($z\sim 1000$). If sampling was larger and convergence was achieved, then $w_1$ would probably have the largest error bar.

\begin{table}[h!]
\centering\small
\begin{tabular}{SSSSSSS} \toprule
    {Parameter} & {Best fit} & {$68\%$ limits} & {$95\%$ limits} & {Planck (TT+lowP)} \\ \midrule
    $\Omega_{b}h^{2}$ & {0.02204} & {$0.02207\pm 0.00015$} & {$0.02207^{+0.00028}_{-0.00028}$} & {$0.02222\pm 0.00023$}  \\
    $\Omega_{c}h^{2}$  & {0.1211} & {$0.1199^{+0.0012}_{-0.0014}$} & {$0.1199^{+0.0027}_{-0.0025}$} & {$0.1197\pm 0.0022$} \\
    $n_{s}$  & {0.9648} & {$0.9641\pm 0.0040$} & {$0.9641^{+0.0078}_{-0.0082}$} & {$0.9655\pm 0.0062$} \\
    ln${(10^{10}A_{s})}$ & {3.0998} & {$3.0989^{+0.0023}_{-0.0018}$} & {$3.0989^{+0.0048}_{-0.0058}$}  & {$3.089\pm 0.036$} \\
    $H_{0}$  & {65.5} & {$66.6\pm 2.4$} & {$66.6^{+4.9}_{-5.2}$} & {$67.31\pm 0.96$} \\
    $\tau$  & {0.0822} & {$0.0820^{+0.0022}_{-0.0029}$} & {$0.0820^{+0.0055}_{-0.0047}$} & {$0.078\pm 0.019$} \\
    {$w_{1}$}  & {-0.89} & {$-0.97^{+0.32}_{-0.40}$} & {$-0.97^{+0.81}_{-0.72}$} & {-} \\
    {$w_{2}$}  & {-1.05} & {$-0.97^{+0.12}_{-0.13}$} & {$-0.97^{+0.27}_{-0.25}$} & {-} \\
    {$w_{3}$}  & {-1.05} & {$-0.99\pm 0.26$} & {$-0.99^{+0.51}_{-0.52}$} & {-} \\
    {$w_{4}$}  & {-1.21} & {$-1.17^{+0.37}_{-0.24}$} & {$-1.17^{+0.58}_{-0.63}$} & {-} \\
    {$w_{5}$}  & {-0.94} & {$-0.83^{+0.39}_{-0.34}$} & {$-0.83^{+0.67}_{-0.69}$} & {-} \\
    {$w_{6}$}  & {-1.10} & {$-1.21^{+0.72}_{-0.52}$} & {$-1.21^{+1.10}_{-1.30}$} & {-} \\ \midrule
    $\Omega_{m}$ & {0.333} & {$0.321^{+0.019}_{-0.023}$} & {$0.321^{+0.052}_{-0.041}$} & {$0.315\pm 0.013$} \\
    $\Omega_{\Lambda}$ & {0.665} & {$0.677^{+0.023}_{-0.019}$} & {$0.677^{+0.041}_{-0.053}$} & {$0.685\pm 0.013$} \\
    $\bar{\chi}^{2}$  & {11268.7} & {$11288.4^{+5.5}_{-8.1}$} & {$11290^{+14}_{-13}$} & {-} \\ \bottomrule
\end{tabular}
\caption{Marginalized estimates from CMB dataset. Best fit, $68\%$ and $95\%$ confidence levels are displayed. In these estimates, PCA method has already been applied in order to obtain uncorrelated $w_i$'s. For comparison, Planck constraints for $\Lambda$CDM model at $1\sigma$ \cite{planck2} are also displayed.}
\label{table_CMB}
\end{table}

\section{BAO+Clocks+CMB+Supernovae}

\begin{figure}[h!]
\centering
\includegraphics[width=1.0\textwidth]{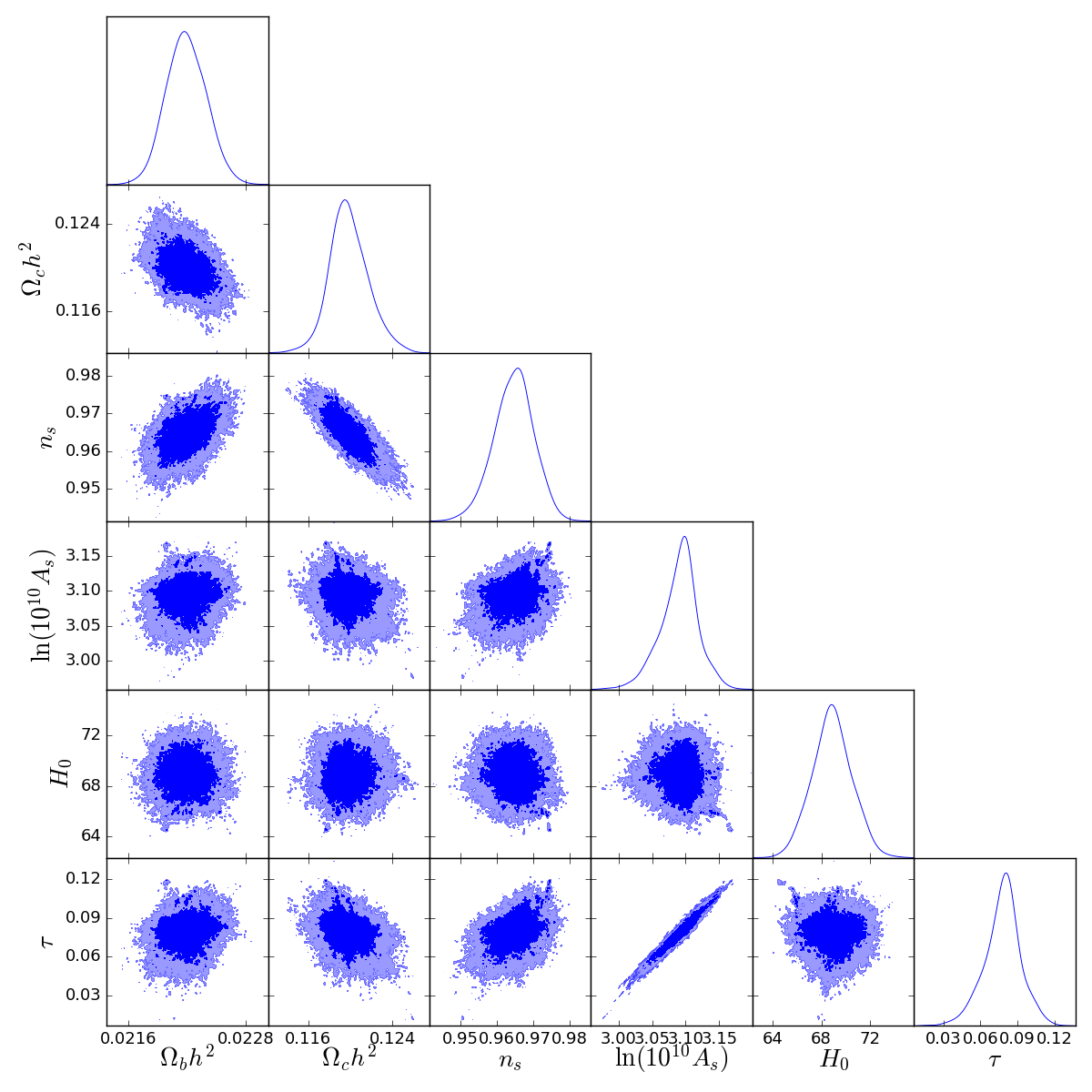}
\caption{1D and 2D cosmological parameter constraints from  BAO, cosmic clocks, CMB and supernovae datasets combination. $68\%$ and $95\%$ confidence levels are displayed.}
\label{BAO_CC_CMB_SN_cosmo}
\end{figure}

\begin{figure}[h!]
\centering
\includegraphics[width=1.0\textwidth]{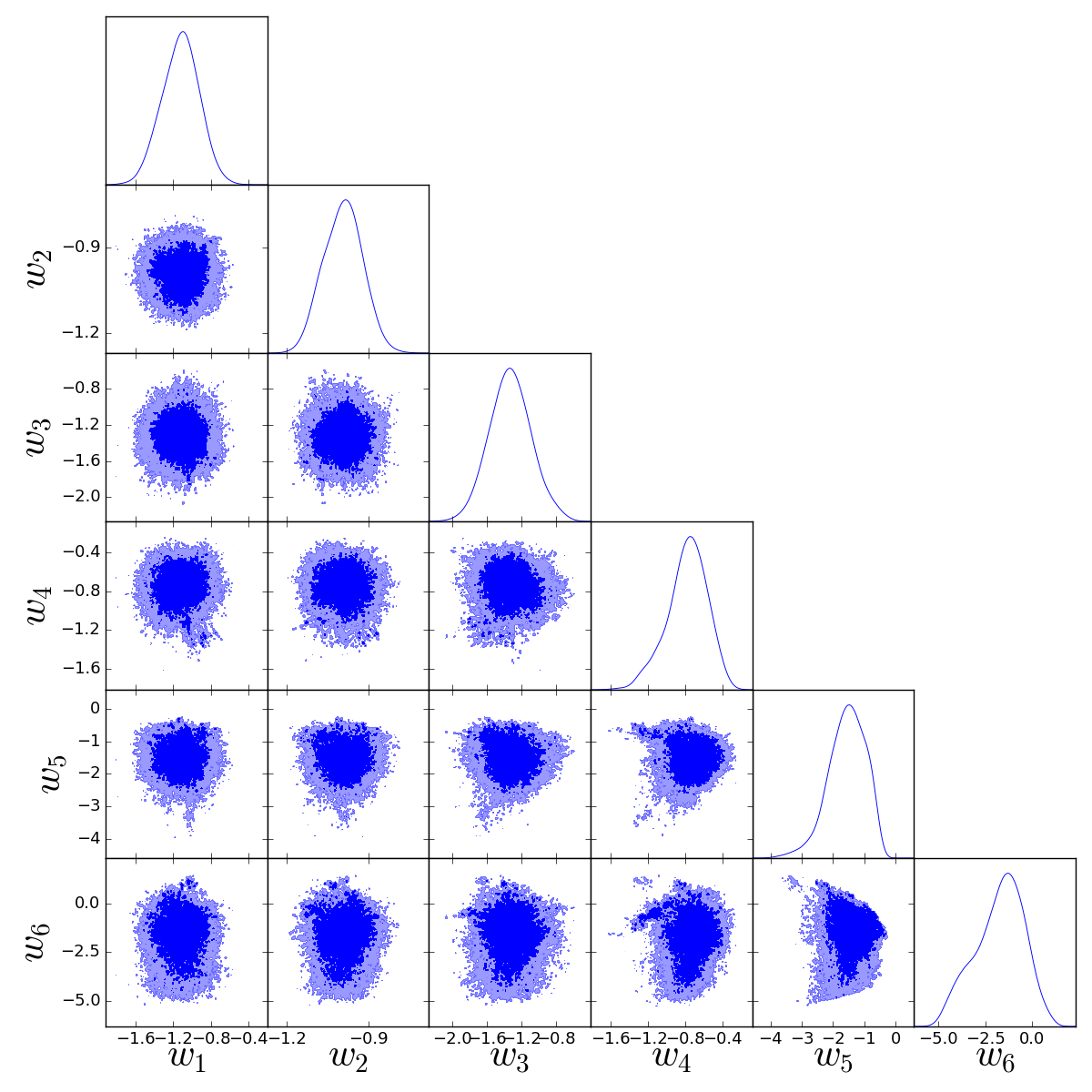}
\caption{1D and 2D dark energy equation of state parameter constraints from BAO, cosmic clocks, CMB and supernovae datasets combination. $68\%$ and $95\%$ confidence levels are displayed.}
\label{BAO_CC_CMB_SN_wis}
\end{figure}

\begin{figure}[h!]
\centering
\includegraphics[width=0.8\textwidth]{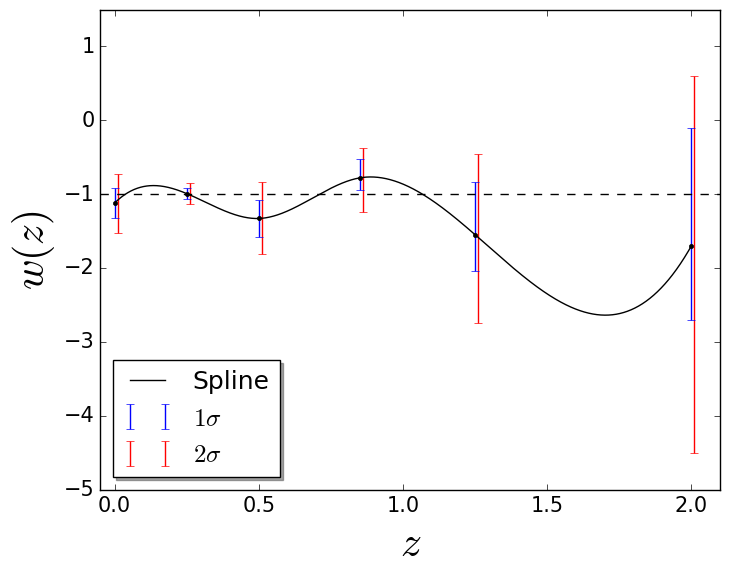}
\caption{Uncorrelated constraints on the dark energy equation of state parameters derived from BAO, cosmic clocks, CMB and supernovae datasets combination. $68\%$ and $95\%$ confidence levels are displayed. Reconstructed $w(z)$ is represented by the dark line using cubic spline interpolation. For better visualization, $95\%$ confidence levels are slightly shifted to the right.}
\label{w_plot_all_data}
\end{figure}

In the main analysis, we combine all datasets: BAO, cosmic clocks, CMB and supernovae. Our parameter estimates are displayed in Fig. \ref{BAO_CC_CMB_SN_cosmo}, Fig. \ref{BAO_CC_CMB_SN_wis}, Fig. \ref{w_plot_all_data} and Table \ref{table_BAO_CC_CMB_SN}. This sampling has converged more than the one from the previous section, although it still has not fully converged yet. As one can see, all cosmological parameters are constrained and consistent with results from Planck \cite{planck2} at $1\sigma$ (Fig. \ref{BAO_CC_CMB_SN_cosmo} and Table \ref{table_BAO_CC_CMB_SN}). Dark energy equation of state parameters are better constrained than in the previous section because we are now using data (BAO, Clocks, Supernovae) that are also able to constrain parameters at lower redshifts, which can be seen in Fig. \ref{BAO_CC_CMB_SN_wis} and Fig. \ref{w_plot_all_data}. On the other hand, this time not all $w_i$'s agree with $\Lambda$CDM model at $1\sigma$: there is a small deviation in $w_3$ and $w_4$, which agree with $\Lambda$CDM model only at $2\sigma$. We believe that this deviation happens because sampling has not fully converged and therefore using a longer chain would solve the problem. As expected, $w_2$ is the dark energy equation of state parameter with the smallest error bar, which is explained due to the fact that the greatest amount of information is around the corresponding redshit ($z=0.25$). Another interesting fact is that even Planck data has not enough power to constrain the last dark energy equation of state parameter, $w_6$, inside our prior (in the last section, apparently all $w_i$'s were constrained, but now we know that this was just because of the lack of convergence). This can be seen in Fig. \ref{BAO_CC_CMB_SN_wis} as a ``cut'' in the 2D plot of $w_5\, \times\,w_6$, which prevents it of forming a ``smooth edge''.

\begin{table}[h!]
\centering \small
\begin{tabular}{SSSSSSS} \toprule
    {Parameter} & {Best fit} & {$68\%$ limits} & {$95\%$ limits} & {Planck (TT+lowP)} \\ \midrule
    $\Omega_{b}h^{2}$ & {0.02215} & {$0.02219\pm 0.00020$} & {$0.02219^{+0.00041}_{-0.00038}$} & {$0.02222\pm 0.00023$}  \\
    $\Omega_{c}h^{2}$  & {0.1199} & {$0.1199\pm 0.0019$} & {$0.1199^{+0.0039}_{-0.0039}$} & {$0.1197\pm 0.0022$} \\
    $n_{s}$  & {0.9639} & {$0.9645\pm 0.0055$} & {$0.965^{+0.011}_{-0.011}$} & {$0.9655\pm 0.0062$} \\
    ln${(10^{10}A_{s})}$ & {3.091} & {$3.094^{+0.024}_{-0.018}$} & {$3.094^{+0.044}_{-0.053}$} & {$3.089\pm 0.036$} \\
    $H_{0}$  & {70.2} & {$69.0\pm 1.7$} & {$69.0^{+3.2}_{-3.3}$} & {$67.31\pm 0.96$} \\
    $\tau$  & {0.079} & {$0.079^{+0.013}_{-0.010}$} & {$0.079^{+0.024}_{-0.028}$} & {$0.078\pm 0.019$} \\
    {$w_{1}$}  & {-0.94} & {$-1.12\pm 0.20$} & {$-1.12^{+0.39}_{-0.40}$} & {-} \\
    {$w_{2}$}  & {-0.958} & {$-0.993\pm 0.072$} & {$-0.99^{+0.14}_{-0.14}$} & {-} \\
    {$w_{3}$}  & {-1.42} & {$-1.33\pm 0.25$} & {$-1.33^{+0.50}_{-0.48}$} & {-} \\
    {$w_{4}$}  & {-0.75} & {$-0.78^{+0.25}_{-0.17}$} & {$-0.78^{+0.41}_{-0.46}$} & {-} \\
    {$w_{5}$}  & {-1.55} & {$-1.55^{+0.71}_{-0.49}$} & {$-1.5^{+1.1}_{-1.2}$} & {-} \\
    {$w_{6}$}  & {-1.13} & {$-1.7^{+1.6}_{-1.0}$} & {$-1.7^{+2.3}_{-2.8}$} & {-} \\ \midrule
    $\Omega_{m}$ & {0.288} & {$0.299\pm 0.015$} & {$0.299^{+0.030}_{-0.029}$} & {$0.315\pm 0.013$}  \\
    $\Omega_{\Lambda}$ & {0.710} & {$0.699\pm 0.015$} & {$0.699^{+0.029}_{-0.030}$} & {$0.685\pm 0.013$} \\
    $\bar{\chi}^{2}$  & {11756.3} & {$11779.1^{+6.1}_{-8.6}$} & {$11780^{+15}_{-14}$} & {-} \\ \bottomrule
\end{tabular}
\caption{Marginalized estimates from BAO+CC+CMB+SN datasets. Best fit, $68\%$ and $95\%$ confidence levels are displayed. In these estimates, PCA method has already been applied in order to obtain uncorrelated $w_i$'s. For comparison, Planck constraints for $\Lambda$CDM model at $1\sigma$ \cite{planck2} are also displayed.}
\label{table_BAO_CC_CMB_SN}
\end{table}

\section{Comparison of datasets, results and models}

\begin{figure}[h!]
\centering
\includegraphics[width=1.0\textwidth]{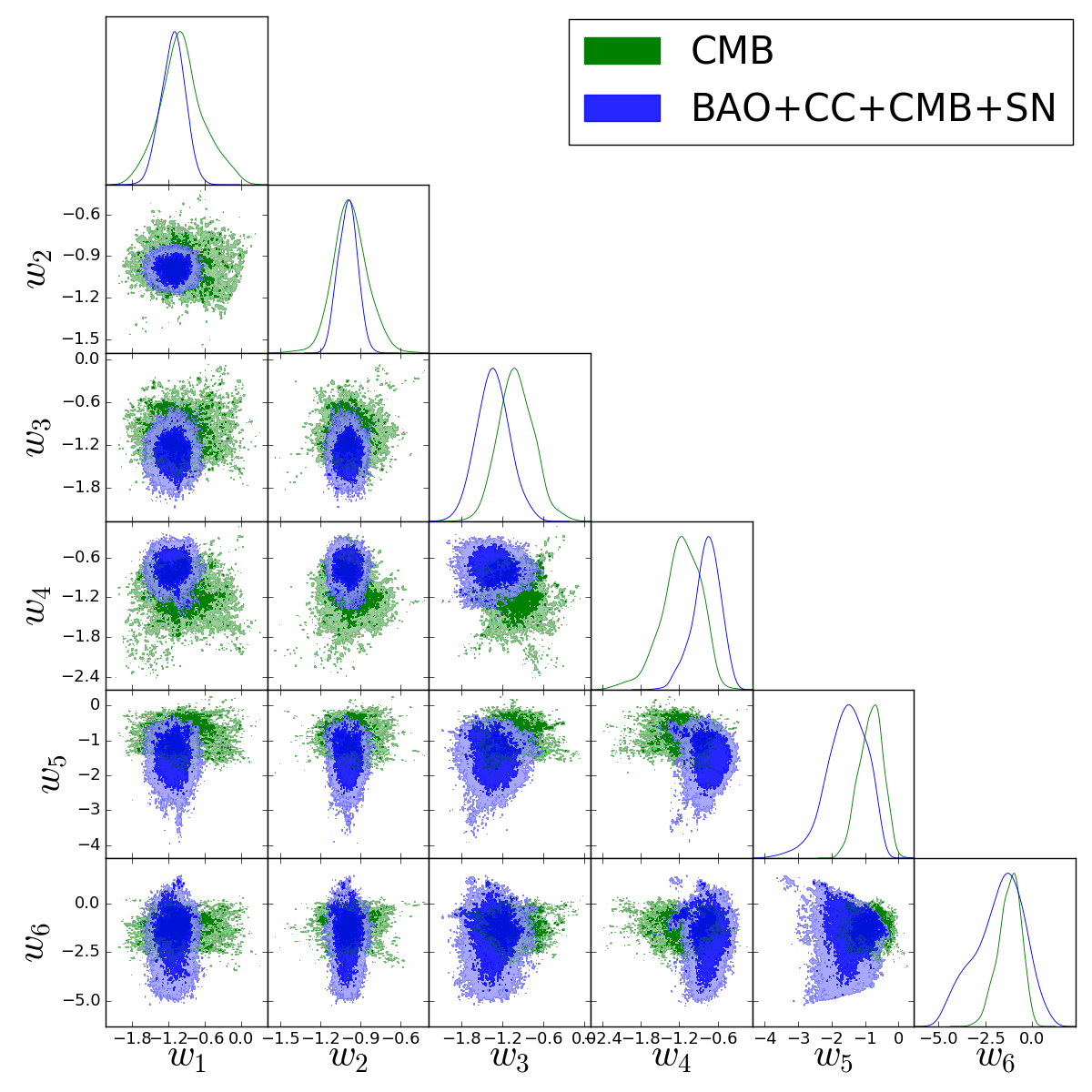}
\caption{Comparison of uncorrelated constraints on the dark energy equation of state parameters for two cases: BAO, cosmic clocks, CMB and supernovae datasets combination (blue) and CMB data alone (green). $68\%$ and $95\%$ confidence levels are displayed.}
\label{comparison_wis}
\end{figure}

As pointed out in the last chapter, the importance of using different datasets is that only the combination of independent data breaks degeneracies and allows us to better constrain parameters. This is very clear when we look at Fig. \ref{comparison_wis}, which shows that the combination of all datasets improves the precision of parameter estimation (for example, in the 2D plot $w_1\, \times\,w_2$ the area estimated using BAO+CC+CMB+SN is smaller than the area estimated using just CMB data alone). We would expect to see the same area in the 2D plot $w_5\, \times\,w_6$ for both BAO+CC+CMB+SN and CMB data alone because at higher redshits, as we have seen, BAO+CC+SN does not have enough power to constrain $w_i$'s. Unfortunately, as both chains are not fully converged, we do not see this happening. The important thing is that BAO+CC+SN combination constrains $w_i$'s better at lower redshifts, while CMB data constrains $w_i$'s better at higher redshifts.

\begin{figure}[h!]
\centering
\includegraphics[width=1.0\textwidth]{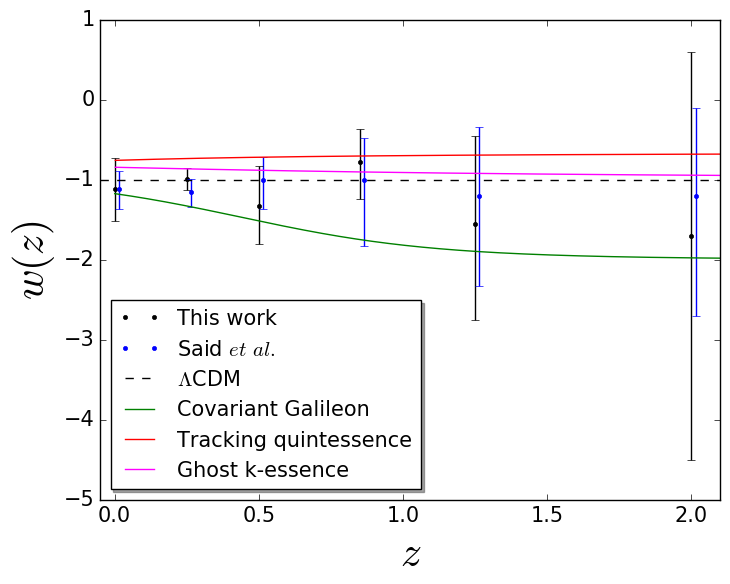}
\caption{Uncorrelated constraints on the dark energy equation of state parameters and some theoretical models for $w(z)$. Black dots and black error bars refer to our results derived from BAO, cosmic clocks, CMB and supernovae datasets combination while blue dots and blue error bars represent Said \textit{et al.} \cite{Said2013} main results. For both, $95\%$ confidence levels are displayed. The theoretical models for $w(z)$ plotted are $\Lambda$CDM model, covariant Galileon model \cite{Tsujikawa2012}, tracking quintessence with power-law potential model \cite{Ratra1988} and dilatonic ghost condensate model (k-essence) \cite{Piazza2004}. Some of these models deviate from our results (and from Said \textit{et al.}) at more than $2\sigma$ confidence levels, indicating possible problems in the theory. For better visualization, results from Said \textit{et al.} are slightly shifted to the right.}
\label{w_plot_Said_models}
\end{figure}

To picture the implications of our results, in Fig. \ref{w_plot_Said_models} we display our estimations of $w_i$'s constraints at $2\sigma$ along with results from Said \textit{et al.} \cite{Said2013}, also at $2\sigma$, and a few alternative models to $\Lambda$CDM. Said \textit{et al.} paper uses the following datasets combination: WMAP(Wilkinson Microwave Anisotropy Probe \cite{Komatsu2011})+SNLS(Supernova Legacy Survey \cite{Riess_2004})+BAO(6 degree Field Galaxy Redshift Survey at $z=0.1$ \cite{Beutler2012} and WiggleZ at $z=0.44, 0.60, 0.73$ \cite{Blake2012})+$H(z)$(dataset from Moresco \textit{et al.} \cite{Moresco_2012}). The constraints from Said \textit{et al.} seem better than our results, despite the fact that the datasets used in our work are more recent (and more precise) and should, at first sight, better constrain the $w_i$ parameters. The only explanation we could think of is that the parametrization for dark energy equation of state may be somehow affecting the results, because while we use a cubic spline interpolation for connecting the $w_i$'s Said \textit{et al.} uses hyperbolic tangent functions for this purpose. Although we have this difficulty, the results from both estimations can be used to rule out or indicate that theoretical models for $w(z)$ present in the literature might be wrong, as stated in Fig. \ref{w_plot_Said_models}.

%% file: Conclusions.tex
\newpage
\chapter*{Conclusions}
\markboth{CONCLUSIONS}{}
\addcontentsline{toc}{chapter}{Conclusions}

The main objective of this work was to analyze how current observations coming from baryon acoustic oscillations, cosmic chronometers, cosmic microwave background anisotropies and type Ia supernovae could better constrain dark energy equation of state and maybe indicate deviations from $\Lambda$CDM model, since although this one seems the best theory to explain different observations, it suffers from intrinsic problems that lack of a consistent explanation.

In the fist chapter, we made a review of the background cosmology necessary to understand the basis of the theory. In the second chapter, we described in details the datasets used in our work and the theory behind them in order to picture how data are obtained and what kind of information can be extracted from them. In the third chapter, we concerned about statistical tools to perform the analyses and in the final chapter we explored the results obtained and its consequences to cosmology.

The analysis of the datasets could show us the effect of certain kinds of experiments in parameter estimation. BAO, cosmic chronometers and supernovae data have the power to constrain parameters of low to intermediate redshift, while CMB data constrain parameters better at higher redshifts. However, we did not have very good results for high redshifts even using Planck data, probably from the fact that dark energy does not have much influence in the early Universe in which CMB photons dominate. Maybe combination with other datasets that possess more information at high redshifts (e.g. quasars, redshift space distortion data, CMB lensing) could improve the constrains.

The results obtained in our parameter estimations using MCMC sampling were strongly influenced by the level of convergence achieved in each combination of datasets. The most reliable analysis is the one performed using BAO+CC+SN, because the Gelman-Rubin criterion adopted is very good ($\hat{R}<1.01$) and the corner plot is visually acceptable. However, the other two analyses might be biased in a negative way because Gelman-Rubin criterion adopted is not good and corner plot is  obviously not very pleasant, for both CMB data alone and BAO+CC+CMB+SN. Despite this problem, it is customary to say that new physics might be presented when we have at least three or four sigmas of discrepancy, and therefore even if our results are biased we do not see possibility for new physics, which tells us that concordance with $\Lambda$CDM still remains.

Our work showed the inherent difficulties related to studying dark energy equation of state. Besides the fact that the datasets used do not have enough power to constrain $w(z)$ completely, we also need to face the problem of the computational cost of using MCMC methods for sampling, which are generally very expensive in cosmology (in our case because CMB theory generated by CAMB code is extremely complex) and generally demand the use of parallel programming with the help of large computer clusters. As an example, in the sampling with BAO+CC+CMB+SN each MCMC point took 3.5 seconds on average to be generated in the fastest computer we used in this work (CPU: Intel Core i7, cores: 8, memory: SSD 16GB), while we would probably need at least $\mathcal{O}(3\times 10^6)$ points to finally achieve convergence. The effect of these problems are seen as huge error bars in the dark energy equation of state parameters (the last $w_i$'s), turning the evolution of the equation of state very hard to predict.

Using \textit{emcee} instead of standard parameter estimations codes for cosmology, such as CosmoMC and Monte Python, was made with the intuit of confirming that we could recover old results even with a different sampler. This was successfully achieved, since we tested simple models such as $\Lambda$CDM and our cosmological parameter estimation really matched results found in literature. This is important to give us much more confidence in theses codes, which are generally used as a ``black box'' in most applications to cosmology.

Due to the computational cost problem, it is worth mentioning some alternative tools to MCMC sampling as possible extensions to this work. One of them is \textit{Gaussian Processes} \cite{GP}, which provides a surprisingly computationally effective method to reconstruct functions (this has already been applied to $w(z)$ \cite{Seikel2012,Wang2017}) from data at low computational cost. The other one is \textit{Variational Inference} \cite{Blei2017}, a tool which has been tested in astrophysics problems \cite{Jain2018} and converts the inference problem into an optimization problem, performing parameter estimation much faster than standard MCMC methods.

As a way of helping cosmology students interested in MCMC methods, all the codes developed in this dissertation can be found at \textit{Github} in reference \cite{git_will}.